\documentclass[journal]{IEEEtran}

\usepackage{amsmath,amsfonts,amssymb}
\usepackage{bm}
\usepackage{array}
\usepackage{booktabs}
\usepackage[caption=false,font=normalsize,labelfont=sf,textfont=sf]{subfig}

\usepackage{textcomp}
\usepackage{stfloats}
\usepackage{url}
\usepackage{graphicx}
\usepackage{tikz}
\usepackage{orcidlink}
\usepackage{hyperref}
\input{xr-supplementary}
\newcommand{\sref}[1]{\ref*{S-#1}}
\usetikzlibrary{arrows.meta,calc}
\colorlet{bodytop}{black!20}
\colorlet{bodyfront}{black!35}
\colorlet{bodyside}{black!45}
\colorlet{bodybite}{black!30}
\colorlet{bodybiteedge}{black!60}
\newcommand{\bodysolid}{%
  \fill[bodyfront] (-1.059,-0.444) -- (0.641,-0.444) -- (0.641,0.151) -- (-1.059,0.151) -- cycle;
  \fill[bodyside] (0.641,-0.444) -- (1.059,-0.151) -- (1.059,-0.092)
    -- (1.016,-0.111) -- (0.975,-0.11) -- (0.937,-0.087) -- (0.904,-0.043) -- (0.877,0.019) -- (0.856,0.097) -- (0.844,0.189) -- (0.84,0.29)
    -- (0.641,0.151) -- cycle;
  \fill[bodytop] (-1.059,0.151) -- (0.641,0.151) -- (0.84,0.29)
    -- (0.739,0.293) -- (0.651,0.302) -- (0.579,0.316) -- (0.525,0.335) -- (0.492,0.358) -- (0.48,0.385) -- (0.491,0.414) -- (0.523,0.444)
    -- (-0.641,0.444) -- cycle;
  \fill[bodybite] (0.84,0.29)
    -- (0.739,0.293) -- (0.651,0.302) -- (0.579,0.316) -- (0.525,0.335) -- (0.492,0.358) -- (0.48,0.385) -- (0.491,0.414) -- (0.523,0.444)
    -- (0.534,0.339) -- (0.564,0.239) -- (0.614,0.146) -- (0.68,0.065) -- (0.761,-0.001) -- (0.854,-0.051) -- (0.954,-0.081) -- (1.059,-0.092)
    -- (1.016,-0.111) -- (0.975,-0.11) -- (0.937,-0.087) -- (0.904,-0.043) -- (0.877,0.019) -- (0.856,0.097) -- (0.844,0.189) -- cycle;
  \draw[line join=round] (-1.059,-0.444) -- (0.641,-0.444) -- (1.059,-0.151) -- (1.059,-0.092);
  \draw[line join=round] (-1.059,-0.444) -- (-1.059,0.151) -- (0.641,0.151) -- (0.641,-0.444);
  \draw[line join=round] (-1.059,0.151) -- (-0.641,0.444) -- (0.523,0.444);
  \draw[line join=round] (0.641,0.151) -- (0.84,0.29);
  \draw (0.84,0.29) -- (0.739,0.293) -- (0.651,0.302) -- (0.579,0.316) -- (0.525,0.335) -- (0.492,0.358) -- (0.48,0.385) -- (0.491,0.414) -- (0.523,0.444);
  \draw (1.059,-0.092) -- (1.016,-0.111) -- (0.975,-0.11) -- (0.937,-0.087) -- (0.904,-0.043) -- (0.877,0.019) -- (0.856,0.097) -- (0.844,0.189) -- (0.84,0.29);
  \draw[bodybiteedge] (0.523,0.444) -- (0.534,0.339) -- (0.564,0.239) -- (0.614,0.146) -- (0.68,0.065) -- (0.761,-0.001) -- (0.854,-0.051) -- (0.954,-0.081) -- (1.059,-0.092);
}

\usepackage[backend=biber,style=ieee,citestyle=ieee-comp,date=year]{biblatex}
\graphicspath{{figures/}}

\newcommand{\ii}{\mathrm{i}}               %
\newcommand{\eps}{\varepsilon_\alpha}      %
\newcommand{\nua}{\nu_\alpha}              %
\newcommand{\zs}{\zeta_\sigma}             %
\newcommand{\zm}{\zeta_\mu}                %
\newcommand{\vE}{\bm{E}}
\newcommand{\vH}{\bm{H}}
\newcommand{\vB}{\bm{B}}
\newcommand{\vJ}{\bm{J}}
\newcommand{\vF}{\bm{F}}
\newcommand{\vU}{\bm{U}}
\newcommand{\vf}{\bm{f}}                   %
\newcommand{\vw}{\bm{w}}                   %
\newcommand{\vx}{\bm{x}}                   %
\newcommand{\vy}{\bm{y}}                   %
\newcommand{\vz}{\bm{z}}                   %
\newcommand{\vxi}{\bm{\xi}}                %
\newcommand{\vm}{\bm{m}}                   %
\newcommand{\vp}{\bm{p}}                   %
\newcommand{\vd}{\bm{d}}                   %
\newcommand{\va}{\bm{a}}                   %
\newcommand{\ehat}{\hat{\bm{e}}}           %
\newcommand{\nhat}{\hat{\bm{n}}}

\newcommand{\cJ}{\bm{\mathcal{J}}}         %
\newcommand{\cK}{\bm{\mathcal{K}}}         %
\newcommand{\Tens}{\bm{T}}                 %
\newcommand{\Prop}{\bm{P}}                 %
\newcommand{\Ret}{\bm{R}}                  %

\newcommand{\Id}{\bm{I}}
\newcommand{\sigs}{\sigma_\mathrm{s}}
\newcommand{\mur}{\mu_\mathrm{r}}
\newcommand{\murs}{\mu_{\mathrm{r,s}}}
\newcommand{\murb}{\mu_{\mathrm{r},*}}
\newcommand{\gams}{\gamma_\mathrm{s}}
\newcommand{\dels}{\delta_\mathrm{s}}
\newcommand{\hs}{h_\mathrm{s}}
\newcommand{\ks}{k_\mathrm{s}}
\newcommand{\Mw}{\mathbf{M}}               %
\newcommand{\Nw}{\mathbf{N}}               %
\newcommand{\TE}{\mathrm{TE}}
\newcommand{\TM}{\mathrm{TM}}
\newcommand{\air}{\mathcal{A}}
\newcommand{\sea}{\mathcal{S}}
\newcommand{\dd}{\,\mathrm{d}}
\newcommand{\curl}{\nabla\!\times}          %
\newcommand{\curlx}{\nabla_{\!x}\!\times}      %
\newcommand{\Jimp}[1][]{\bm{J}_{\mathrm{x}\if\relax\detokenize{#1}\relax\else,#1\fi}}
\newcommand{\Mmat}[1][]{\bm{M}_{\mathrm{x}\if\relax\detokenize{#1}\relax\else,#1\fi}}
\newcommand{\Hcurl}{H(\mathrm{curl})}

\begin{document}

\title{A Diffusive T-Matrix Unifying the Electric and Magnetic Polarizability Tensors in Conducting Layered Media}

\author{Fraser~Williams\orcidlink{0000-0002-5422-2827}, Dhammika~Jayalath\orcidlink{0000-0001-6130-0275}, Clinton~Fookes\orcidlink{0000-0002-8515-6324}, Ajay~Pandey\orcidlink{0000-0002-6599-745X}%
\thanks{This work has been submitted to the IEEE for possible publication. Copyright may be transferred without notice, after which this version may no longer be accessible.}}

\markboth{Williams \emph{et al.}: A Diffusive T-Matrix Unifying the Electric and Magnetic Polarizability Tensors}%
{Williams \emph{et al.}: A Diffusive T-Matrix Unifying the Electric and Magnetic Polarizability Tensors}

\maketitle

\begin{abstract}
Electromagnetic induction is used to detect and classify buried and submerged objects, such as unexploded ordnance in soil and on the seabed. A small object is commonly described by its magnetic polarizability tensor (MPT), which is independent of its position and of the sensor. The MPT assumes that the surrounding medium does not conduct. In seawater at extremely low frequency, or in clay at kilohertz frequencies, this assumption fails at depths comparable to the skin depth. The object then also diverts the current induced in the medium, a galvanic response that the MPT omits. We describe the object instead by its transition matrix (T-matrix), which keeps the independence from position and sensor and includes both responses. It is computed once per medium and frequency from a single finite element problem, for any shape and materials. Its inductive block reduces to the MPT when the medium conducts weakly, and its galvanic block reduces to the electric polarizability tensor in the static limit. We test the model against full finite element solutions for a conducting box, non-magnetic and permeable, in a layered sea and in clay over bedrock. Two orders beyond the dipole, it is accurate to within $1.4\%$ over two decades of frequency and for sizes up to $0.7$ of the depth, whereas the error of the MPT carried into the conducting medium reaches $91\%$. Blocks coupling the two responses, contained in neither tensor, contribute up to $13\%$.
\end{abstract}


\section{Introduction}
\label{sec:intro}
\IEEEPARstart{E}{lectromagnetic} induction is widely used to detect and characterise objects whose electromagnetic properties differ from those of their surroundings. A transmitter drives a low-frequency magnetic field, the object perturbs it through its eddy currents and magnetisation, and a receiver measures the perturbation, which depends on the instrument and its position as well as on the object. For a small object, the description of the object alone is the magnetic polarizability tensor (MPT), the leading term of an asymptotic expansion in the size of the object~\cite{ammariTargetDetectionCharacterization2014}, whose higher-order terms are carried by generalised tensors~\cite{ledgerGeneralisedMagneticPolarizability2018}. The MPT depends on the shape, materials and frequency of the object, but not on its position or on the coils used to sense it~\cite{ledgerCharacterizingShapeMaterial2015}. Its spectral signature has been used to classify objects in metal detection and security screening~\cite{ledgerIdentificationMetallicObjects2021,makkonenKNNClassificationMetallic2014}, and related dipole descriptions are used to discriminate unexploded ordnance~\cite{pasionDiscriminationAlgorithmUXO2001}. Many of these objects lie in soil or seawater, whose conductivity can alter the measured response~\cite{sanfilipoEffectsSeawaterEMI2005}, yet the theory behind the tensor assumes a host that does not conduct.

The MPT has been extended to a conducting host by carrying the free-space tensor into the Green's function of a conducting half space~\cite{ledgerCharacterisingBuriedObjects2025} and of a layered sea~\cite{songTransientElectromagneticScattering2016}. Geophysics, where the host is always a conductor, shows that these extensions capture only part of the response. An object in a conducting host responds to a transmitter in two ways~\cite{westPHYSICSELECTROMAGNETICINDUCTION1991}: inductively, through eddy currents circulating within the object and through its magnetisation, and galvanically, by diverting the current that flows in the host. The extensions of the MPT describe the object by a magnetic dipole~\cite{ledgerCharacterisingBuriedObjects2025,songTransientElectromagneticScattering2016}, and so retain only the first. Geophysical models retain both, treating a compact conductor in a layered ground~\cite{wardElectromagneticTheoryGeophysical1988} by integral equations for the full current it scatters~\cite{hohmannThreedimensionalInducedPolarization1975,wannamakerElectromagneticModelingThreedimensional1984}, but each source requires a new solution, which gives up the source independence that makes the MPT useful. Approximations that do not depend on the source~\cite{westSimpleParametricModel1985,nortonEddycurrentCurrentchannelingResponse2005} neglect the induction of the host across the object. The galvanic response has a tensor of its own: the perturbation of a steady current by a conducting object is described by the polarization tensor of Pólya and Szegő and its generalisations~\cite{polyaIsoperimetricInequalitiesMathematical1951,ammariPolarizationMomentTensors2007}, which are used to classify objects by electrolocation~\cite{ammariShapeRecognitionClassification2014}. In a host that both conducts and induces, neither tensor alone describes the object.

A description that is independent of the source and yet retains the host in full exists in wave scattering, where an object is characterised by its transition matrix, or T-matrix~\cite{watermanMatrixFormulationElectromagnetic1965,mishchenkoScatteringAbsorptionEmission2002}. It can be computed by finite elements~\cite{demesyScatteringMatrixArbitrarily2018}, and with Sommerfeld integrals it describes an object near a planar interface or between layers through the reflection of the object's own field~\cite{egelSMUTHIPythonPackage2021}. Kristensson and Ström formulated the T-matrix of an inhomogeneity beneath a planar interface~\cite{kristenssonScatteringBuriedInhomogeneities1978,kristenssonElectromagneticScatteringBuried1980}, and Kristensson extended it to a lossy ground~\cite{kristenssonElectromagneticScatteringBuried1979}, but by the null-field method, which limits it to objects built from homogeneous regions~\cite{kristenssonScatteringElectromagneticWaves2016}. All of this work concerns propagating waves, in which the displacement current dominates and the object can resonate. The opposite limit arises at extremely low frequency (ELF) in seawater, or at kilohertz frequencies in clay, where the conduction current far exceeds the displacement current. Fields then diffuse over the skin depth $\delta$, no object resonates, and the air above is static. Two ratios govern an object of size $\alpha$ at depth $D$: $D/\delta$ sets the share of the galvanic response, and $\alpha/D$ the weight of the higher-order terms and of the field reflected by the layers. We consider objects for which neither is small, at depths comparable to the skin depth, such as $50$\,m in seawater at $10$\,Hz, and with sizes a few tenths of the depth.

In this paper we carry the T-matrix into the diffusive limit for an object in a layered conducting ground. Expressed in the transverse-electric (TE) and transverse-magnetic (TM) modes of the host, its diagonal blocks are the inductive and galvanic responses. In a weakly conducting host the TE block reduces to the MPT, and in the static limit the TM block reduces to the electric polarizability tensor, so that the two tensors developed separately, with their generalisations, are limits of one T-matrix. Both responses scale as the cube of the size of the object, but within a skin depth the galvanic one decays one power of distance more slowly, so its share is set by the depth against the skin depth and not by the size. The off-diagonal blocks couple the two responses and belong to neither tensor; for an object with a centre of symmetry they make up the whole of the first correction beyond the dipole. The T-matrix is computed once per object, host and frequency from a single finite element problem and a reciprocity integral, for highly conducting, permeable and spatially varying materials of arbitrary shape. The point source of the host, carried through the layers, then gives the field at the sensor, by reciprocity the drive from the transmitter, and the reflection of the object's own field by the layers.

Section~\ref{sec:model} sets out the problem and the structure of the model, Section~\ref{sec:tensor} defines and computes the T-matrix, and Section~\ref{sec:surface} carries it through the layers. Section~\ref{sec:properties} gives its limits, cross blocks and range laws, and Section~\ref{sec:numerics} compares the model with full finite element solutions.

\section{Problem and Structure of the Model}
\label{sec:model}

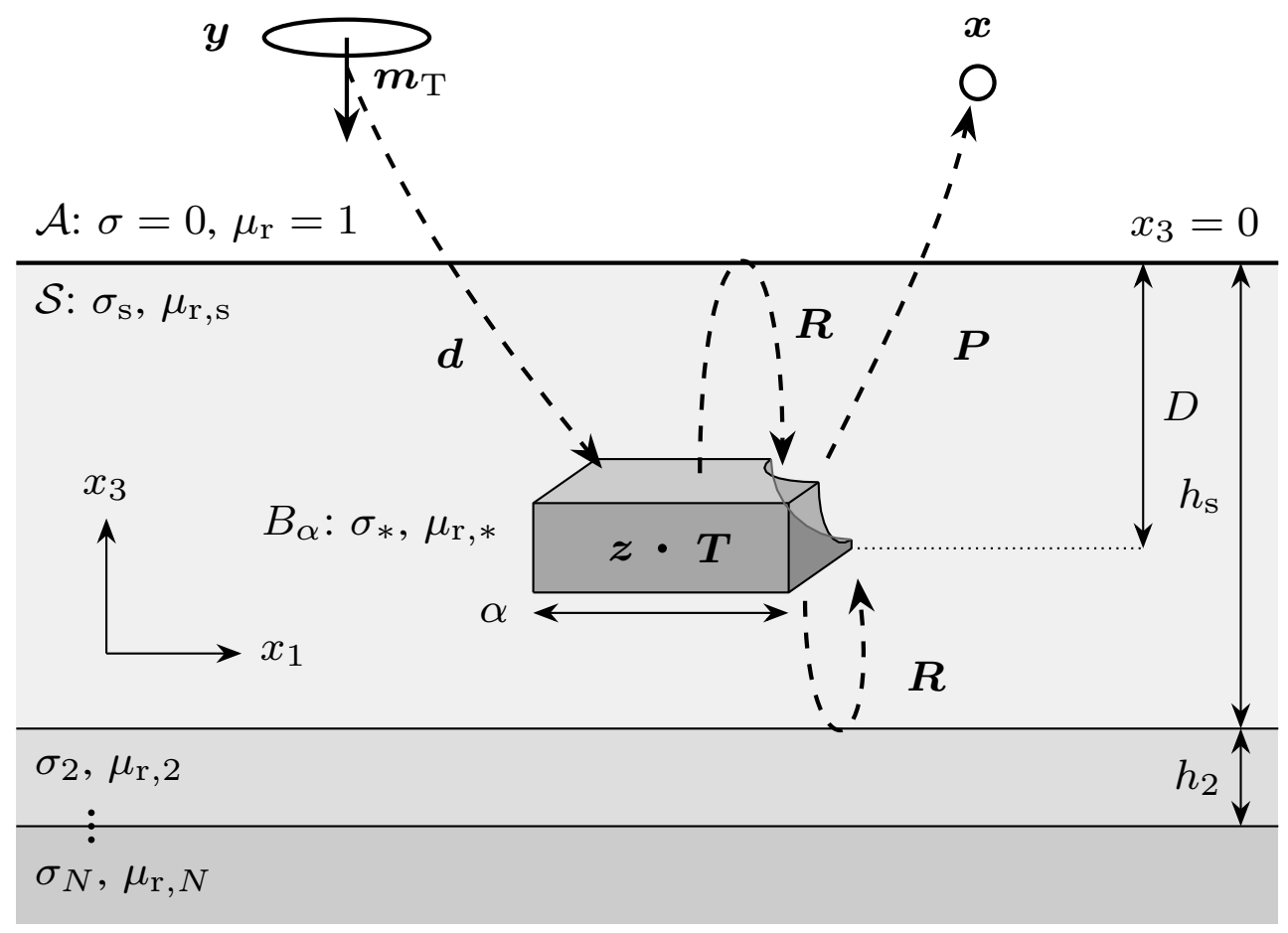
\begin{figure}[!t]
\centering
\colorlet{sceneair}{white}
\colorlet{scenehost}{black!6}
\colorlet{scenelayer}{black!13}
\colorlet{scenebase}{black!20}
\colorlet{sceneinterface}{black}
\colorlet{scenedrive}{black}
\colorlet{sceneprop}{black}
\colorlet{sceneret}{black}
\begin{tikzpicture}[>=Stealth, x=1cm, y=1cm, font=\footnotesize]
  \fill[sceneair]   (-4.2,0) rectangle (4.2,1.75);
  \fill[scenehost]  (-4.2,-3.1) rectangle (4.2,0);
  \fill[scenelayer] (-4.2,-3.75) rectangle (4.2,-3.1);
  \fill[scenebase]  (-4.2,-4.4) rectangle (4.2,-3.75);
  \draw[thick, sceneinterface] (-4.2,0) -- (4.2,0);
  \draw[sceneinterface] (-4.2,-3.1) -- (4.2,-3.1);
  \draw[sceneinterface] (-4.2,-3.75) -- (4.2,-3.75);
  \begin{scope}[sceneinterface]
    \node[anchor=south west] at (-4.2,0.02) {$\air$: $\sigma=0$, $\mur=1$};
    \node[anchor=north west] at (-4.2,-0.02) {$\sea$: $\sigs$, $\murs$};
    \node[anchor=south east] at (4.2,0.02) {$x_3=0$};
    \node[anchor=west] at (-4.2,-3.38) {$\sigma_2$, $\mu_{\mathrm{r},2}$};
    \foreach \y in {-3.66,-3.75,-3.84} \fill (-3.7,\y) circle (0.022);
    \node[anchor=west] at (-4.2,-4.12) {$\sigma_N$, $\mu_{\mathrm{r},N}$};
  \end{scope}
  \draw[<->] (3.95,0) -- (3.95,-3.1) node[midway, left] {$\hs$};
  \draw[<->] (3.95,-3.1) -- (3.95,-3.75) node[midway, left] {$h_2$};
  \draw[->] (-3.6,-2.6) -- (-3.6,-1.7) node[above] {$x_3$};
  \draw[->] (-3.6,-2.6) -- (-2.7,-2.6) node[right] {$x_1$};
  \draw[thick] (-2.0,1.5) ellipse (0.55 and 0.12);
  \draw[->, thick] (-2.0,1.5) -- (-2.0,0.8);
  \node[anchor=west] at (-1.95,1.2) {$\vm_\mathrm{T}$};
  \node[anchor=east] at (-2.65,1.5) {$\vy$};
  \draw[thick, fill=white] (2.2,1.2) circle (0.11);
  \node[anchor=south] at (2.2,1.38) {$\vx$};
  \begin{scope}[shift={(0.3,-1.75)}]
    \bodysolid
  \end{scope}
  \fill (0.1,-1.9) circle (0.03);
  \node[anchor=east] at (0.04,-1.92) {$\vz$};
  \node[anchor=west] at (0.2,-1.9) {$\Tens$};
  \node[anchor=east] at (-0.85,-1.75) {$B_\alpha$: $\sigma_*$, $\murb$};
  \draw[densely dotted] (1.4,-1.9) -- (3.3,-1.9);
  \draw[<->] (3.3,-1.9) -- (3.3,0) node[midway, right] {$D$};
  \draw[<->] (-0.76,-2.33) -- (0.94,-2.33);
  \node[anchor=east] at (-0.8,-2.33) {$\alpha$};
  \draw[->, thick, dashed, scenedrive]
    (-2.0,1.3) .. controls (-1.5,0.1) and (-0.8,-0.75) .. (-0.3,-1.37);
  \node[anchor=east, scenedrive] at (-1.1,-0.6) {$\vd$};
  \draw[->, thick, dashed, sceneprop]
    (1.2,-1.3) .. controls (1.55,-0.6) and (2.0,0.3) .. (2.15,1.05);
  \node[anchor=west, sceneprop] at (1.9,-0.55) {$\Prop$};
  \draw[->, thick, dashed, sceneret]
    (0.35,-1.4) .. controls (0.35,0.45) and (0.9,0.45) .. (0.9,-1.35);
  \node[anchor=west, sceneret] at (0.85,-0.4) {$\Ret$};
  \draw[->, thick, dashed, sceneret]
    (1.05,-2.25) .. controls (1.05,-3.4) and (1.55,-3.4) .. (1.4,-2.1);
  \node[anchor=west, sceneret] at (1.6,-2.75) {$\Ret$};
\end{tikzpicture}
\caption{Geometry and structure of the model. The object lies in the top layer of a horizontally layered medium. The transmitter's field reaches the object as the drive $\vd$, the object responds through the T-matrix $\Tens$, and the response reaches the sensor through the propagator $\Prop$. Its reflections at the surface and at the interfaces below return to the object through $\Ret$.}
\label{fig:scene}
\end{figure}

Fig.~\ref{fig:scene} shows the configuration. Positions are given in Cartesian coordinates $\vx=(x_1,x_2,x_3)$, with $x_3$ directed upward and the surface of the ground at $x_3=0$. The upper half space $\air=\{x_3>0\}$ is air. Below the surface lies a stack of $N$ horizontal layers, each homogeneous. The top layer $\sea=\{-\hs<x_3<0\}$ is the host, such as the sea or a layer of clay, of conductivity $\sigs$, relative permeability $\murs$ and thickness $\hs$. Layer $j=2,\dots,N$ has conductivity $\sigma_j$, relative permeability $\mu_{\mathrm{r},j}$ and thickness $h_j$, and the last is a half space. The object $B_\alpha\subset\sea$ is a bounded domain of arbitrary shape and characteristic size $\alpha$, centred at $\vz$. Its centre lies at a depth $D$ below the surface, so that $z_3=-D$, and at a height $\hs-D$ above the bottom of the host. Its conductivity $\sigma_*$ and relative permeability $\murb$ differ from those of $\sea$ and may vary within the object. The transmitter is a point magnetic dipole of moment $\vm_\mathrm{T}$ at $\vy\in\air$, and the sensor, which records the magnetic field, is at $\vx\in\air$. A single conducting half space is the case $N=1$, $\hs\to\infty$, to which each result below reduces.

Without the object, the transmitter produces the magnetic field $\vH_0(\vx)$, which includes the response of the layers. With the object centred at $\vz$, the field is $\vH_\alpha(\vx,\vz)$. The quantity modelled here is the perturbation due to the object,
\begin{equation}
  \Delta \vH(\vx,\vz) := \vH_\alpha(\vx,\vz) - \vH_0(\vx) .
  \label{eq:perturbation}
\end{equation}

\subsection{Governing Equations}
\label{sec:governing}

The transmitter operates at a single angular frequency $\omega$, and fields are written as $\operatorname{Re}\{\vF(\vx)e^{-\ii\omega t}\}$ and represented by their complex amplitudes $\vF$. For linear materials, with $\vB=\mu_0\mur\vH$ and $\vJ=\sigma\vE$, the ratio of displacement to conduction current is $\omega\epsilon/\sigma$, with $\epsilon$ the permittivity, which at extremely low frequency is of order $10^{-8}$ in seawater and $10^{-16}$ in steel. The displacement current is therefore neglected in all conductors; replacing $\sigma$ by $\sigma-\ii\omega\epsilon$ would restore it without changing the structure of the model. In every layer and in the object Maxwell's equations then reduce to the eddy-current equations
\begin{equation}
  \curl  \vE = \ii\omega\mu_0\mur \vH ,
  \qquad
  \curl  \vH = \sigma \vE ,
  \label{eq:eddy}
\end{equation}
whose solutions diffuse over the skin depth $\delta=\sqrt{2/\omega\mu_0\mur\sigma}$. In the air there is no conduction current, the displacement current is negligible because every range $R$ of interest satisfies $k_0R\ll1$ with $k_0=\omega/c$, and $\mur=1$, so that outside the transmitter
\begin{equation}
  \curl  \vH = 0 ,
  \qquad
  \nabla\cdot \vH = 0 ,
  \label{eq:air}
\end{equation}
and $\vH=-\nabla\chi$ with $\nabla^2\chi=0$. The skin depth in $\sea$, about $80$\,m in seawater at $10$\,Hz, is comparable to the size and depth of the object and the range of the sensor, so no term of \eqref{eq:eddy} may be neglected there. In the object it may be far smaller than the object, $8$\,mm in steel at the same frequency.

\subsection{Scaling to the Unit Object}
\label{sec:scaling}
About the centre of the object, the scaled coordinate $\vxi$ and its gradient are
\begin{equation}
  \vxi = \frac{\vx - \vz}{\alpha} ,
  \qquad
  \nabla_\xi = \alpha\,\nabla_x ,
  \label{eq:scaling}
\end{equation}
so that in the coordinate $\vxi$ the object occupies a unit-size domain $B$ centred at the origin. Hereafter $\nabla$ denotes $\nabla_\xi$ unless stated otherwise. In the scaled coordinate, \eqref{eq:eddy} becomes
\begin{equation}
  \curl  \vH = \alpha\sigma \vE ,
  \qquad
  \curl  \vE = \ii\omega\mu_0\mur\,\alpha\,\vH .
  \label{eq:scaledpair}
\end{equation}
Where $\sigma$ and $\mur$ are constant, taking the curl of either equation in \eqref{eq:scaledpair} and substituting the other gives, since both fields are divergence-free, the vector Helmholtz equation
\begin{equation}
  \nabla^2 \vF + \kappa^2 \vF = 0 ,
  \qquad
  \kappa^2 := \ii\,\omega\mu_0\mur\sigma\alpha^2 ,
  \label{eq:helmholtz}
\end{equation}
for $\vF\in\{\vE,\vH\}$. The coefficient is dimensionless, and its value in each conductor defines an induction number,
\begin{equation}
  \nua := \omega\mu_0\murb\sigma_*\alpha^2 ,
  \qquad
  \eps := \omega\mu_0\murs\sigs\alpha^2 = 2\Bigl(\frac{\alpha}{\dels}\Bigr)^{2} ,
  \label{eq:groups}
\end{equation}
so that $\kappa^2=\ii\eps$ in $\sea$ and $\kappa^2=\ii\nua$ in a homogeneous object, with the nominal $\sigma_*$ and $\murb$ of an object whose materials vary.
The parameter $\nua$ measures induction within the object; $\eps$ measures induction in $\sea$ over the length scale of the object. With the conductivity and permeability contrasts of the object to $\sea$,
\begin{equation}
  \zs := \frac{\sigma_*}{\sigs} ,
  \qquad
  \zm := \frac{\murb}{\murs} ,
  \label{eq:contrasts}
\end{equation}
the two are related by $\nua=\zs\zm\eps$, and the T-matrix of Section~\ref{sec:tensor} depends on the frequency, the size and the materials only through the three groups $(\eps,\zs,\zm)$. A change of size, a common change of both conductivities and a common change of both permeabilities therefore each act as a change of frequency.

\subsection{Structure of the Model}
\label{sec:structure}
The model derived in Sections~\ref{sec:tensor} and~\ref{sec:surface} has the form
\begin{equation}
  \Delta \vH(\vx,\vz) = \Prop(\vx,\vz)\,\Tens\,(\Id-\Ret\Tens)^{-1}\,\vd(\vz) ,
  \label{eq:shape}
\end{equation}
a product of four factors, each a block matrix over the degrees of the transverse electric (TE) and transverse magnetic (TM) vector spherical modes of $\sea$ about the centre of the object. The drive $\vd(\vz)$ holds the coefficients of the transmitter's field in regular modes about the object. The T-matrix $\Tens$ maps them to the coefficients of the response in outgoing modes, with blocks coupling TE and TM modes. The reflection matrix $\Ret$ returns the response to the object through its reflections at the surface and at the interfaces below, and the propagator $\Prop(\vx,\vz)$ carries it to the sensor. Only $\Tens$ depends on the object, and only $\Tens$ requires a finite element problem to be solved, once for each object, host and frequency.

\section{The T-Matrix}
\label{sec:tensor}

\subsection{Modes of the Host}
\label{sec:sea}

In $\sea$, $\kappa^2=\ii\eps$, with the root $\kappa=(1+\ii)\sqrt{\eps/2}$ so that the outgoing solution $e^{\ii\kappa r}$ decays with distance. A source-free field satisfies \eqref{eq:helmholtz} together with $\nabla\cdot \vF=0$, which leaves two scalar degrees of freedom per mode. Following Stratton~\cite{strattonElectromagneticTheory2007}, both are expressed through solutions $\psi$ of the scalar Helmholtz equation $(\nabla^2+\kappa^2)\psi=0$ and the operators
\begin{equation}
  \Mw[\psi] := \curl(\vxi\psi) ,
  \qquad
  \Nw[\psi] := \curl\curl(\vxi\psi) ,
  \label{eq:MN}
\end{equation}
so that $\Nw=\curl\Mw$. Both satisfy \eqref{eq:helmholtz}, both are divergence-free, and
\begin{equation}
  \curl \Nw[\psi] = \kappa^2\,\Mw[\psi] ,
  \label{eq:closure}
\end{equation}
so the pair is closed under the curl. Since $\Mw[\psi]=\nabla\psi\times\vxi$, $\Mw$ is tangential to spheres centred on the object, whereas $\Nw$ has a radial component.

A field of $\sea$ is a pair $(\vE,\vH)$ satisfying \eqref{eq:scaledpair} with $\sigs$ and $\murs$. By \eqref{eq:MN} and \eqref{eq:closure} exactly two such pairs are built on a given $\psi$,
\begin{align}
  \TE:\quad & \vE = e_{\TE}\,\Mw[\psi] ,
  & \vH &= \Nw[\psi] ,
  \label{eq:TE}\\
  \TM:\quad & \vE = e_{\TM}\,\Nw[\psi] ,
  & \vH &= \Mw[\psi] ,
  \label{eq:TM}
\end{align}
each normalised by its magnetic field, with electric amplitudes $e_{\TE}:=\ii\omega\mu_0\murs\alpha$ and $e_{\TM}:=1/\alpha\sigs$. The transverse-electric (TE) mode has tangential $\vE$ and radial $\vH$, and carries the inductive response of the object; the transverse-magnetic (TM) mode has tangential $\vH$ and radial $\vE$, and carries the galvanic response, the current conducted through $\sea$. Whether a mode is TE or TM is its type, indexed by $c$ where needed.

Fields are expanded in modes centred on the object, built on the Schmidt quasi-normalised~\cite{winchGeomagnetismSchmidtQuasinormalization2005} real spherical harmonics $Y^p_m(\theta,\phi)$ of degree $m$ and order $|p|\le m$ (Section~\sref{sec:s-harmonics}), with $(r,\theta,\phi)$ the spherical coordinates of $\vxi$. The solid harmonic $P^p_m:=r^m\,Y^p_m$ is a polynomial in $\vxi$; at degree one $P^1_1=\xi_1$, $P^{-1}_1=\xi_2$ and $P^0_1=\xi_3$, so quantities of degree one are Cartesian vectors. For a polynomial $F(\vxi)$, $F(\nabla)$ is the operator obtained by replacing each $\xi_i$ with $\partial_i$, and $F(\nabla_x)$ likewise with $\partial_{x_i}$. The outgoing scalar mode of degree $m$ and order $p$ is a derivative of the outgoing spherical wave,
\begin{align}
  \Psi^p_m(\vxi) &:= \frac{(-1)^m}{(2m-1)!!}\,P^p_m(\nabla)\,\frac{e^{\ii\kappa r}}{r}
  \nonumber\\
  &\phantom{:}= \frac{\ii\,\kappa^{m+1}}{(2m-1)!!}\,h^{(1)}_m(\kappa r)\,Y^p_m(\theta,\phi) ,
  \label{eq:Psi}
\end{align}
where the second line follows from Hobson's theorem~\cite{hobsonTheorySphericalEllipsoidal1931} (Section~\sref{sec:s-vectormodes}) and $h^{(1)}_m$ is the spherical Hankel function, singular at the origin. The regular scalar mode of degree $n$ and order $q$ is the same derivative of the standing spherical wave,
\begin{align}
  \Phi^q_n(\vxi) &:= \frac{(-1)^n(2n+1)!!}{\kappa^{2n+1}}\,P^q_n(\nabla)\,\frac{\sin\kappa r}{r}
  \nonumber\\
  &\phantom{:}= \frac{(2n+1)!!}{\kappa^{n}}\,j_n(\kappa r)\,Y^q_n(\theta,\phi) ,
  \label{eq:Phi}
\end{align}
with $j_n$ the spherical Bessel function, regular at the origin. The prefactors are chosen so that as $\kappa\to0$ the modes reduce to $r^nY^q_n$ and $Y^p_m/r^{m+1}$.

Every source-free field of $\sea$ is a combination of TE and TM modes built on \eqref{eq:Psi}--\eqref{eq:Phi}~\cite{strattonElectromagneticTheory2007}. The applied field is finite at the centre and expands in regular modes,
\begin{equation}
  \vH_0 = \sum_{n,q}\Bigl[\,d^{(n)}_{\TE,q}\,\Nw[\Phi^q_n] + d^{(n)}_{\TM,q}\,\Mw[\Phi^q_n]\Bigr] ,
  \label{eq:ambient}
\end{equation}
with $\vE_0$ following from \eqref{eq:TE}--\eqref{eq:TM}. The $2(2n+1)$ coefficients $d^{(n)}_{\TE,q}$, $d^{(n)}_{\TM,q}$ form the degree-$n$ drive. Beyond the circumscribed radius of the object, the largest distance from its centre to a point of it, the response decays away from the object and expands in outgoing modes,
\begin{equation}
  \Delta\vH = \sum_{m,p}\Bigl[\,a^{(m)}_{\TE,p}\,\Nw[\Psi^p_m] + a^{(m)}_{\TM,p}\,\Mw[\Psi^p_m]\Bigr] ,
  \label{eq:response}
\end{equation}
with response coefficients $a^{(m)}_{\TE,p}$, $a^{(m)}_{\TM,p}$. Since \eqref{eq:ambient} converges only out to the nearer of the surface and the interface beneath, the model requires the circumscribed radius to be less than both $D$ and $\hs-D$.

\subsection{The Transmission Problem and the Excess Sources}
\label{sec:object}
Inside the object $\sigma$ and $\mur$ vary, so $\nabla\cdot(\sigma \vE)=0$ no longer implies $\nabla\cdot \vE=0$ and no Helmholtz equation holds. Eliminating $\vH$ from \eqref{eq:scaledpair} without moving coefficients through the operators gives instead the curl-curl equation
\begin{equation}
  \curl \bigl(\mur^{-1}\,\curl  \vE\bigr) = g\,\vE ,
  \qquad
  g := \ii\,\omega\mu_0\sigma\alpha^2 ,
  \label{eq:curlcurl}
\end{equation}
which holds across material jumps in the weak sense below. The coefficient $g$ is $\kappa^2/\mur$: it equals $\ii\nua/\murb$ in a homogeneous object and $\ii\eps/\murs$ outside it.

The electric field is split as $\vE=\vE_0+\Delta\vE$, where the ambient field $\vE_0$ satisfies \eqref{eq:curlcurl} with the materials of $\sea$ everywhere and $\Delta\vE$ is the perturbation due to the object. Subtracting the two equations,
\begin{align}
  \curl \bigl(\mur^{-1}\curl  \Delta\vE\bigr) - g\,\Delta\vE &= \vf ,
  \label{eq:pert}\\
  \vf := -\curl \bigl[(\mur^{-1}-\murs^{-1})\curl  \vE_0\bigr] &+ \bigl(g-\ii\eps/\murs\bigr)\vE_0 ,
  \nonumber
\end{align}
where the load $\vf$ is supported on $B$ because both material contrasts vanish outside it. The first term of $\vf$ is the permeability contrast acting on the ambient magnetic field, and the second is the conductivity contrast acting on the ambient electric field.

Since the coefficients of \eqref{eq:pert} are discontinuous, it is posed in weak form on a bounded domain $\Omega\supset B$. In this form, the transmission problem, the perturbation $\Delta\vE$ lies in $\Hcurl(\Omega)$, its tangential trace vanishes on $\partial\Omega$, and it satisfies
\begin{multline}
  \int_\Omega \mur^{-1}\curl\Delta\vE\cdot\curl\vw
  - \int_\Omega g\,\Delta\vE\cdot \vw \\
  = -\int_B (\mur^{-1}-\murs^{-1})\curl\vE_0\cdot\curl\vw
  + \int_B \bigl(g-\ii\eps/\murs\bigr)\vE_0\cdot \vw
  \label{eq:weak}
\end{multline}
for all test fields $\vw$ in the same space. It is discretised by edge elements, which admit any shape and material profile without special treatment, with the drives evaluated on $B$ to first order in $\kappa^2$ (Section~\sref{sec:s-discretisation}).

The right side of \eqref{eq:weak} is linear in the ambient field, which by \eqref{eq:ambient} is a combination of regular modes. The transmission problem is therefore solved not for any particular transmitter field, but once per basis drive. Each basis drive is a single regular mode, \eqref{eq:TE} or \eqref{eq:TM} with $\psi=\Phi^q_n$, of degree $n$ and order $q$, with unit coefficient. Its solution gives the field in the object. The operator on the left of \eqref{eq:weak} does not depend on the drive. It is therefore assembled and factorised once, and each basis drive costs one back-substitution.

The field in the object is converted into sources in $\sea$ by splitting each material into that of $\sea$ plus contrast. The excess current density and the excess magnetisation, the current and magnetisation in the object beyond those the host would carry in its place, are formed from the solved field as
\begin{equation}
  \Jimp := (\sigma-\sigs)\,\vE ,
  \qquad
  \Mmat := \Bigl(\frac{\mur}{\murs}-1\Bigr)\vH ,
  \label{eq:densities}
\end{equation}
which vanish outside the object, since both contrasts do, and satisfy $\sigma \vE=\sigs \vE+\Jimp$ and $\mur\vH=\murs(\vH+\Mmat)$ exactly. The perturbation then satisfies, on all of space,
\begin{align}
  \curl  \Delta\vH &= \alpha\sigs \Delta\vE + \cJ ,
  & \cJ &:= \alpha\,\Jimp ,
  \label{eq:withsources}\\
  \curl  \Delta\vE &= \ii\omega\mu_0\murs\alpha\,\Delta\vH + \cK ,
  & \cK &:= \ii\omega\mu_0\murs\alpha\,\Mmat ,
  \nonumber
\end{align}
Maxwell's equations in uniform $\sea$ with two sources supported on the object: the excess current, which enters Amp\`ere's law, and the excess magnetisation, which enters Faraday's law.

\subsection{Extraction by Reciprocity}
\label{sec:extract}
Outside the object the perturbation is the outgoing expansion \eqref{eq:response} with unknown coefficients $\va$, and its sources $\cJ$, $\cK$ are known functions on $B$. The coefficients are extracted by projection using the Lorentz reciprocity theorem~\cite{harringtonTimeharmonicElectromagneticFields2001}. For two fields in the same medium, which need not be uniform, satisfying \eqref{eq:withsources} with sources $(\cJ_1,\cK_1)$ and $(\cJ_2,\cK_2)$,
\begin{multline}
  \nabla\cdot\bigl(\vE_1\times \vH_2 - \vE_2\times \vH_1\bigr) \\
  = \bigl(\vE_2\cdot\cJ_1 + \vH_2\cdot\cK_1\bigr)
  - \bigl(\vE_1\cdot\cJ_2 + \vH_1\cdot\cK_2\bigr) .
  \label{eq:reciprocity}
\end{multline}

Field 1 is the perturbation $(\Delta\vE,\Delta\vH)$ produced by one basis drive, the regular mode of type $c'$, degree $n$ and order $q$: its sources $(\cJ^{(n)}_{c',q},\cK^{(n)}_{c',q})$ are \eqref{eq:withsources} with the densities \eqref{eq:densities} formed from the field that this drive produces in the object. Field 2 is a probe: the regular mode of type $c$, degree $m$ and order $p$, with fields $(\vE_2,\vH_2)$ built on $\Phi^p_m$ by \eqref{eq:TE} or \eqref{eq:TM}, with no sources. Integrating \eqref{eq:reciprocity} over a ball $r\le R_0$ enclosing the object, the probe is source-free inside the ball, and
\begin{multline}
  \oint_{r=R_0}\bigl(\Delta\vE\times \vH_2 - \vE_2\times \Delta\vH\bigr)\cdot\nhat\dd S \\
  = \int_B\bigl(\vE_2\cdot\cJ^{(n)}_{c',q} + \vH_2\cdot\cK^{(n)}_{c',q}\bigr)\dd\vxi .
  \label{eq:master}
\end{multline}
An outgoing mode is inadmissible as a probe since it is singular at the origin. By orthogonality of the spherical harmonics and the Wronskian of $j_m$ and $h^{(1)}_m$, the left side reduces to a constant times $a^{(m)}_{c,p}$, independent of $R_0$ (Section~\sref{sec:s-extraction}), and this coefficient per unit of the drive $d^{(n)}_{c',q}$ is by definition an entry of the T-matrix,
\begin{equation}
  \Tens^{(m,n)}_{c,c'}[p,q] = \frac{\beta_m}{e_c}\int_B\bigl(\vE_2\cdot\cJ^{(n)}_{c',q} + \vH_2\cdot\cK^{(n)}_{c',q}\bigr)\dd\vxi ,
  \label{eq:reaction}
\end{equation}
the reaction of the probe with the sources the drive induces, where $\beta_m:=1/4\pi m(m+1)$ follows from the normalisation of $Y^p_m$, and $e_c$ is the electric amplitude \eqref{eq:TE}--\eqref{eq:TM} of the probe's type. Every entry is an integral over the object of known functions, with $(c,m,p)$ the probed mode and $(c',n,q)$ the basis drive.

The ambient field is the combination \eqref{eq:ambient} of basis drives with coefficients $d^{(n)}_{c',q}$, and the chain from a basis drive through \eqref{eq:weak} and \eqref{eq:densities} to \eqref{eq:reaction} is linear, so the response to the transmitter's field is, grouped by type and degree with the orders as matrix indices,
\begin{equation}
  \begin{pmatrix} \va^{(m)}_{\TE}\\[2pt] \va^{(m)}_{\TM} \end{pmatrix}
  = \sum_{n\ge1}
  \begin{pmatrix}
    \Tens^{(m,n)}_{\TE,\TE} & \Tens^{(m,n)}_{\TE,\TM}\\[2pt]
    \Tens^{(m,n)}_{\TM,\TE} & \Tens^{(m,n)}_{\TM,\TM}
  \end{pmatrix}
  \begin{pmatrix} \vd^{(n)}_{\TE}\\[2pt] \vd^{(n)}_{\TM} \end{pmatrix} ,
  \label{eq:tensor}
\end{equation}
the T-matrix of the object, each block a $(2m+1)\times(2n+1)$ matrix over the orders with the entries \eqref{eq:reaction}. The diagonal blocks are the inductive and galvanic responses and the off-diagonal blocks the cross-couplings. The T-matrix is independent of the object's position, the layers of the ground and the instruments, since the transmission problem \eqref{eq:weak} sees only the object and the host about it. For a fixed host and $\nua$ the contribution of the $(m,n)$ block to the field at a sensor is $O(\alpha^{m+n+1})$ for the diagonal blocks and $O(\alpha^{m+n+2})$ for the cross blocks, which convert a current moment into a magnetic moment across the size of the object, or the reverse (Section~\sref{sec:s-order}). The sum over the blocks is therefore an expansion in the size of the object, in which a block has the order $m+n$, or $m+n+1$ if it is a cross block, and its truncation is controlled.

Substituting the probe's fields \eqref{eq:TE}--\eqref{eq:TM} and the sources \eqref{eq:withsources} into \eqref{eq:reaction} splits each entry into two integrals over the object, with $\Jimp[c']$, $\Mmat[c']$ the densities \eqref{eq:densities} of the solution for the drive $(c',n,q)$:
\begin{align}
  \Tens^{(m,n)}_{\TE,c'}[p,q] &= \beta_m\Bigl[\alpha\int_B\Jimp[c']\cdot\Mw[\Phi^p_m]\dd\vxi
  \nonumber\\
  &\qquad\qquad{}+ \int_B\Mmat[c']\cdot\Nw[\Phi^p_m]\dd\vxi\Bigr] ,
  \label{eq:integrandTE}\\
  \Tens^{(m,n)}_{\TM,c'}[p,q] &= \beta_m\Bigl[\alpha\int_B\Jimp[c']\cdot\Nw[\Phi^p_m]\dd\vxi
  \nonumber\\
  &\qquad\qquad{}+ \kappa^2\int_B\Mmat[c']\cdot\Mw[\Phi^p_m]\dd\vxi\Bigr] .
  \label{eq:integrandTM}
\end{align}
In \eqref{eq:integrandTE} the first term is the eddy-current mechanism, in which the circulating part of the excess current forms a current loop and radiates a magnetic multipole, and the second is magnetic polarisation, in which the magnetised object radiates a magnetic multipole directly. In \eqref{eq:integrandTM} the first term is galvanic conduction, in which current driven through the object returns through $\sea$, and the second is the current that the magnetised object induces in $\sea$, which carries the factor $\kappa^2=\ii\eps$ and vanishes as $\omega\to0$. At degree one the response is a pair of point dipoles at the centre, with magnetic moment $\vm$ and current moment $\vp$ (Section~\sref{sec:s-dipoles}),
\begin{equation}
  \va^{(1)}_{\TE} = \frac{\vm}{4\pi\alpha^3} ,
  \qquad
  \va^{(1)}_{\TM} = \frac{\vp}{4\pi\alpha^2} .
  \label{eq:response1}
\end{equation}
Fig.~\sref{fig:blocks} of the supplementary material shows the four blocks for the object of Section~\ref{sec:numerics}.

The conductivity and permeability contrasts of the object size the entries of the T-matrix through the densities \eqref{eq:densities}, but the host decides which entries exist. TE modes carry magnetic flux through the host and TM modes carry current, which a non-conducting host cannot carry. For $\sigs=0$ the block $\Tens_{\TM,\TE}$ vanishes despite a conductivity contrast, and no TM drive exists to excite $\Tens_{\TE,\TM}$ or $\Tens_{\TM,\TM}$ (Section~\sref{sec:s-nonconducting}). What remains is $\Tens_{\TE,\TE}$, the eddy currents and the magnetic polarisation, neither of which needs the host to conduct. A weakly conducting, permeable soil admits all four mechanisms, the TM ones in proportion to its conductivity.

The T-matrix is also symmetric under exchange of probe and drive, since the probes of \eqref{eq:reaction} are the same modes as the basis drives. Applied to the total fields of two basis drives, \eqref{eq:reciprocity} gives (Section~\sref{sec:s-symmetry})
\begin{equation}
  \frac{e_c}{\beta_m}\,\Tens^{(m,n)}_{c,c'} = \frac{e_{c'}}{\beta_n}\,\bigl(\Tens^{(n,m)}_{c',c}\bigr)^{\!\top} .
  \label{eq:symmetry}
\end{equation}
The diagonal blocks $\Tens^{(m,m)}_{c,c}$ are therefore symmetric at every degree, as the MPT is. Since $e_{\TE}/e_{\TM}=\kappa^2$, the cross blocks are one block, $\Tens^{(n,m)}_{\TM,\TE}=\kappa^2(\beta_n/\beta_m)\bigl(\Tens^{(m,n)}_{\TE,\TM}\bigr)^{\!\top}$, and only one needs computing.

\section{Propagation Through Layered Media}
\label{sec:surface}
The T-matrix describes the object in an unbounded host. The layers enter through the propagator, the drive and the reflection matrix of \eqref{eq:shape}, which are all built from the point source of the host, carried through the layers. They are given below and derived in Section~\sref{sec:propagation} of the supplementary material.

\subsection{Transmission Through the Layers}
\label{sec:sommerfeld}
The layers are crossed by transforming in the horizontal coordinates, as in the Green's functions of half spaces and of layered media~\cite{wardElectromagneticTheoryGeophysical1988,ledgerCharacterisingBuriedObjects2025}. By \eqref{eq:Psi} every outgoing mode is a derivative of the outgoing point source of the host, which the Sommerfeld identity~\cite{sommerfeldarnoldUberAusbreitungWellen1909} expands in horizontal modes,
\begin{align}
  u(\vx,\vz) &:= \frac{e^{\ii\ks|\vx-\vz|}}{|\vx-\vz|}
  \nonumber\\
  &\phantom{:}= \int_0^\infty \frac{k_\rho}{\gams}\,J_0(k_\rho\rho)\,e^{-\gams|x_3-z_3|}\dd k_\rho ,
  \label{eq:sommerfeld}
\end{align}
with $\ks:=\kappa/\alpha$ the wavenumber of the host, $k_\rho$ the horizontal wavenumber, $\rho$ the horizontal distance from $\vx$ to $\vz$, and $\gams$ the vertical propagation constant of the host, given below. The layers are invariant under horizontal translation, so each horizontal mode crosses them with its horizontal dependence unchanged, and only its vertical profile is altered.

In each layer a horizontal mode is fixed by two scalars, the vertical magnetic field $H_3$ and the vertical current $\sigma E_3$. Within a layer $\nabla\cdot\vH=0$ fixes the divergent part of the horizontal field from $H_3$, and the vertical current $\sigma E_3=(\curlx\vH)_3$ fixes the rest, so that
\begin{equation}
  \vH = \ehat_3H_3 + \frac{1}{k_\rho^2}\Bigl[\nabla_\rho\,\partial_{x_3}H_3 + \nabla_\rho(\sigma E_3)\times\ehat_3\Bigr] ,
  \label{eq:fromscalars}
\end{equation}
with $\nabla_\rho:=(\partial_{x_1},\partial_{x_2},0)$ the horizontal gradient. Taking $u$ once as the vertical field and once as the vertical current, \eqref{eq:fromscalars} gives the vector point sources of the unbounded host,
\begin{equation}
  \vU := \bigl(\ehat_3+k_\rho^{-2}\nabla_\rho\,\partial_{x_3}\bigr)\,u ,
  \qquad
  \vU' := k_\rho^{-2}\,\nabla_\rho u\times\ehat_3 .
  \label{eq:Ufree}
\end{equation}

At an interface $H_3$ and $\sigma E_3$ are reflected independently. In layer $j$ both vary vertically with the propagation constant $\gamma_j$, and their reflections are set by the ratios $Y_j$ for $H_3$ and $Y'_j$ for $\sigma E_3$,
\begin{equation}
  \gamma_j = \sqrt{k_\rho^2-\ii\omega\mu_0\mu_{\mathrm{r},j}\sigma_j} ,
  \quad
  Y_j = \frac{\gamma_j}{\mu_{\mathrm{r},j}} ,
  \quad
  Y'_j = \frac{\gamma_j}{\sigma_j} ,
  \label{eq:admittance}
\end{equation}
with $\gams:=\gamma_1$ in the host. At the surface above the host, where $Y=k_\rho$ in the air, $H_3$ is reflected and transmitted with the coefficients
\begin{equation}
  \Gamma_0(k_\rho) = \frac{\gams-\murs k_\rho}{\gams+\murs k_\rho} ,
  \qquad
  \tau(k_\rho) = \frac{2\murs\gams}{\gams+\murs k_\rho} .
  \label{eq:RT}
\end{equation}
The vertical current is reflected by the same rule with $Y'$ in place of $Y$. In the non-conducting air $Y'=\gamma/\sigma$ is unbounded, so its reflection coefficient is $\Gamma'_0=-1$: no current crosses into the air, and only $H_3$ reaches it.
The layers beneath reflect each scalar with a generalised reflection coefficient, $\Gamma_1$ for $H_3$ and $\Gamma'_1$ for the vertical current. Both follow from the recursion
\begin{equation}
  \begin{gathered}
  \Gamma_j = \frac{\Gamma_{j,j+1}+\Gamma_{j+1}\,e^{-2\gamma_{j+1}h_{j+1}}}{1+\Gamma_{j,j+1}\Gamma_{j+1}\,e^{-2\gamma_{j+1}h_{j+1}}} ,
  \\
  \Gamma_{j,j+1} = \frac{Y_j-Y_{j+1}}{Y_j+Y_{j+1}} ,
  \end{gathered}
  \label{eq:recursion}
\end{equation}
taken upward from $\Gamma_N=0$ in the half space at the bottom to $\Gamma_1$, and with $Y'_j$ in place of $Y_j$ for $\Gamma'_1$.

Each horizontal mode of $u$ has a rising half, which climbs from the object to the surface, and a falling half, which reaches the interface beneath and returns with $\Gamma_1$. Both then reflect between the surface and the interface beneath, and their sum at the surface is the layer factor
\begin{equation}
  L(k_\rho,z_3) := \frac{e^{\gams z_3}+\Gamma_1\,e^{-\gams(2\hs+z_3)}}{1-\Gamma_0\Gamma_1\,e^{-2\gams \hs}} ,
  \label{eq:layerfactor}
\end{equation}
where $\Gamma_0$ and $\Gamma_1$ are the reflections of $H_3$ above and beneath the host, \eqref{eq:RT} and \eqref{eq:recursion}. In the air each horizontal mode continues with the factor $\tau\,e^{-k_\rho x_3}$, so $L\,\tau\,e^{-k_\rho x_3}$ replaces $e^{-\gams|x_3-z_3|}$ in \eqref{eq:sommerfeld} and gives the vertical field of $u$ in the air, the transmitted point source
\begin{equation}
  u_\tau(\vx,\vz) := \int_0^\infty\frac{k_\rho}{\gams}\,J_0(k_\rho\rho)\,L\,\tau\,e^{-k_\rho x_3}\dd k_\rho .
  \label{eq:transmittedsource}
\end{equation}
The same expression with the transmission $\tau'=1+\Gamma'_0$ of the vertical current in place of $\tau$ is denoted $u'_\tau$. With $u_\tau$ and $u'_\tau$ in place of $u$, \eqref{eq:Ufree} gives the vector point sources in the air. There $\partial_{x_3}$ inserts $-k_\rho$, and $\Gamma'_0=-1$ makes $\tau'=0$, so that
\begin{equation}
  \vU_\tau = \bigl(\ehat_3-k_\rho^{-1}\nabla_\rho\bigr)\,u_\tau ,
  \qquad
  \vU'_\tau = \bm{0} .
  \label{eq:kernel}
\end{equation}

\subsection{The Propagator and the Drive}
\label{sec:propagator}
By \eqref{eq:Psi} and \eqref{eq:MN} the outgoing mode of type $c$, degree $m$ and order $p$ is a derivative of the point source. The derivatives are taken in the position $\vz$ of the source rather than at the field point. In the host the two are equivalent, since $u$ depends on $\vx-\vz$ only; across the layers only derivatives in $\vz$ remain valid, since the layers do not depend on $\vz$ and the derivatives then commute with propagation through them. The field of the mode is
\begin{equation}
  \vH = Q^{(m)}_{c,p}(-\nabla_z)\,\vU + {Q'}^{(m)}_{c,p}(-\nabla_z)\,\vU' ,
  \label{eq:Qpoly}
\end{equation}
for two polynomials $Q^{(m)}_{c,p}$ and ${Q'}^{(m)}_{c,p}$ fixed by the mode. Since the vertical component of $\vU$ is $u$ and $\vU'$ is horizontal, they are read off the vertical components of the magnetic field \eqref{eq:TE}--\eqref{eq:TM} of the mode built on $\Psi^p_m$ and of its curl. By \eqref{eq:scaling}, \eqref{eq:Psi} is $\Psi^p_m=c_m\,\alpha^{m+1}P^p_m(\nabla_x)\,u$ with $c_m:=(-1)^m/(2m-1)!!$. On $\Psi^p_m$ the derivative $\vxi\times\nabla$ in $\Mw$ acts on $P^p_m$ alone, and the curl in $\Nw$ becomes a factor $\vxi\times$ (Section~\sref{sec:s-vertical}), so the vertical fields are $Q^{(m)}_{c,p}(\nabla_x)\,u$ with the polynomials
\begin{align}
  Q^{(m)}_{\TM,p}(\vxi) &= -c_m\,\alpha^{m+1}\,\ehat_3\cdot\bigl(\vxi\times\nabla P^p_m(\vxi)\bigr) ,
  \nonumber\\
  Q^{(m)}_{\TE,p}(\vxi) &= -c_m\,\alpha^{m+2}\,\ehat_3\cdot\bigl(\vxi\times(\vxi\times\nabla P^p_m(\vxi))\bigr) ,
  \label{eq:Q}
\end{align}
of degree $m$ for a TM mode and $m+1$ for a TE mode, so that $Q(\nabla_x)$ acts on $u$ with constant coefficients. By \eqref{eq:closure} the curl of a field of one type is a field of the other, so the primed polynomials are the same two, exchanged,
\begin{equation}
  {Q'}^{(m)}_{\TM,p}(\vxi) = \alpha^{-1}\,Q^{(m)}_{\TE,p}(\vxi) ,
  \quad
  {Q'}^{(m)}_{\TE,p}(\vxi) = \alpha\ks^2\,Q^{(m)}_{\TM,p}(\vxi) .
  \label{eq:Qprime}
\end{equation}

Applied to the vector point sources in the air, the same derivatives give the field there of the mode. Collected over the orders into the rows $\bm{Q}^{(m)}_c$ of the $Q^{(m)}_{c,p}$ and ${\bm{Q}'}^{(m)}_c$ likewise, they give the fields of all the modes of one type and degree, which form the propagator block of \eqref{eq:shape}. Since $\vU'_\tau=\bm0$ by \eqref{eq:kernel}, only the unprimed rows remain,
\begin{align}
  \Prop^{(m)}_c &= \bm{Q}^{(m)}_c(-\nabla_z)\,\vU_\tau + {\bm{Q}'}^{(m)}_c(-\nabla_z)\,\vU'_\tau
  \nonumber\\
  &= \bm{Q}^{(m)}_c(-\nabla_z)\,\vU_\tau ,
  \label{eq:propentries}
\end{align}
a $3\times(2m+1)$ matrix whose column $p$ is the field of the mode of order $p$.
The perturbation at the sensor is then $\Delta\vH=\sum_{m,c}\Prop^{(m)}_c\va^{(m)}_c$. The axially symmetric TM order $p=0$ is the exception: $P^0_m$ is symmetric about the vertical, so $\ehat_3\cdot(\vxi\times\nabla P^0_m)=\partial_\phi P^0_m=0$ and by \eqref{eq:Q} its vertical field vanishes. A vertical current dipole and its axially symmetric multipoles are invisible from the air, and by reciprocity cannot be driven from it.

The drive is the propagator read in the opposite direction. Reciprocity between the field of the transmitter, a coil at $\vy$ with moment $\vm_\mathrm{T}$, and the field of each outgoing mode gives
\begin{equation}
  \vd^{(n)}_c = \frac{\ii\omega\mu_0\,\beta_n}{\alpha^2e_c}\;\Prop^{(n)}_c(\vy,\vz)^{\top}\,\vm_\mathrm{T} ,
  \label{eq:drivereciprocity}
\end{equation}
with $\beta_n$ the constant of \eqref{eq:reaction}, so the drive at every degree is the transposed propagator and no field of the transmitter in $\sea$ need be computed (Section~\sref{sec:s-drive}). At degree one the drive is the ambient field at the centre, with magnetic field $\vH_0(\vz)$ and current density $\sigs\vE_0(\vz)$,
\begin{equation}
  \vd^{(1)}_{\TE} = \tfrac12\,\vH_0(\vz) ,
  \qquad
  \vd^{(1)}_{\TM} = \tfrac12\,\alpha\sigs\,\vE_0(\vz) .
  \label{eq:drive1}
\end{equation}

\subsection{The Reflection Matrix}
\label{sec:return}
The object response is itself incident upon the layers and its reflection returns to the object as a field regular about $\vz$, which drives the object as the transmitter's field does. With $\Ret$ the operator that takes the outgoing coefficients of a response to the regular coefficients of its reflection, summing over every number of reflections gives
\begin{equation}
  \va = \Tens\sum_{\ell\ge0}(\Ret\Tens)^{\ell}\,\vd = \Tens\,(\Id-\Ret\Tens)^{-1}\vd .
  \label{eq:return}
\end{equation}

The reflection matrix is built as the propagator is, with the transmission into the air replaced by the reflections at the surface and at the interface beneath. Following both halves of each horizontal mode of $u$ through the reflections back into the host gives, at a field point $\vx$ in the host, the vertical profile
\begin{multline}
  g_\Gamma(x_3,z_3) := \Gamma_0\,e^{\gams(x_3+z_3)}+\Gamma_1\,e^{-\gams(x_3+z_3+2\hs)}\\
  +\Gamma_0\Gamma_1\,e^{-\gams(2\hs+x_3-z_3)}+\Gamma_0\Gamma_1\,e^{-\gams(2\hs-x_3+z_3)} .
  \label{eq:gGamma}
\end{multline}
Each term is a path from the object to the field point, with a factor $e^{-\gams\lambda}$ for its length $\lambda$ and a reflection coefficient for each boundary it meets: via the surface, via the interface beneath, and via both in either order. With the further round trips between the boundaries, as in the layer factor \eqref{eq:layerfactor}, $g_\Gamma$ replaces $e^{-\gams|x_3-z_3|}$ in \eqref{eq:sommerfeld} and gives the reflected point source
\begin{equation}
  u_{\Gamma}(\vx,\vz) := \int_0^\infty\frac{k_\rho}{\gams}\,J_0(k_\rho\rho)\,
  \frac{g_\Gamma(x_3,z_3)}{1-\Gamma_0\Gamma_1\,e^{-2\gams \hs}}\dd k_\rho .
  \label{eq:reflectedsource}
\end{equation} The same expression with the reflections $\Gamma'_0=-1$ and $\Gamma'_1$ of the vertical current is denoted $u'_\Gamma$. With $u_\Gamma$ and $u'_\Gamma$ in place of $u$, \eqref{eq:Ufree} gives the reflected vector point sources $\vU_\Gamma$ and $\vU'_\Gamma$, and by \eqref{eq:Qpoly} the reflected fields of all the modes of type $c$ and degree $m$ are, like the propagator block,
\begin{equation}
  \bm{H}^{\Gamma}_c = \bm{Q}^{(m)}_c(-\nabla_z)\,\vU_\Gamma + {\bm{Q}'}^{(m)}_c(-\nabla_z)\,\vU'_\Gamma .
  \label{eq:reflectedfield}
\end{equation}
Column $p$ of $\bm{H}^{\Gamma}_c$, the reflected field $\vH^{\Gamma}_{c,p}$ of the mode of order $p$, has no sources in the host. Within a distance $\min(D,\hs-D)$ of $\vz$ it is therefore a sum of regular modes, as the ambient field \eqref{eq:ambient} is, and its coefficients are column $p$ of the reflection matrix.

Each coefficient of a regular expansion is a derivative of the field at the centre: applied at $\vz$, the operator $(\nabla P^q_n)(\nabla_x)$ extracts the TE coefficient of degree $n$ and order $q$ from the field and the TM coefficient from its curl. The entries are therefore (Section~\sref{sec:s-return})
\begin{align}
  \Ret^{(n,m)}_{\TE,c}[q,p] &= w_n\,\alpha^{n-1}\bigl[(\nabla P^q_n)(\nabla_x)\cdot\vH^{\Gamma}_{c,p}\bigr](\vz) ,
  \nonumber\\
  \Ret^{(n,m)}_{\TM,c}[q,p] &= w_n\,\alpha^{n-1}\bigl[(\nabla P^q_n)(\nabla_x)\cdot\curl\vH^{\Gamma}_{c,p}\bigr](\vz) ,
  \label{eq:retentries}
\end{align}
with $w_n:=1/[n(n+1)(2n-1)!!]$. At degree one $\nabla P^q_1=\ehat_q$ and $w_1=\tfrac12$, so that, as for the drive \eqref{eq:drive1},
\begin{equation}
  \Ret^{(1,m)}_{\TE,c} = \tfrac12\,\bm{H}^{\Gamma}_c(\vz) ,
  \qquad
  \Ret^{(1,m)}_{\TM,c} = \tfrac12\,\curl\bm{H}^{\Gamma}_c(\vz) .
  \label{eq:ret1}
\end{equation}
By \eqref{eq:reflectedfield} the derivatives act on $\vU_\Gamma$ and $\vU'_\Gamma$ inside the integral over $k_\rho$, where they reduce to the polynomials \eqref{eq:Q} of degree $n$ acting on $u_\Gamma$ and $u'_\Gamma$, and at $\rho=0$ they give polynomials in $k_\rho$ and $\gams$, so each entry is one integral over $k_\rho$. The reflection matrix depends only on the layers and the position of the centre, and couples only orders of the same $|p|$.

The size of the reflection matrix follows from the exponents of \eqref{eq:gGamma}: each term of $u_\Gamma$ is a point source at an image of $\vz$, in the surface, in the interface beneath, or in both, and the $(n,m)$ block of $\Ret$ is $O\bigl((\alpha/2D)^{m+n+1}\bigr)$ from the surface and $O\bigl((\alpha/2(\hs-D))^{m+n+1}\bigr)$ from beneath, times the reflection coefficients (Section~\sref{sec:s-reflectionsize}). The expansion about the images converges under the condition of Section~\ref{sec:sea} on the circumscribed radius; an object nearer an interface, or straddling one, lies outside the model. At degree one the reflection is a relative correction of $O\bigl((\alpha/2D)^3\bigr)$ and $O\bigl((\alpha/2(\hs-D))^3\bigr)$, smaller than the next degree of the T-matrix unless the interface beneath is near.

\subsection{The Assembled Model}
\label{sec:assembled}
The degrees are collected by type into the row $\Prop_c:=\bigl(\Prop^{(1)}_c\;\Prop^{(2)}_c\;\cdots\bigr)$, the column $\vd_c$ of the $\vd^{(n)}_c$, and the square matrices $\Tens_{c,c'}$ and $\Ret_{c,c'}$ of the blocks $\Tens^{(m,n)}_{c,c'}$ and $\Ret^{(n,m)}_{c,c'}$. Over the two types these form
\begin{equation*}
  \begin{gathered}
  \Prop = \begin{pmatrix}\Prop_{\TE} & \Prop_{\TM}\end{pmatrix} ,
  \qquad
  \vd = \begin{pmatrix}\vd_{\TE}\\ \vd_{\TM}\end{pmatrix} ,
  \\
  \Tens = \begin{pmatrix}\Tens_{\TE,\TE} & \Tens_{\TE,\TM}\\ \Tens_{\TM,\TE} & \Tens_{\TM,\TM}\end{pmatrix} ,
  \enspace
  \Ret = \begin{pmatrix}\Ret_{\TE,\TE} & \Ret_{\TE,\TM}\\ \Ret_{\TM,\TE} & \Ret_{\TM,\TM}\end{pmatrix} ,
  \end{gathered}
\end{equation*}
and the response \eqref{eq:return} with $\Delta\vH=\Prop\va$ gives the model of Section~\ref{sec:structure},
\begin{equation}
  \Delta \vH(\vx,\vz) = \Prop(\vx,\vz)\,\Tens\,(\Id-\Ret\Tens)^{-1}\,\vd(\vz) .
  \tag{\ref*{eq:shape}}
\end{equation}

The propagator, the drive and the reflection matrix depend only on the layers, the centre and the coils, so moving the coil or the sensor changes only one-dimensional integrals, and changing the object changes only the T-matrix.

The model is reciprocal, as the physical problem is: exchanging the transmitter and a receiving coil, positions and moments, leaves the signal unchanged. By \eqref{eq:drivereciprocity} the signal at a receiving coil is a bilinear form in the propagators to the two coils, which is symmetric under their exchange (Section~\sref{sec:s-reciprocity}) because $\Tens$ satisfies \eqref{eq:symmetry} and $\Ret$ satisfies its counterpart (Section~\sref{sec:s-retsymmetry}). Truncation by order (Section~\sref{sec:s-truncation}) keeps the block $(n,m)$ of the types $(c',c)$ whenever it keeps the block $(m,n)$ of the types $(c,c')$, so every truncation of Section~\ref{sec:numerics} is reciprocal as well.

\section{Structure and Limits of the T-Matrix}
\label{sec:properties}
The diagonal blocks of the T-matrix reduce to two established tensors, the TE block to the MPT $\mathcal{M}$ and the TM block to the electric polarizability tensor $\mathcal{P}$, which describes how an object perturbs a steady current; the cross blocks belong to neither.

\subsection{The MPT and the Polarization Tensor as Limits}
\label{sec:classical}
The inductive block reduces to the MPT when the host conducts weakly, $\eps\to0$ at fixed $\nua$, in which the T-matrix may be evaluated at $\eps=0$ with an error of $O(\eps)$. The drive of degree one is then the uniform magnetic field $\vH_0$, and the coefficient $\ii\eps/\murs$ of \eqref{eq:curlcurl} outside the object vanishes. The transmission problem \eqref{eq:weak} then reduces to the eddy-current problem for the object in a non-conducting exterior, which defines the MPT $\mathcal{M}$ through $\vm=\mathcal{M}\vH_0(\vz)$~\cite{ammariTargetDetectionCharacterization2014,ledgerCharacterizingShapeMaterial2015}. With the moment \eqref{eq:response1} and the drive \eqref{eq:drive1} (Section~\sref{sec:s-mpt}),
\begin{equation}
  \mathcal{M} = 2\pi\alpha^3\,\Tens^{(1,1)}_{\TE,\TE}\big|_{\eps=0} .
  \label{eq:mpt}
\end{equation}
The galvanic block reduces to the electric polarizability tensor as $\omega\to0$ at a fixed contrast $\zs$, for a homogeneous object of conductivity $\sigma_*$, so that $\eps$ and $\nua$ vanish together. The drive of degree one is then the uniform electric field $\vE_0$, and the coefficient $g$ of \eqref{eq:curlcurl} vanishes inside the object and outside it, so the electric field is curl-free. The transmission problem \eqref{eq:weak} then reduces to the conduction problem for the object in a host of conductivity $\sigs$, which defines the electric polarizability tensor $\mathcal{P}(\sigma_*/\sigs)$, the polarization tensor of Pólya and Szegő at the conductivity contrast, through $\vp=\mathcal{P}\sigs\vE_0(\vz)$~\cite{polyaIsoperimetricInequalitiesMathematical1951,ammariPolarizationMomentTensors2007}. As $\mathcal{P}$ scales with the volume of the object (Section~\sref{sec:s-pst}),
\begin{equation}
  \mathcal{P}(\zs) = 2\pi\alpha^3\,\Tens^{(1,1)}_{\TM,\TM}\big|_{\omega=0} .
  \label{eq:pst}
\end{equation}

At higher degrees the drives and probes are the gradients of $P^q_n$ and $P^p_m$, and the same two reductions give, in the harmonic basis, the generalised MPTs (GMPTs)~\cite{ledgerGeneralisedMagneticPolarizability2018} from the inductive blocks and the generalised polarization tensors (GPTs)~\cite{ammariPolarizationMomentTensors2007} from the galvanic blocks (Section~\sref{sec:s-higher}). In statics the two are one tensor: for a non-conducting object $\mathcal{M}=\mathcal{P}(\zm)$, as Ledger and Lionheart show in free space~\cite{ledgerCharacterizingShapeMaterial2015}, the permeability contrast in place of the conductivity contrast.

The contrast fixes the sign and size of the galvanic response: an object of high contrast draws the host current through itself, and one of low contrast excludes it and reverses the sign of $\mathcal{P}$, as observed for coated objects in seawater~\cite{sanfilipoEffectsSeawaterEMI2005}.

\subsection{The Cross Blocks}
\label{sec:cross}
The off-diagonal blocks have no counterpart in either tensor. $\mathcal{M}$ describes only the inductive response to a TE drive, and $\mathcal{P}$ only the galvanic response to a TM drive. Each cross block is the other response to one of these drives. Under a TM drive, current conducted through an object without symmetry partly circulates and so forms a magnetic dipole, which is $\Tens_{\TE,\TM}$. $\mathcal{P}$ is defined through the electric potential alone, so it has no term for this dipole. Under a TE drive, the induced electric field drives current through the object and forms a current dipole, which is $\Tens_{\TM,\TE}$. $\mathcal{M}$ has no term for it either, since a non-conducting host carries no TM response (Section~\ref{sec:extract}). By \eqref{eq:symmetry} only one of the two need be computed.

Which cross blocks exist depends on the symmetry of the object. An object has a centre of symmetry if its shape and materials are unchanged by the inversion $\vxi\to-\vxi$ through its centre, as for a box or an ellipsoid. The inversion reverses the electric field, and since $\vH$ follows from the curl of $\vE$ it leaves the magnetic field unchanged, so a field $(\vE(\vxi),\vH(\vxi))$ maps to $(-\vE(-\vxi),\vH(-\vxi))$. By \eqref{eq:MN} and the parity $(-1)^n$ of $P^q_n$, it multiplies the TE modes of degree $n$ by $(-1)^{n+1}$ and the TM modes by $(-1)^n$. An object with a centre of symmetry therefore has diagonal blocks only where $m+n$ is even and cross blocks only where $m+n$ is odd. It has no cross blocks of degree one, and every block that remains has even order, so the first correction beyond the dipole, of order four, joins the cross blocks $(2,1)$ and $(1,2)$ to the diagonal blocks of total degree four. A model assembled from the GMPTs and GPTs lacks the cross part of that correction, whereas a truncation by order retains it.

\subsection{Range Laws in the Near Field}
\label{sec:rangelaws}
Within a skin depth of the object the transmitted point source \eqref{eq:transmittedsource} is static, $\gams\approx k_\rho$, and each derivative of it costs one power of the range $R:=|\vx-\vz|$. In this limit $\Gamma_0$ and $\Gamma_1$ vanish for non-magnetic layers, so the layers do not alter the count. Since $Q^{(m)}_{c,p}$ in \eqref{eq:propentries} has degree $m$ for a TM mode and $m+1$ for a TE mode (Section~\sref{sec:s-rangelaws}),
\begin{equation}
  \bigl|\Prop^{(m)}_{\TM}\bigr| \sim \Bigl(\frac{\alpha}{R}\Bigr)^{m+1} ,
  \qquad
  \bigl|\Prop^{(m)}_{\TE}\bigr| \sim \Bigl(\frac{\alpha}{R}\Bigr)^{m+2} ,
  \label{eq:rangelaw}
\end{equation}
and the galvanic dipole decays as $R^{-2}$ where the inductive dipole decays as $R^{-3}$. The drive obeys the same count from a transmitter at distance $R_\mathrm{T}:=|\vy-\vz|$ from the object, with $|\vd^{(1)}_{\TM}|/|\vd^{(1)}_{\TE}|\sim\eps R_\mathrm{T}/\alpha$ by \eqref{eq:drivereciprocity}, since $e_{\TE}/e_{\TM}=\kappa^2$. The galvanic dipole term $\Delta\vH_{\TM}$ therefore stands to the inductive one $\Delta\vH_{\TE}$ as
\begin{equation}
  \frac{|\Delta\vH_{\TM}|}{|\Delta\vH_{\TE}|}
  \sim \frac{|\mathcal{P}|}{|\mathcal{M}|}\,\frac{2R_\mathrm{T}R}{\dels^2} ,
  \label{eq:ratio}
\end{equation}
where the two diagonal blocks, which depend only weakly on $\eps$, are replaced by their limits \eqref{eq:mpt} and \eqref{eq:pst}. The ratio does not depend on $\alpha$: $\mathcal{M}$ and $\mathcal{P}$ are both $O(\alpha^3)$, and otherwise depend on the shape of the object and on $\nua$ and $\sigma_*/\sigs$ respectively. Since neither range is shorter than $D$, \eqref{eq:ratio} is $\gtrsim(D/\dels)^2$, and it grows with the range of the sensor. Beyond a skin depth the contributing wavenumbers satisfy $k_\rho\dels\ll1$, where $\gams\approx(1-\ii)/\dels$; the source-side derivative $\partial_{z_3}$ then inserts this constant in place of a power of $1/R$, and the ratio stops growing.

\subsection{Reduction to the Buried and Free-Space Models}
\label{sec:reductions}
For a small object in a weakly conducting soil, Ledger and Lionheart~\cite{ledgerCharacterisingBuriedObjects2025} obtain the leading term
\begin{equation}
  \Delta \vH(\vx,\vz) = \nabla_x\nabla_x G_\mathrm{s}(\vx,\vz)\,\mathcal{M}\,\vH_0(\vz) ,
  \label{eq:buried}
\end{equation}
where $\mathcal{M}$ is the MPT of the object in free space and $G_\mathrm{s}$ is a half-space kernel. For the half space this kernel is the transmitted point source $u_\tau/4\pi$ (Section~\sref{sec:s-rangelaws}), and the layer factor \eqref{eq:layerfactor} carries it through the layers. Equation \eqref{eq:buried} is then the TE term of \eqref{eq:shape} at degree one, without reflection, with two changes: the block is replaced by its limit \eqref{eq:mpt}, and the propagator $\Prop^{(1)}_{\TE}/4\pi\alpha^3$ by $\nabla_x\nabla_xG_\mathrm{s}$.

Each step from \eqref{eq:shape} to \eqref{eq:buried} is small in one of the two factors of
\begin{equation}
  \eps = \Bigl(\frac{\alpha}{D}\Bigr)^2\cdot2\Bigl(\frac{D}{\dels}\Bigr)^2 .
  \label{eq:epsfactor}
\end{equation}
The first, the size of the object against its depth, removes the higher degrees, smaller by $\alpha/D$, and the reflection, of $O\bigl((\alpha/2D)^3\bigr)$ and $O\bigl((\alpha/2(\hs-D))^3\bigr)$, and takes $\eps\to0$, which gives \eqref{eq:mpt}. The second, the induction of the host over the depth, controls the other two steps. Dropping the three blocks with a galvanic drive or response neglects a share \eqref{eq:ratio} $\gtrsim(D/\dels)^2$, and moving the vertical derivative of a horizontal magnetic moment from the source, where it inserts $\gams$, to the sensor, where it inserts $k_\rho$, changes the field by $O\bigl((R/\dels)^2\bigr)$ (Section~\sref{sec:s-rangelaws}). For the example of Section~\ref{sec:intro}, $D/\dels\approx0.6$, so neither step is small. As $\sigs\to0$ with $\murs=1$, every step except the truncation in degree becomes exact, and \eqref{eq:buried} becomes the free-space model of~\cite{ammariTargetDetectionCharacterization2014,ledgerCharacterizingShapeMaterial2015}.

\begin{figure*}[!t]
\centering
\includegraphics{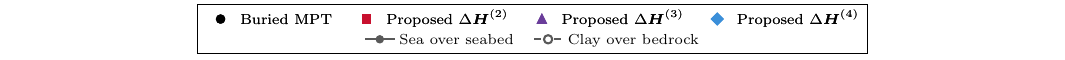}\\[2pt]
\subfloat[]{\includegraphics{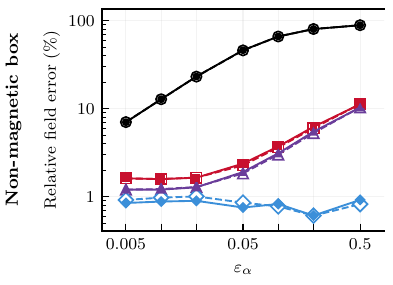}\label{fig:erroreps}}%
\subfloat[]{\includegraphics{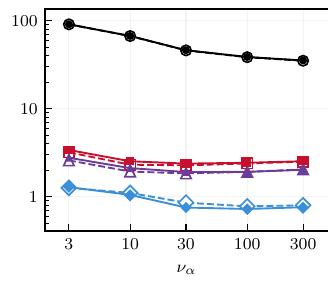}\label{fig:errornu}}%
\subfloat[]{\includegraphics{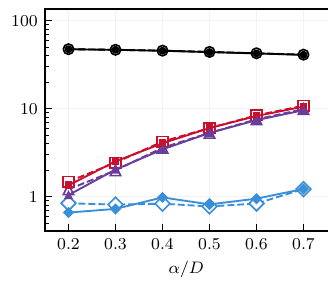}\label{fig:errorsize}}\\
\subfloat[]{\includegraphics{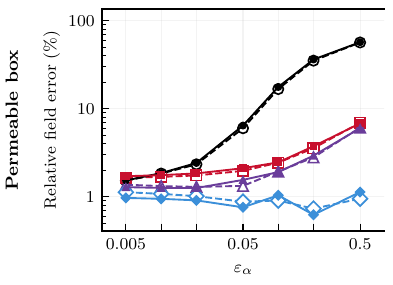}\label{fig:errorepsperm}}%
\subfloat[]{\includegraphics{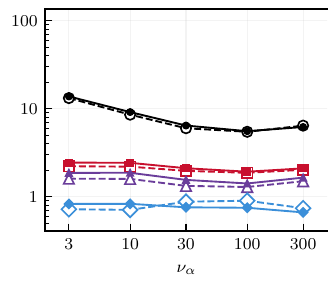}\label{fig:errornuperm}}%
\subfloat[]{\includegraphics{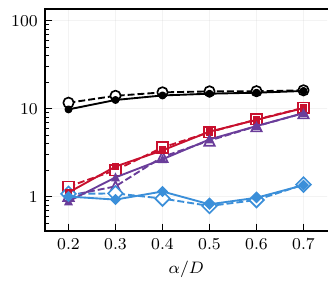}\label{fig:errorsizeperm}}
\caption{Relative field error over the sensor ring in the sea and on land, (a)--(c)~for the non-magnetic box and (d)--(f)~for the permeable box, swept from the base case $\eps=0.05$, $\nua=30$, $\alpha/D=0.3$. (a), (d)~Against $\eps$, from $0.005$ to $0.5$ by the frequency, with $\sigma_*$ scaled to hold $\nua$. (b), (e)~Against $\nua$, from $3$ to $300$ by $\sigma_*$. (c), (f)~Against $\alpha/D$, from $0.2$ to $0.7$ at the base frequency, so that $D/\dels$ is fixed and $\eps$ grows as $\alpha^2$, from $0.022$ to $0.27$.}
\label{fig:errors}
\end{figure*}

\section{Numerical Experiments}
\label{sec:numerics}
The model is tested against full finite element solutions of the eddy-current problem for an object in two layered media, a shallow sea and a layer of clay on land, in the arrangement of Fig.~\ref{fig:scene} with the parameters of Table~\ref{tab:setup}. The reference solves for the perturbation directly, and the drive $\vd$ is computed exactly from the coil by \eqref{eq:drivereciprocity}, so that every factor of the model is under test (Section~\sref{sec:reference}). The error of a model is its relative field error over the sensor ring of Table~\ref{tab:setup}, $\lVert\Delta \vH_{\mathrm{model}}-\Delta \vH_{\mathrm{ref}}\rVert/\lVert\Delta \vH_{\mathrm{ref}}\rVert$, with each field stacked over its complex components at every ring sensor.

The object is a conducting box with one corner removed, in two versions. At the base case the non-magnetic box has the contrasts \eqref{eq:contrasts} $\zs=600$ and $\zm=1$, and the galvanic response carries most of its perturbation. The permeable box has $\zs=6$ and $\zm=100$, and so the same $\nua=\zs\zm\eps$; its magnetisation, its eddy currents and the host current it diverts all enter the perturbation. In both media the interface beneath the top layer lies $0.4D$ below its centre, a little beyond its circumscribed radius of $0.37D$, so that the perturbation over the sensor ring differs by $10\%$ from that over a half space of the top layer. From the base case, the first two sweeps change $\eps$ and $\nua$ alone, and the third enlarges the object at a fixed depth and frequency, which by \eqref{eq:epsfactor} holds $D/\dels$ and lets $\eps$ grow as $\alpha^2$. For this sweep the top layer is deepened to $100$\,m in the sea, over $50$\,m of sediment, and to $20$\,m on land, so that the largest object, of circumscribed radius $0.85D$, clears the interface beneath it as it clears the surface.

\begin{table}[!t]
\caption{Parameters of the numerical experiments.}
\label{tab:setup}
\centering
\setlength{\tabcolsep}{4pt}
\begin{tabular}{@{}lll@{}}
\toprule
 & Sea & Land\\
\midrule
Top layer & Seawater, $4$\,S/m, $70$\,m & Clay, $0.1$\,S/m, $10$\,m\\
Beneath & Sediment, $1$\,S/m, $10$\,m & Bedrock, $0.002$\,S/m\\
 & Rock, $0.02$\,S/m &\\
Object & \multicolumn{2}{l}{Notched box, half-sides $\alpha$, $0.6\alpha$, $0.35\alpha$}\\
 & $\alpha=15$\,m, $D=50$\,m & $\alpha=2.1$\,m, $D=7.1$\,m\\
 & $\sigma_*=2400$\,S/m, $\murb=1$ & $\sigma_*=60$\,S/m, $\murb=1$\\
 & $\sigma_*=24$\,S/m, $\murb=100$ & $\sigma_*=0.6$\,S/m, $\murb=100$\\
Frequency & $7.0$\,Hz & $13.8$\,kHz\\
Transmitter & \multicolumn{2}{l}{Coil of diameter $0.02D$ at altitude $D$, offset $0.8D$}\\
Sensors & \multicolumn{2}{l}{Ring of $17$ points at altitude $D$: above the centre,}\\
 & \multicolumn{2}{l}{and $8$ on each circle of radius $D$ and $2D$ about it}\\
\bottomrule
\end{tabular}
\end{table}

Four models are compared. The baseline is the buried MPT of~\cite{ledgerCharacterisingBuriedObjects2025}, the free-space polarizability tensor with the sensor-side kernel of \eqref{eq:buried}, driven by the same field. The other three are \eqref{eq:shape} truncated by order: $\Delta \vH^{(K)}$ keeps the blocks of $\Tens$ of order at most $K$ and those of $\Ret$ with $n+m\le K$, every block that contributes up to $O(\alpha^{K+1})$ (Section~\sref{sec:s-truncation}). The dipole model $\Delta \vH^{(2)}$ keeps the diagonal blocks $(1,1)$, $\Delta \vH^{(3)}$ adds the cross blocks $(1,1)$ and the diagonal blocks $(2,1)$ and $(1,2)$, and $\Delta \vH^{(4)}$ adds the cross blocks of total degree three and the diagonal blocks of total degree four. The gain from each component of the model is shown in Section~\sref{sec:s-ladder}, where the components are added to the baseline one at a time.

The buried MPT fails in proportion to the share \eqref{eq:ratio} of the galvanic response that it neglects. That share dominates the perturbation of the non-magnetic box, for which the error of the buried MPT reaches $91\%$. In the permeable box the magnetisation strengthens the inductive response, and at the lowest $\eps$, where the host barely conducts over the scale of the object, the buried MPT is as accurate as the dipole model, the limit in which it was derived. For both boxes its error grows with $\eps$ (Figs.~\ref{fig:erroreps} and~\ref{fig:errorepsperm}), falls as $\nua$ strengthens the inductive response (Figs.~\ref{fig:errornu} and~\ref{fig:errornuperm}), and is set by the depth rather than by the size of the object (Figs.~\ref{fig:errorsize} and~\ref{fig:errorsizeperm}).

With the galvanic response kept, the dipole model $\Delta \vH^{(2)}$ is limited by truncation, and its error grows to about $10\%$ wherever the drive varies across the object, at the highest $\eps$ and for the largest object. The order three, whose terms arise from the notch alone (Section~\ref{sec:cross}), changes this little. The order four removes the truncation error: $\Delta \vH^{(4)}$ holds to within $1.4\%$ in every case for both boxes, and only with the reflection matrix, without which its error reaches $2.3\%$.

What remains is the error of the reference, not of the model. Refining the reference and the transmission problem together at the base case lowers the error of $\Delta \vH^{(4)}$ to $0.1\%$ for the non-magnetic box and $0.2\%$ for the permeable one, while each further order changes the field by less than $0.1\%$ (Section~\sref{sec:s-budget}).

The T-matrix also carries what the established tensors cannot. Its galvanic dipole block agrees with $\mathcal{P}$ of Section~\ref{sec:properties} for the non-magnetic box while $\eps$ is small, to within $0.8\%$ in its real part, but departs from it as the host induces across the object, by $14\%$ at $\eps=0.5$; for the permeable box it departs by up to $27\%$ already at the base $\eps$, where $\zm\eps=5$ is not small. The cross blocks of total degree three, which belong to neither tensor, carry up to $13\%$ of the perturbation.

In every sweep the errors of the model on land follow those in the sea to within $0.4$ percentage points: the two grounds differ in the layers beneath, but in both a resistive layer lies close below the object and turns back almost all of the current that reaches it.

\section{Conclusion}
\label{sec:conclusion}
This paper characterises a compact object in a conducting, permeable layered medium independently of its surroundings, as the MPT does in free space, through a T-matrix that maps the regular TE and TM modes at the object to outgoing modes of both types, so that its inductive and galvanic responses are carried side by side. The T-matrix is computed from one transmission problem and reciprocity integrals, for any shape and materials, and is the only factor of the model that depends on the object; the drive, the propagator and the reflection matrix follow from the point source of the host, carried through the layers.

The diagonal blocks contain the established tensors as limits: the inductive block is the MPT in a weakly conducting host, and the galvanic block is the electric polarizability tensor in the static limit. The two responses enter at the same order in the size of the object, so higher-order tensors cannot supply the galvanic one. The off-diagonal blocks belong to neither tensor, and for an object with a centre of symmetry they enter the first correction beyond the dipole together with the diagonal blocks of total degree four. Against full finite element solutions for a conducting object, non-magnetic and permeable, in a layered sea and in clay over bedrock, the model holds to within $1.4\%$ at two orders beyond the dipole, against errors of up to $91\%$ for the buried MPT, and most of what remains is the error of the reference.

The layers are assumed flat, and the object is confined to one of them, further from each interface than its circumscribed radius; an object that straddles an interface is not described. The removal of the primary field and the response of the host, both of which reach the sensor without the object, is not addressed, nor is the recovery of the T-matrix, and through it the object, from measured fields.

\section*{Acknowledgments}
We acknowledge support from the Queensland University of Technology (QUT) through its provision of subsidised access to specialised research infrastructure including HPC and the eResearch team at QUT for providing expertise in enablement of this project.

\printbibliography

@book{ammariPolarizationMomentTensors2007,
  title = {Polarization and {{Moment Tensors}}},
  author = {Ammari, Habib and Kang, Hyeonbae},
  date = {2007},
  series = {Applied {{Mathematical Sciences}}},
  volume = {162},
  publisher = {Springer},
  location = {New York, NY},
  doi = {10.1007/978-0-387-71566-7}
}

@article{ammariShapeRecognitionClassification2014,
  title = {Shape Recognition and Classification in Electro-Sensing},
  author = {Ammari, Habib and Boulier, Thomas and Garnier, Josselin and Wang, Han},
  date = {2014-08-12},
  journaltitle = {Proceedings of the National Academy of Sciences},
  volume = {111},
  number = {32},
  pages = {11652--11657},
  doi = {10.1073/pnas.1406513111}
}

@article{ammariTargetDetectionCharacterization2014,
  title = {Target Detection and Characterization from Electromagnetic Induction Data},
  author = {Ammari, Habib and Chen, Junqing and Chen, Zhiming and Garnier, Josselin and Volkov, Darko},
  date = {2014-01-01},
  journaltitle = {Journal de Mathématiques Pures et Appliquées},
  volume = {101},
  number = {1},
  pages = {54--75},
  doi = {10.1016/j.matpur.2013.05.002}
}

@article{demesyScatteringMatrixArbitrarily2018,
  title = {Scattering Matrix of Arbitrarily Shaped Objects: Combining Finite Elements and Vector Partial Waves},
  author = {Demésy, Guillaume and Auger, Jean-Claude and Stout, Brian},
  date = {2018-08-01},
  journaltitle = {Journal of the Optical Society of America A},
  volume = {35},
  number = {8},
  pages = {1401--1409},
  publisher = {Optica Publishing Group},
  doi = {10.1364/JOSAA.35.001401}
}

@article{egelSMUTHIPythonPackage2021,
  title = {{{SMUTHI}}: {{A}} Python Package for the Simulation of Light Scattering by Multiple Particles near or between Planar Interfaces},
  author = {Egel, Amos and Czajkowski, Krzysztof M. and Theobald, Dominik and Ladutenko, Konstantin and Kuznetsov, Alexey S. and Pattelli, Lorenzo},
  date = {2021-10},
  journaltitle = {Journal of Quantitative Spectroscopy and Radiative Transfer},
  volume = {273},
  pages = {107846},
  doi = {10.1016/j.jqsrt.2021.107846}
}

@book{harringtonTimeharmonicElectromagneticFields2001,
  title = {Time-Harmonic Electromagnetic Fields},
  author = {Harrington, Roger F.},
  date = {2001},
  publisher = {Wiley-IEEE Press},
  location = {New York},
  doi = {10.1109/9780470546710}
}

@book{hobsonTheorySphericalEllipsoidal1931,
  title = {The {{Theory}} of {{Spherical}} and {{Ellipsoidal Harmonics}}},
  author = {Hobson, E. W.},
  date = {1931},
  publisher = {CUP Archive}
}

@article{hohmannThreedimensionalInducedPolarization1975,
  title = {Three-Dimensional Induced Polarization and Electromagnetic Modeling},
  author = {Hohmann, G. W.},
  date = {1975-04-01},
  journaltitle = {Geophysics},
  volume = {40},
  number = {2},
  pages = {309--324},
  doi = {10.1190/1.1440527}
}

@report{kristenssonElectromagneticScatteringBuried1979,
  type = {Technical Report},
  title = {Electromagnetic Scattering from a Buried Three-Dimensional Inhomogeneity in a Lossy Ground},
  author = {Kristensson, Gerhard},
  date = {1979},
  number = {79-29},
  institution = {Institute of Theoretical Physics, Chalmers University of Technology},
  location = {Göteborg}
}

@article{kristenssonElectromagneticScatteringBuried1980,
  title = {Electromagnetic Scattering from Buried Inhomogeneities—a General Three-Dimensional Formalism},
  author = {Kristensson, Gerhard},
  date = {1980-07-01},
  journaltitle = {Journal of Applied Physics},
  volume = {51},
  number = {7},
  pages = {3486--3500},
  doi = {10.1063/1.328201}
}

@article{kristenssonScatteringBuriedInhomogeneities1978,
  title = {Scattering from Buried Inhomogeneities—a General Three-Dimensional Formalism},
  author = {Kristensson, Gerhard and Ström, Staffan},
  date = {1978-09-01},
  journaltitle = {The Journal of the Acoustical Society of America},
  volume = {64},
  number = {3},
  pages = {917--936},
  doi = {10.1121/1.382052}
}

@book{kristenssonScatteringElectromagneticWaves2016,
  title = {Scattering of {{Electromagnetic Waves}} by {{Obstacles}}},
  author = {Kristensson, Gerhard},
  date = {2016-07-22},
  publisher = {{Institution of Engineering and Technology}},
  doi = {10.1049/SBEW524E}
}

@unpublished{ledgerCharacterisingBuriedObjects2025,
  title = {Characterising Buried Objects in Metal Detection},
  author = {Ledger, P. D. and Lionheart, W. R. B.},
  date = {2025},
  eprint = {2507.04450},
  eprinttype = {arXiv},
  eprintclass = {math.AP},
  publisher = {arXiv},
  doi = {10.48550/arXiv.2507.04450}
}

@article{ledgerCharacterizingShapeMaterial2015,
  title = {Characterizing the Shape and Material Properties of Hidden Targets from Magnetic Induction Data},
  author = {Ledger, Paul D. and Lionheart, William R. B.},
  date = {2015-12-01},
  journaltitle = {IMA Journal of Applied Mathematics},
  volume = {80},
  number = {6},
  pages = {1776--1798},
  doi = {10.1093/imamat/hxv015}
}

@article{ledgerGeneralisedMagneticPolarizability2018,
  title = {Generalised Magnetic Polarizability Tensors},
  author = {Ledger, P. D. and Lionheart, W. R. B.},
  date = {2018},
  journaltitle = {Mathematical Methods in the Applied Sciences},
  volume = {41},
  number = {8},
  pages = {3175--3196},
  doi = {10.1002/mma.4809}
}

@article{ledgerIdentificationMetallicObjects2021,
  title = {Identification of Metallic Objects Using Spectral Magnetic Polarizability Tensor Signatures: {{Object}} Characterisation and Invariants},
  author = {Ledger, Paul D. and Wilson, Ben A. and Amad, Alan A. S. and Lionheart, William R. B.},
  date = {2021},
  journaltitle = {International Journal for Numerical Methods in Engineering},
  volume = {122},
  number = {15},
  pages = {3941--3984},
  doi = {10.1002/nme.6688}
}

@article{makkonenKNNClassificationMetallic2014,
  title = {{{KNN}} Classification of Metallic Targets Using the Magnetic Polarizability Tensor},
  author = {Makkonen, J and Marsh, L A and Vihonen, J and Järvi, A and Armitage, D W and Visa, A and Peyton, A J},
  date = {2014-03},
  journaltitle = {Measurement Science and Technology},
  volume = {25},
  number = {5},
  pages = {055105},
  publisher = {IOP Publishing},
  doi = {10.1088/0957-0233/25/5/055105}
}

@book{mishchenkoScatteringAbsorptionEmission2002,
  title = {Scattering, Absorption, and Emission of Light by Small Particles},
  author = {Mishchenko, Michael I. and Travis, Larry D. and Lacis, Andrew A.},
  date = {2002},
  publisher = {Cambridge University Press},
  location = {Cambridge ; New York}
}

@article{nortonEddycurrentCurrentchannelingResponse2005,
  title = {Eddy-Current and Current-Channeling Response to Spheroidal Anomalies},
  author = {Norton, S.J. and SanFilipo, W.A. and Won, I.J.},
  date = {2005-10},
  journaltitle = {IEEE Transactions on Geoscience and Remote Sensing},
  volume = {43},
  number = {10},
  pages = {2200--2209},
  doi = {10.1109/TGRS.2005.856641}
}

@book{olverNISTHandbookMathematical2010,
  title = {{{NIST}} Handbook of Mathematical Functions},
  editor = {Olver, Frank W. J. and {National Institute of Standards and Technology (U.S.)}},
  date = {2010},
  publisher = {Cambridge University Press : NIST},
  location = {Cambridge ; New York}
}

@article{pasionDiscriminationAlgorithmUXO2001,
  title = {A {{Discrimination Algorithm}} for {{UXO Using Time Domain Electromagnetics}}},
  author = {Pasion, Leonard R. and Oldenburg, Douglas W.},
  date = {2001-06-01},
  journaltitle = {Journal of Environmental and Engineering Geophysics},
  volume = {6},
  number = {2},
  pages = {91--102},
  doi = {10.4133/JEEG6.2.91}
}

@book{polyaIsoperimetricInequalitiesMathematical1951,
  title = {Isoperimetric {{Inequalities}} in {{Mathematical Physics}}},
  author = {Pólya, G. and Szegő, G.},
  date = {1951-12-31},
  publisher = {Princeton University Press},
  doi = {10.1515/9781400882663}
}

@inproceedings{sanfilipoEffectsSeawaterEMI2005,
  title = {The Effects of Seawater on the {{EMI}} Response of {{UXO}}},
  booktitle = {Proceedings of {{OCEANS}} 2005 {{MTS}}/{{IEEE}}},
  author = {SanFilipo, B. and Norton, S. and Won, I.J.},
  date = {2005-09},
  pages = {607-614 Vol. 1},
  doi = {10.1109/OCEANS.2005.1639821},
  eventtitle = {{{OCEANS}} 2005 {{MTS}}/{{IEEE}}}
}

@article{sommerfeldarnoldUberAusbreitungWellen1909,
  title = {Über Die {{Ausbreitung}} Der {{Wellen}} in Der Drahtlosen {{Telegraphie}}},
  author = {Sommerfeld, A.},
  date = {1909},
  journaltitle = {Annalen der Physik},
  volume = {333},
  number = {4},
  pages = {665--736},
  doi = {10.1002/andp.19093330402}
}

@article{songTransientElectromagneticScattering2016,
  title = {Transient {{Electromagnetic Scattering}} of a {{Metallic Object Buried}} in {{Underwater Sediments}}},
  author = {Song, Lin-Ping and Billings, S. D. and Pasion, L. R. and Oldenburg, Douglas W.},
  date = {2016-02},
  journaltitle = {IEEE Transactions on Geoscience and Remote Sensing},
  volume = {54},
  number = {2},
  pages = {1091--1102},
  doi = {10.1109/TGRS.2015.2473851}
}

@book{strattonElectromagneticTheory2007,
  title = {Electromagnetic {{Theory}}},
  author = {Stratton, Julius Adams},
  date = {1941},
  publisher = {John Wiley \& Sons}
}

@article{wannamakerElectromagneticModelingThreedimensional1984,
  title = {Electromagnetic Modeling of Three-Dimensional Bodies in Layered Earths Using Integral Equations},
  author = {Wannamaker, Philip E. and Hohmann, Gerald W. and SanFilipo, William A.},
  date = {1984-01-01},
  journaltitle = {Geophysics},
  volume = {49},
  number = {1},
  pages = {60--74},
  doi = {10.1190/1.1441562}
}

@incollection{wardElectromagneticTheoryGeophysical1988,
  title = {Electromagnetic {{Theory}} for {{Geophysical Applications}}},
  booktitle = {Electromagnetic {{Methods}} in {{Applied Geophysics}}: {{Volume}} 1, {{Theory}}},
  author = {Ward, Stanley H. and Hohmann, Gerald W.},
  editor = {Nabighian, Misac N.},
  date = {1988-01},
  pages = {130--311},
  publisher = {Society of Exploration Geophysicists},
  doi = {10.1190/1.9781560802631.ch4}
}

@article{watermanMatrixFormulationElectromagnetic1965,
  title = {Matrix Formulation of Electromagnetic Scattering},
  author = {Waterman, P. C.},
  date = {1965},
  journaltitle = {Proceedings of the IEEE},
  volume = {53},
  number = {8},
  pages = {805--812},
  doi = {10.1109/PROC.1965.4058}
}

@incollection{westPHYSICSELECTROMAGNETICINDUCTION1991,
  title = {Physics of the Electromagnetic Induction Exploration Method},
  booktitle = {Electromagnetic {{Methods}} in {{Applied Geophysics}}: {{Volume}} 2, {{Application}}, {{Parts A}} and {{B}}},
  author = {West, G. F. and Macnae, J. C.},
  editor = {Nabighian, Misac N. and Corbett, John D.},
  date = {1991},
  pages = {5--46},
  publisher = {Society of Exploration Geophysicists},
  doi = {10.1190/1.9781560802686.ch1}
}

@article{westSimpleParametricModel1985,
  title = {A Simple Parametric Model for the Electromagnetic Response of an Anomalous Body in a Host Medium},
  author = {West, G. F. and Edwards, R. N.},
  date = {1985-12-01},
  journaltitle = {Geophysics},
  volume = {50},
  number = {12},
  pages = {2542--2557},
  doi = {10.1190/1.1441883}
}

@article{winchGeomagnetismSchmidtQuasinormalization2005,
  title = {Geomagnetism and {{Schmidt}} Quasi-Normalization},
  author = {Winch, D. E. and Ivers, D. J. and Turner, J. P. R. and Stening, R. J.},
  date = {2005-02},
  journaltitle = {Geophysical Journal International},
  volume = {160},
  number = {2},
  pages = {487--504},
  doi = {10.1111/j.1365-246X.2004.02472.x}
}

\newcommand{\biophoto}[1]{%
  \IfFileExists{figures/bio/#1.jpg}%
    {\includegraphics[width=1in,height=1.25in,clip,keepaspectratio]{bio/#1}}%
    {\fbox{\parbox[c][1.15in][c]{0.9in}{\centering\scriptsize photo}}}}

\begin{IEEEbiography}[{\biophoto{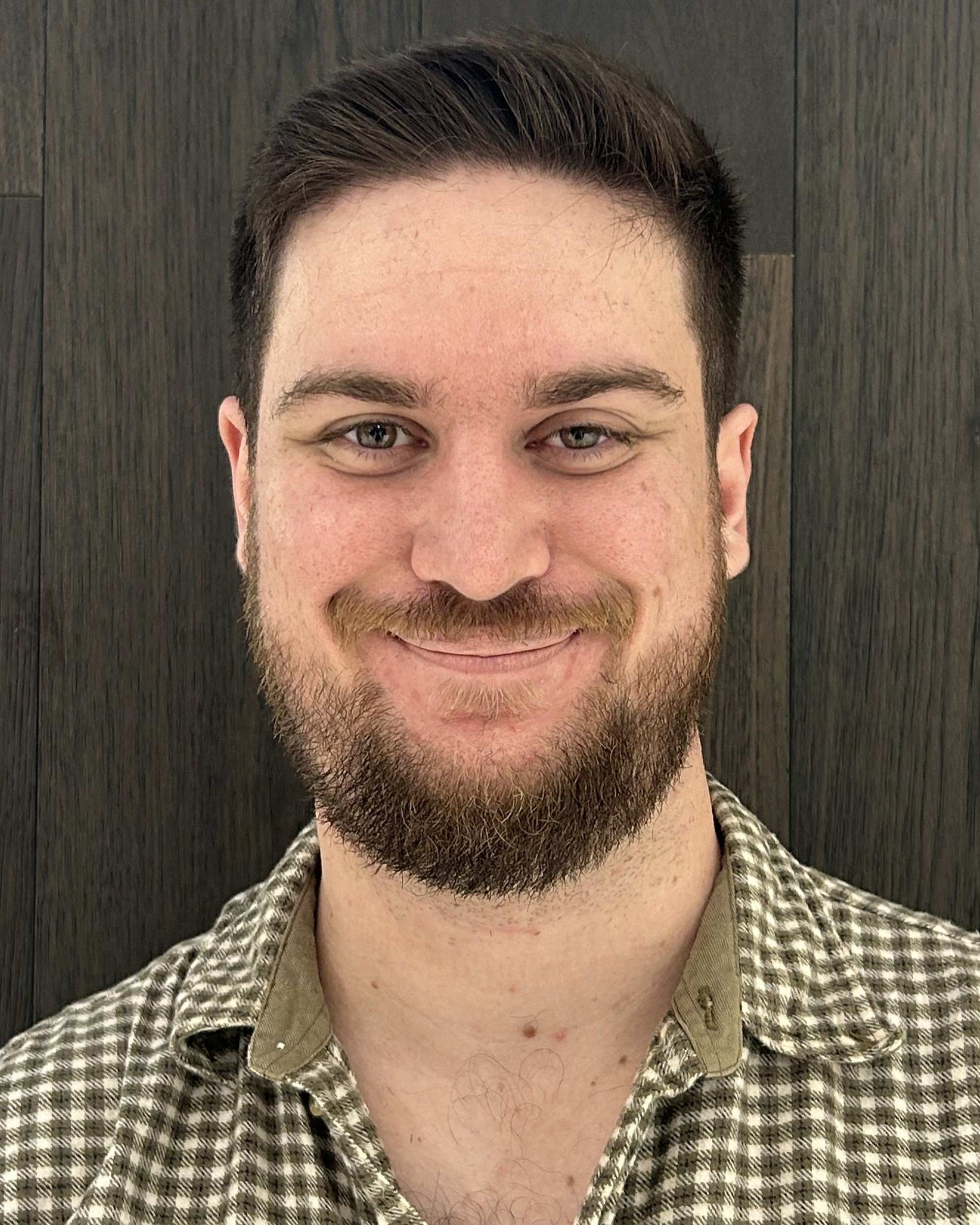}}]{Fraser Williams}
received the B.Eng. degree in electrical engineering and the B.Sc. degree in physics, both in 2021, and the Ph.D. degree in electrical engineering in 2026, all from the Queensland University of Technology, Brisbane, QLD, Australia.

He is currently an Associate Lecturer with the School of Electrical Engineering and Robotics, Queensland University of Technology, Brisbane, QLD, Australia. His research interests include electromagnetic modeling, inverse electromagnetics, and statistical signal processing.
\end{IEEEbiography}

\begin{IEEEbiography}[{\biophoto{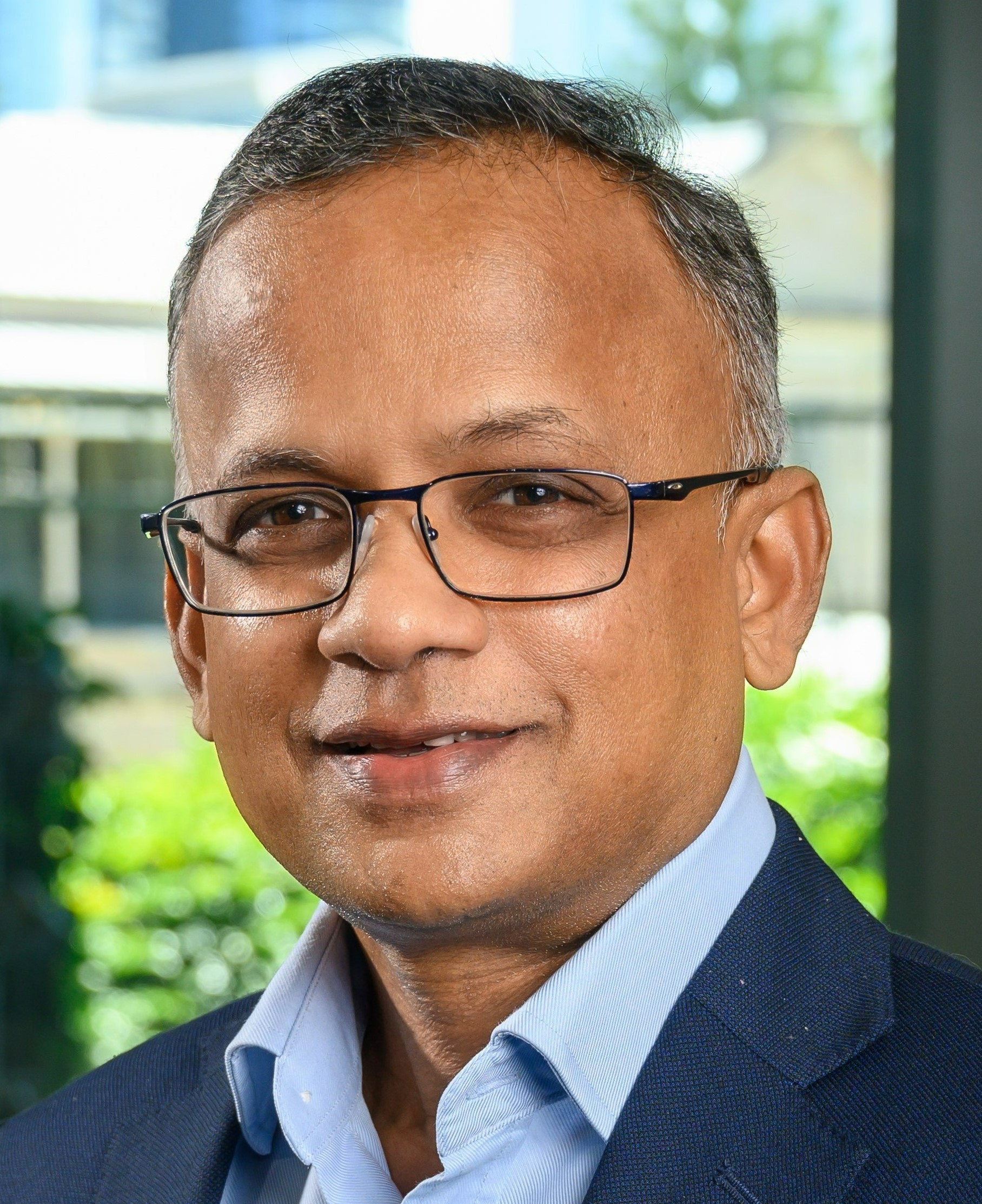}}]{Dhammika Jayalath}
(Senior Member, IEEE) received the B.Sc. degree in electronics and telecommunications engineering from the University of Moratuwa, Moratuwa, Sri Lanka, in 1994, the M.Eng. degree in telecommunications from the Asian Institute of Technology, Pathum Thani, Thailand, in 1998, and the Ph.D. degree in wireless communications from Monash University, Melbourne, VIC, Australia, in 2002.

He is currently an Associate Professor with the School of Electrical Engineering and Robotics, Queensland University of Technology, Brisbane, QLD, Australia, which he joined in 2007. He was previously a Fellow with the Australian National University, Canberra, ACT, Australia, and a Senior Researcher with National ICT Australia. His research interests include communications and signal processing, in particular machine learning for communication systems, physical-layer security, and signal design for robust communications.
\end{IEEEbiography}

\begin{IEEEbiography}[{\biophoto{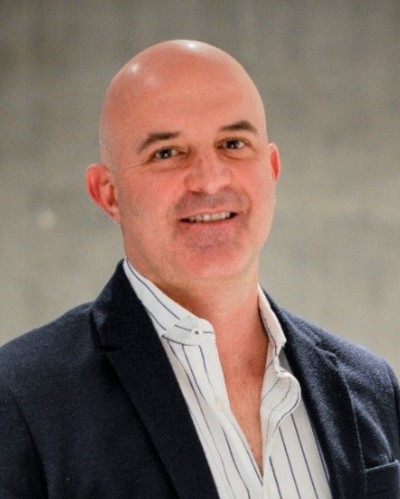}}]{Clinton Fookes}
(Senior Member, IEEE) received the B.Eng. degree in aerospace avionics in 1999, the Ph.D. degree in computer vision in 2004, and the M.B.A. degree in 2011, all from the Queensland University of Technology, Brisbane, QLD, Australia.

He is currently a Professor and the Head of the School of Electrical Engineering and Robotics, Queensland University of Technology. His research interests include computer vision, machine learning, signal processing, and artificial intelligence.

He serves on the editorial boards of the \textit{IEEE Transactions on Image Processing} and \textit{Pattern Recognition}, and previously served on that of the \textit{IEEE Transactions on Information Forensics and Security}. He is a Fellow of the International Association for Pattern Recognition and of the Australian Academy of Technological Sciences and Engineering, and a Senior Fulbright Scholar. His awards include the Australian Institute of Policy and Science Young Tall Poppy Award, an Australian Museum Eureka Prize, an Engineers Australia Engineering Excellence Award, and the Australian Defence Scientist of the Year.
\end{IEEEbiography}

\begin{IEEEbiography}[{\biophoto{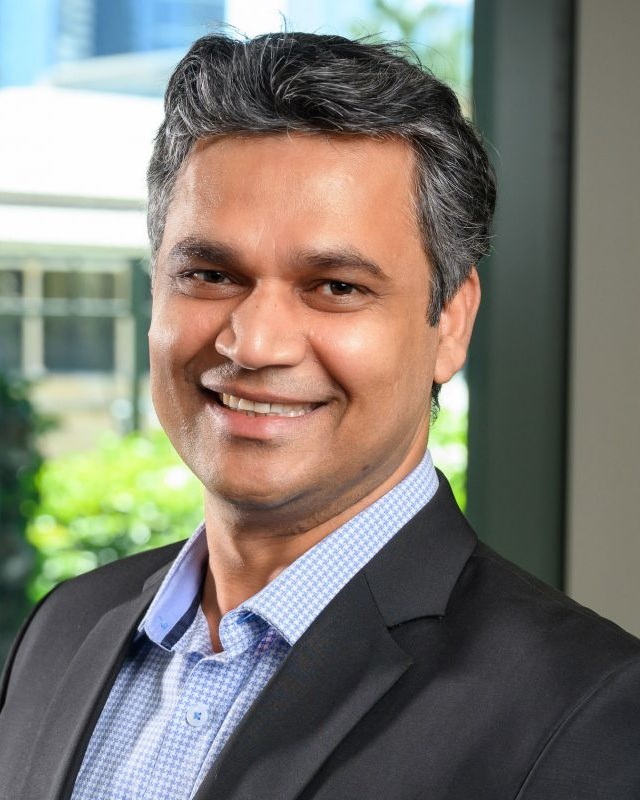}}]{Ajay Pandey}
received the M.Sc. degree in physics from the University of Allahabad, Prayagraj, India, in 2001, the M.Tech. degree in laser science and applications from Devi Ahilya University, Indore, India, in 2003, and the Ph.D. degree in physics from the University of Angers, Angers, France, in 2007, for work on molecular optoelectronics.

He is currently a Professor with the School of Electrical Engineering and Robotics, Queensland University of Technology, Brisbane, QLD, Australia. His research interests include photon upconversion, single-photon emission and detection, and integrated photonic platforms for quantum sensing.
\end{IEEEbiography}

\vfill

\end{document}


\title{Supplementary Material: A Diffusive T-Matrix Unifying the Electric and Magnetic Polarizability Tensors in Conducting Layered Media}

\author{Fraser~Williams\orcidlink{0000-0002-5422-2827}, Dhammika~Jayalath\orcidlink{0000-0001-6130-0275}, Clinton~Fookes\orcidlink{0000-0002-8515-6324}, Ajay~Pandey\orcidlink{0000-0002-6599-745X}}

\markboth{Williams \emph{et al.}: A Diffusive T-Matrix Unifying the Electric and Magnetic Polarizability Tensors}%
{Williams \emph{et al.}: A Diffusive T-Matrix Unifying the Electric and Magnetic Polarizability Tensors}

\maketitle

\section{The T-Matrix}
\label{sec:s-tmatrix}
This section lists the harmonics and the properties of the vector modes, gives the discretisation of the transmission problem and the reduction of the extraction integral, finds the order of each block in the size of the object, relates the response of degree one to point dipoles, shows the four blocks for the object of the numerical experiments, shows which blocks survive in a non-conducting host, and proves the symmetry \meqref{eq:symmetry} of the T-matrix under exchange of probe and drive.

\subsection{Real Solid Harmonics}
\label{sec:s-harmonics}
The solid harmonics of Section~\mref{sec:sea} are $P^p_m(\vxi)=r^mY^p_m(\theta,\phi)$, with $\theta$ the polar angle of $\vxi$ from $\ehat_3$, $\phi$ its azimuth from $\ehat_1$, and
\begin{equation}
  Y^p_m(\theta,\phi) = \Theta^{|p|}_m(\theta)\times
  \begin{cases}
    \cos p\phi , & p\ge0 ,\\
    \sin|p|\phi , & p<0 .
  \end{cases}
  \label{eq:schmidt}
\end{equation}
Here
\begin{equation*}
  \Theta^k_m(\theta) = s^k_m\,\sin^k\theta\,\frac{1}{2^mm!}\,\frac{\mathrm{d}^{m+k}}{\mathrm{d}x^{m+k}}(x^2-1)^m\Big|_{x=\cos\theta} ,
  \qquad
  s^k_m = \Bigl[(2-\delta_{k0})\frac{(m-k)!}{(m+k)!}\Bigr]^{1/2} ,
\end{equation*}
is the associated Legendre function in the Schmidt quasi-normalisation of geomagnetism~\cite{winchGeomagnetismSchmidtQuasinormalization2005}, built by Rodrigues' formula~\cite[eq.~(14.7.13)]{olverNISTHandbookMathematical2010}. Then $\oint(Y^p_m)^2\dd\Omega=4\pi/(2m+1)$ for every order, which gives the constant $\beta_m=1/4\pi m(m+1)$ of \meqref{eq:reaction}. Table~\ref{tab:harmonics} lists the harmonics of degrees one to three, which are all the paper uses, both as functions of angle and as polynomials in $\vxi=(\xi_1,\xi_2,\xi_3)$.

\begin{table}[!htb]
\caption{Schmidt quasi-normalised real spherical harmonics $Y^p_m(\theta,\phi)$ and solid harmonics $P^p_m(\vxi)=r^mY^p_m$ of degrees one to three, with the constant $\beta_m$ of \meqref{eq:reaction}. Every harmonic has $\oint(Y^p_m)^2\dd\Omega=4\pi/(2m+1)$.}
\label{tab:harmonics}
\centering
\renewcommand{\arraystretch}{1.2}
\begin{tabular}{@{}cccll@{}}
\toprule
$m$ & $\beta_m$ & $p$ & $Y^p_m(\theta,\phi)$ & $P^p_m(\vxi)$\\
\midrule
$1$ & $1/8\pi$ & $0$ & $\cos\theta$ & $\xi_3$\\
 & & $1$ & $\sin\theta\cos\phi$ & $\xi_1$\\
 & & $-1$ & $\sin\theta\sin\phi$ & $\xi_2$\\
\midrule
$2$ & $1/24\pi$ & $0$ & $\tfrac12(3\cos^2\theta-1)$ & $\tfrac12(2\xi_3^2-\xi_1^2-\xi_2^2)$\\
 & & $1$ & $\sqrt3\,\sin\theta\cos\theta\cos\phi$ & $\sqrt3\,\xi_1\xi_3$\\
 & & $-1$ & $\sqrt3\,\sin\theta\cos\theta\sin\phi$ & $\sqrt3\,\xi_2\xi_3$\\
 & & $2$ & $\tfrac{\sqrt3}{2}\sin^2\theta\cos2\phi$ & $\tfrac{\sqrt3}{2}(\xi_1^2-\xi_2^2)$\\
 & & $-2$ & $\tfrac{\sqrt3}{2}\sin^2\theta\sin2\phi$ & $\sqrt3\,\xi_1\xi_2$\\
\midrule
$3$ & $1/48\pi$ & $0$ & $\tfrac12\cos\theta\,(5\cos^2\theta-3)$ & $\tfrac12\xi_3(2\xi_3^2-3\xi_1^2-3\xi_2^2)$\\
 & & $1$ & $\tfrac{\sqrt6}{4}\sin\theta\,(5\cos^2\theta-1)\cos\phi$ & $\tfrac{\sqrt6}{4}\xi_1(4\xi_3^2-\xi_1^2-\xi_2^2)$\\
 & & $-1$ & $\tfrac{\sqrt6}{4}\sin\theta\,(5\cos^2\theta-1)\sin\phi$ & $\tfrac{\sqrt6}{4}\xi_2(4\xi_3^2-\xi_1^2-\xi_2^2)$\\
 & & $2$ & $\tfrac{\sqrt{15}}{2}\cos\theta\sin^2\theta\cos2\phi$ & $\tfrac{\sqrt{15}}{2}\xi_3(\xi_1^2-\xi_2^2)$\\
 & & $-2$ & $\tfrac{\sqrt{15}}{2}\cos\theta\sin^2\theta\sin2\phi$ & $\sqrt{15}\,\xi_1\xi_2\xi_3$\\
 & & $3$ & $\tfrac{\sqrt{10}}{4}\sin^3\theta\cos3\phi$ & $\tfrac{\sqrt{10}}{4}\xi_1(\xi_1^2-3\xi_2^2)$\\
 & & $-3$ & $\tfrac{\sqrt{10}}{4}\sin^3\theta\sin3\phi$ & $\tfrac{\sqrt{10}}{4}\xi_2(3\xi_1^2-\xi_2^2)$\\
\bottomrule
\end{tabular}
\end{table}

\subsection{The Vector Modes}
\label{sec:s-vectormodes}
This section collects the properties of the modes of Section~\mref{sec:sea} that the derivations below use: that they are fields of the host, the closed forms of the scalar modes, their parts on a sphere about the centre, their static forms, and an orthogonality of the solid harmonics.

\emph{Closure under the curl.} Both $\Mw[\psi]=\curl(\vxi\psi)$ and $\Nw[\psi]=\curl\Mw[\psi]$ are curls, so both are divergence-free. Each component of $\vxi\psi$ satisfies
\begin{equation*}
  \nabla^2(\xi_i\psi) = \xi_i\,\nabla^2\psi + 2\,\partial_i\psi ,
\end{equation*}
and the curl of the gradient $2\nabla\psi$ vanishes, so $\nabla^2\Mw[\psi]=\curl(\vxi\,\nabla^2\psi)$. If $(\nabla^2+\kappa^2)\psi=0$, then $\Mw[\psi]$ therefore satisfies \meqref{eq:helmholtz}, and since the curl commutes with $\nabla^2$, so does $\Nw[\psi]$. Then $\curl\Nw=\nabla(\nabla\cdot\Mw)-\nabla^2\Mw=\kappa^2\Mw$, which is \meqref{eq:closure}.

\emph{Fields of the host.} In the host $\kappa^2=\ii\omega\mu_0\murs\sigs\alpha^2=e_{\TE}\,\alpha\sigs$. With \meqref{eq:closure}, the TE pair of \meqref{eq:TE} and the TM pair of \meqref{eq:TM} give
\begin{equation*}
  \begin{aligned}
  \TE:&\quad \curl\vE=e_{\TE}\Nw=\ii\omega\mu_0\murs\alpha\vH , &
  \curl\vH&=\kappa^2\Mw=\alpha\sigs\vE ,\\
  \TM:&\quad \curl\vE=e_{\TM}\kappa^2\Mw=\ii\omega\mu_0\murs\alpha\vH , &
  \curl\vH&=\Nw=\alpha\sigs\vE ,
  \end{aligned}
\end{equation*}
so both are fields of the host, \meqref{eq:scaledpair}.

\emph{The scalar modes.} For a polynomial $P$, the operator $P(\nabla)$ replaces each $\xi_i$ in $P$ with $\partial_i$. For a harmonic polynomial $P$ homogeneous of degree $m$ and a function $f$ of $r$ alone, Hobson's theorem~\cite{hobsonTheorySphericalEllipsoidal1931} gives
\begin{equation*}
  P(\nabla)f(r) = P(\vxi)\Bigl(\frac{1}{r}\frac{\mathrm{d}}{\mathrm{d}r}\Bigr)^mf(r) .
\end{equation*}
With $e^{\ii\kappa r}/r=\ii\kappa\,h^{(1)}_0(\kappa r)$, $\sin\kappa r/r=\kappa\,j_0(\kappa r)$, and
\begin{equation*}
  \Bigl(\frac{1}{r}\frac{\mathrm{d}}{\mathrm{d}r}\Bigr)^m \zeta_0(\kappa r) = (-1)^m\kappa^{2m}\,\frac{\zeta_m(\kappa r)}{(\kappa r)^m} ,
\end{equation*}
with $\zeta_m$ either $j_m$ or $h^{(1)}_m$, and with $P(\vxi)=r^mY$, this gives
\begin{equation*}
  \Psi^p_m = \frac{\ii\,\kappa^{m+1}}{(2m-1)!!}\,h^{(1)}_m(\kappa r)\,Y^p_m ,
  \qquad
  \Phi^q_n = \frac{(2n+1)!!}{\kappa^n}\,j_n(\kappa r)\,Y^q_n ,
\end{equation*}
the second lines of \meqref{eq:Psi} and \meqref{eq:Phi}. As $\kappa r\to0$, $j_n(\kappa r)\to(\kappa r)^n/(2n+1)!!$ and $h^{(1)}_m(\kappa r)\to-\ii(2m-1)!!/(\kappa r)^{m+1}$, which gives the static modes $r^nY^q_n$ and $Y^p_m/r^{m+1}$.

\emph{On a sphere.} With $\psi=f(r)Y^p_m$ and $\nabla_1$ the gradient on the unit sphere, $\nabla\psi=(\mathrm{d}f/\mathrm{d}r)\,Y^p_m\,\xihat+r^{-1}f\,\nabla_1Y^p_m$, so
\begin{equation}
  \Mw[fY^p_m] = -f\,\xihat\times\nabla_1Y^p_m ,
  \qquad
  \Nw[fY^p_m] = \frac{m(m+1)}{r}\,fY^p_m\,\xihat + \frac{1}{r}\frac{\mathrm{d}(rf)}{\mathrm{d}r}\,\nabla_1Y^p_m ,
  \label{eq:tangential}
\end{equation}
the radial part of $\Nw$ following from the Helmholtz equation. Expanding the curl in $\Nw[\psi]=\curl(\nabla\psi\times\vxi)$ leaves
\begin{equation*}
  \xihat\cdot\Nw[\psi] = -\frac{1}{r}\nabla_1^2\psi ,
  \qquad
  \nabla_1^2Y^p_m = -m(m+1)\,Y^p_m ,
\end{equation*}
with $\nabla_1^2$ the Laplacian on the unit sphere. The mode $\Mw$ is tangential to the sphere, and the tangential part of $\Nw$ is a gradient on it.

\emph{The static modes.} At $\kappa=0$ the regular modes are built on the solid harmonics. For $P$ harmonic and homogeneous of degree $n$, $\nabla^2P=0$ and $(\vxi\cdot\nabla)\nabla P=(n-1)\nabla P$. The second is Euler's relation for homogeneous functions. A function $f$ homogeneous of degree $k$ is $r^k$ times a function of angle, and $\vxi\cdot\nabla$ is the radial derivative $r\partial_r$, so
\begin{equation*}
  \vxi\cdot\nabla f = r\,\partial_rf = k\,f ,
\end{equation*}
here with $f=\nabla P$ and $k=n-1$. Hence
\begin{equation*}
  \curl(\vxi\times\nabla P) = \vxi\,\nabla^2P-2\nabla P-(\vxi\cdot\nabla)\nabla P = -(n+1)\nabla P ,
\end{equation*}
and
\begin{equation}
  \Mw[P^q_n] = \nabla P^q_n\times\vxi ,
  \qquad
  \Nw[P^q_n] = (n+1)\,\nabla P^q_n .
  \label{eq:staticregular}
\end{equation}
The same identity with $\nabla\Psi$ homogeneous of degree $-(m+2)$ gives $\Nw[\Psi^p_m]=-m\nabla\Psi^p_m$ at $\kappa=0$. A static TE field is thus the gradient of a harmonic potential, as is a static TM electric field.

\emph{Orthogonality of the solid harmonics.} For two harmonic polynomials $P_1$ and $P_2$ of degree $m$,
\begin{equation*}
  P_1(\nabla)\,P_2 = \frac{(2m+1)!!}{4\pi}\oint P_1P_2\dd\Omega ,
\end{equation*}
since both sides are bilinear forms on the harmonics of degree $m$ that are unchanged by a common rotation of $P_1$ and $P_2$, and are therefore proportional, with the ratio fixed by $P_1=(\xi_1+\ii\xi_2)^m$ and $P_2=(\xi_1-\ii\xi_2)^m$. Hence, with the normalisation \eqref{eq:schmidt},
\begin{equation}
  P^p_m(\nabla)\,P^q_m = (2m-1)!!\,\delta_{pq} .
  \label{eq:fischer}
\end{equation}

\subsection{Discretisation of the Transmission Problem}
\label{sec:s-discretisation}
Edge elements discretise $\Hcurl$ and enforce only the tangential continuity of $\vE$, as is correct across $\partial B$, where the normal component jumps; the continuity of tangential $\vH$ and of normal $\sigma \vE$ are natural conditions of \meqref{eq:weak}. An arbitrary shape and material profile therefore require no special treatment, and since the perturbation decays exponentially away from the object, $\Omega$ need extend only a few object lengths beyond $B$.

The drives and the probes are needed only on $B$, where $\kappa r$ is small, and they are evaluated there to first order in $\kappa^2$. The radial factor of \meqref{eq:Phi} is the series
\begin{equation*}
  \frac{(2n+1)!!\,j_n(\kappa r)}{(\kappa r)^n} = 1 - \frac{\kappa^2r^2}{2(2n+3)} + \frac{\kappa^4r^4}{8(2n+3)(2n+5)} - \cdots ,
\end{equation*}
of which the first two terms are kept, $\Phi^q_n\approx P^q_n\bigl(1-\kappa^2r^2/2(2n+3)\bigr)$. The first term omitted has relative size $\eps^2r^4/8(2n+3)(2n+5)$, at most $2\times10^{-3}$ over the object of Section~\mref{sec:numerics} at the largest $\eps=0.5$ there, and less at higher degree. The TM drive takes its curl as $\kappa^2\Mw[\Phi]$, which \meqref{eq:closure} gives for the exact mode and which the truncated mode satisfies to the order kept. The operator of \meqref{eq:weak} needs no regularisation of the gradient fields, which the curl-curl term does not see, when the object and the host both conduct: the conduction term sees them, and the operator is nonsingular for every $\eps>0$. A regularisation $10^{-6}\int\Delta\vE\cdot\vw$ is applied only over a region that does not conduct. In a conducting host the gradient fields carry the galvanic current, on which such a term would compete with the conduction $\eps/\murs$ and change the T-matrix by about $10^{-6}\murs/\eps$.

\subsection{The Extraction Integral}
\label{sec:s-extraction}
Only the tangential parts of the fields enter the integrand of \meqref{eq:master}: with $\vE_t$ and $\vH_t$ the parts of $\vE$ and $\vH$ tangential to the sphere, $(\vE\times\vH)\cdot\xihat=(\vE_t\times\vH_t)\cdot\xihat$, since every term with a radial factor is perpendicular to $\xihat$. For $\psi_1=fY_1$ and $\psi_2=gY_2$, with $Y_1$ and $Y_2$ spherical harmonics, the pairings follow from the tangential parts \eqref{eq:tangential} and two identities for vectors $\bm t_1$ and $\bm t_2$ tangential to the sphere,
\begin{equation*}
  \bigl((\xihat\times\bm t_1)\times\bm t_2\bigr)\cdot\xihat = -\bm t_1\cdot\bm t_2 ,
  \qquad
  \bigl((\xihat\times\bm t_1)\times(\xihat\times\bm t_2)\bigr)\cdot\xihat = \xihat\cdot(\bm t_1\times\bm t_2) .
\end{equation*}
The four pairings are
\begin{align*}
  \bigl(\Mw[\psi_1]\times\Nw[\psi_2]\bigr)\cdot\xihat &= \frac{f}{r}\frac{\mathrm{d}(rg)}{\mathrm{d}r}\,\nabla_1Y_1\cdot\nabla_1Y_2 , &
  \bigl(\Nw[\psi_1]\times\Mw[\psi_2]\bigr)\cdot\xihat &= -\frac{g}{r}\frac{\mathrm{d}(rf)}{\mathrm{d}r}\,\nabla_1Y_1\cdot\nabla_1Y_2 ,\\
  \bigl(\Mw[\psi_1]\times\Mw[\psi_2]\bigr)\cdot\xihat &= fg\,\xihat\cdot(\nabla_1Y_1\times\nabla_1Y_2) , &
  \bigl(\Nw[\psi_1]\times\Nw[\psi_2]\bigr)\cdot\xihat &= \frac{1}{r^2}\frac{\mathrm{d}(rf)}{\mathrm{d}r}\frac{\mathrm{d}(rg)}{\mathrm{d}r}\,\xihat\cdot(\nabla_1Y_1\times\nabla_1Y_2) .
\end{align*}
The electric field is $\Mw$ for one type and $\Nw$ for the other, so the pairings $\Mw\times\Mw$ and $\Nw\times\Nw$ are those between a TE and a TM mode. Both carry $\xihat\cdot(\nabla_1Y_1\times\nabla_1Y_2)$, the surface divergence of $Y_1\nabla_1Y_2\times\xihat$, whose integral over the sphere vanishes. The pairings $\Mw\times\Nw$ and $\Nw\times\Mw$, between modes of one type, carry $\nabla_1Y_1\cdot\nabla_1Y_2$, and for $Y_2$ of degree $m$, by the normalisation \eqref{eq:schmidt},
\begin{equation*}
  \oint\nabla_1Y_1\cdot\nabla_1Y_2\dd\Omega = m(m+1)\oint Y_1Y_2\dd\Omega =
  \begin{cases}
    4\pi m(m+1)/(2m+1) , & Y_1=Y_2 ,\\
    0 , & \text{otherwise} .
  \end{cases}
\end{equation*}
A probe of type $c$ therefore reads only the outgoing coefficients of its own type, degree and order. Writing $\Psi^p_m=\tilde h\,Y^p_m$ and $\Phi^p_m=\tilde j\,Y^p_m$, the pairing on $r=R_0$ of an outgoing mode with the regular mode of the same type, degree and order is, for either type,
\begin{equation}
  \oint_{r=R_0}\bigl(\vE_\Psi\times\vH_\Phi-\vE_\Phi\times\vH_\Psi\bigr)\cdot\xihat\dd S
  = e_c\,\frac{4\pi m(m+1)}{2m+1}\,R_0^2\Bigl(\tilde h\,\frac{\mathrm{d}\tilde j}{\mathrm{d}r}-\tilde j\,\frac{\mathrm{d}\tilde h}{\mathrm{d}r}\Bigr) ,
  \label{eq:pairing}
\end{equation}
the terms $\tilde h\tilde j/r$ cancelling. By \meqref{eq:Psi}--\meqref{eq:Phi}, $\tilde h\tilde j=\ii\kappa(2m+1)\,h^{(1)}_m(\kappa r)\,j_m(\kappa r)$, and the Wronskian of the radial functions is
\begin{equation*}
  j_m(\kappa r)\,\frac{\mathrm{d}}{\mathrm{d}r}h^{(1)}_m(\kappa r)-h^{(1)}_m(\kappa r)\,\frac{\mathrm{d}}{\mathrm{d}r}j_m(\kappa r) = \frac{\ii}{\kappa r^2} .
\end{equation*}
The bracket of \eqref{eq:pairing} is therefore $(2m+1)/r^2$, and $R_0^2$ times it is $2m+1$ on the sphere $r=R_0$, whatever its radius. The pairing is therefore $4\pi m(m+1)\,e_c=e_c/\beta_m$ per unit outgoing coefficient. The left side of \meqref{eq:master} is therefore $(e_c/\beta_m)\,a^{(m)}_{c,p}$, and per unit drive coefficient
\begin{equation*}
  \Tens^{(m,n)}_{c,c'}[p,q] = \frac{\beta_m}{e_c}\int_B\bigl(\vE_2\cdot\cJ^{(n)}_{c',q} + \vH_2\cdot\cK^{(n)}_{c',q}\bigr)\dd\vxi ,
\end{equation*}
which is \meqref{eq:reaction}. Two regular modes, or two outgoing modes, pair to zero, since their radial factors are the same function.

\subsection{The Order of a Block}
\label{sec:s-order}
For a fixed host, frequency and coils, the propagator and the drive depend on the size of the object only through explicit powers of $\alpha$. The T-matrix depends on it only through $\eps$ and $\nua$. The count is made at fixed $\nua$, as in the size sweep of the main paper, so that $\eps\propto\alpha^2$ is the one group that changes. As $\eps\to0$ every block tends to a finite limit, and its dependence on $\eps$ is a relative correction of $O(\eps)$. The exception is $\Tens_{\TM,\TE}$, which by \meqref{eq:symmetry} is $\kappa^2=\ii\eps$ times the transpose of $\Tens_{\TE,\TM}$, and so carries the factor $\eps\propto\alpha^2$.

By \meqref{eq:Q} the propagator block $\Prop^{(m)}_c$ carries $\alpha^{m+1}$, and one further $\alpha$ for a TE mode. By \meqref{eq:drivereciprocity} the drive is $\Prop^{(n)}_c(\vy,\vz)^\top$ divided by $\alpha^2e_c$, with $e_{\TE}\propto\alpha$ and $e_{\TM}\propto\alpha^{-1}$, which gives the powers of \meqref{eq:drive1} at every degree. The contribution $\Prop^{(m)}_c\Tens^{(m,n)}_{c,c'}\vd^{(n)}_{c'}$ of a block takes the propagator of its response type $c$ and the drive of its drive type $c'$, and its factors carry the powers
\begin{equation*}
  \begin{array}{lccccc}
  \toprule
  \text{Block } (c,c') & \Prop^{(m)}_c & \Tens^{(m,n)}_{c,c'} & \vd^{(n)}_{c'} & \text{Contribution} & \text{Order}\\
  \midrule
  \TE,\TE & \alpha^{m+2} & 1 & \alpha^{n-1} & \alpha^{m+n+1} & m+n\\
  \TM,\TM & \alpha^{m+1} & 1 & \alpha^{n} & \alpha^{m+n+1} & m+n\\
  \TE,\TM & \alpha^{m+2} & 1 & \alpha^{n} & \alpha^{m+n+2} & m+n+1\\
  \TM,\TE & \alpha^{m+1} & \alpha^2 & \alpha^{n-1} & \alpha^{m+n+2} & m+n+1\\
  \bottomrule
  \end{array}
\end{equation*}
times factors independent of $\alpha$, so that a diagonal block contributes $O(\alpha^{m+n+1})$ and a cross block $O(\alpha^{m+n+2})$. At degree one the extra power is the size of the object across which a cross block converts one moment into the other: the circulating part of a current $\vp$ conducted through the object has a magnetic moment of $O(\alpha|\vp|)$. The order of a block, in the last column, is the power of $\alpha$ in its contribution less one, so that the diagonal dipole blocks have order two.

\subsection{The Dipole Moments}
\label{sec:s-dipoles}
At degree one the outgoing modes are the fields of point dipoles at every $\kappa$. With $g:=e^{\ii\kappa r}/r$, \meqref{eq:Psi} gives $\Psi^q_1=-\partial_qg$. Since $g$ depends on $r$ alone, $\partial_qg$ is $\xi_q$ times a function of $r$, whose gradient is parallel to $\vxi$, so
\begin{equation*}
  \partial_qg = \frac{1}{r}\frac{\mathrm{d}g}{\mathrm{d}r}\,\xi_q ,
  \qquad
  \nabla\partial_qg\times\vxi = \frac{1}{r}\frac{\mathrm{d}g}{\mathrm{d}r}\,\ehat_q\times\vxi = \ehat_q\times\nabla g ,
\end{equation*}
and by \meqref{eq:MN}, with $\Mw[\psi]=\nabla\psi\times\vxi$,
\begin{equation}
  \Mw[\Psi^q_1] = \curl(g\,\ehat_q) ,
  \qquad
  \Nw[\Psi^q_1] = \curl\curl(g\,\ehat_q) = (\nabla\nabla+\kappa^2)(g\,\ehat_q) ,
  \label{eq:dipolemodes}
\end{equation}
the second because $(\nabla^2+\kappa^2)g=0$ away from the centre. By \meqref{eq:TE}--\meqref{eq:TM} the response of degree one is therefore
\begin{equation*}
  \Delta\vH = (\nabla\nabla+\kappa^2)\bigl(g\,\va^{(1)}_{\TE}\bigr) + \curl\bigl(g\,\va^{(1)}_{\TM}\bigr) .
\end{equation*}

A point magnetic dipole $\vm$ and a point current dipole $\vp$ at $\vz$ in $\sea$ have the magnetic fields
\begin{equation*}
  \vH_{\vm} = (\nabla_x\nabla_x+\ks^2)(\vm\,G) ,
  \qquad
  \vH_{\vp} = \nabla_x\times(\vp\,G) ,
  \qquad
  G = \frac{e^{\ii\ks R}}{4\pi R} ,
\end{equation*}
with $R=|\vx-\vz|$. In the scaled coordinates \meqref{eq:scaling}, $R=\alpha r$, $\ks R=\kappa r$ and $\nabla=\alpha\nabla_x$, so $G=g/4\pi\alpha$ and
\begin{equation*}
  \vH_{\vm} = \frac{(\nabla\nabla+\kappa^2)(g\,\vm)}{4\pi\alpha^3} ,
  \qquad
  \vH_{\vp} = \frac{\curl(g\,\vp)}{4\pi\alpha^2} .
\end{equation*}
These are the response of degree one exactly when
\begin{equation}
  \va^{(1)}_{\TE} = \frac{\vm}{4\pi\alpha^3} ,
  \qquad
  \va^{(1)}_{\TM} = \frac{\vp}{4\pi\alpha^2} ,
  \label{eq:dipolemoments}
\end{equation}
which is \meqref{eq:response1} and defines $\vm$ and $\vp$ at every $\kappa$, not only in the limits.

\subsection{The Four Blocks of the Object}
\label{sec:s-blocks}
\begin{figure*}[!htb]
\centering
\subfloat[]{\includegraphics[width=0.48\textwidth]{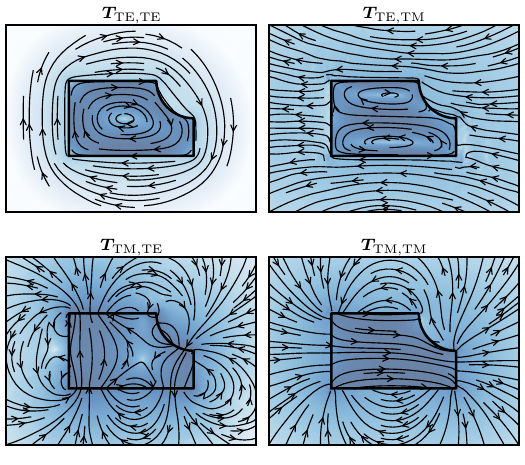}\label{fig:mechanisms}}
\hfil
\subfloat[]{\includegraphics[width=0.48\textwidth]{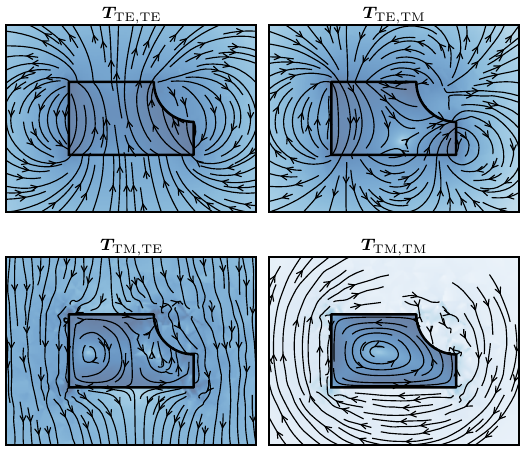}\label{fig:magnetic}}
\caption{The four blocks of \meqref{eq:tensor} for a box with one corner removed, on a section through the notch. Rows are the inductive and galvanic parts of the response and columns are the drives $\vd_{\TE}$ and $\vd_{\TM}$; shading is the magnitude, logarithmic and normalised per panel. (a)~Perturbation current $\Delta\vJ$ for a conductivity contrast alone, $\sigma_*=30\sigs$, at $\eps=1$ and $\nua=30$. (b)~Perturbation flux density $\Delta\vB/\mu_0\murs$ for a permeability contrast alone, $\murb=20$ in a host of $\murs=2$, at $\eps=1$ and $\nua=10$. The galvanic row exists only because the host conducts: in (a) it is current that crosses the surface of the object and closes through the host, and in (b) it is due to current that the magnetised object induces in the host, the $\kappa^2$ term of \meqref{eq:integrandTM}.}
\label{fig:blocks}
\end{figure*}

Fig.~\ref{fig:blocks} shows the four blocks of \meqref{eq:tensor} for the object of the numerical experiments, a box with one corner removed, from the conductivity contrast alone and from the permeability contrast alone.

\subsection{A Non-Conducting Host}
\label{sec:s-nonconducting}
In a non-conducting host only the block $\Tens_{\TE,\TE}$ survives, whatever the contrasts of the object (Section~\mref{sec:extract}). The block $\Tens_{\TM,\TE}$ is taken first, through the galvanic term of \meqref{eq:integrandTM}. With $\sigs=0$, $\kappa=0$ in the host, so the series $(2m+1)!!\,j_m(\kappa r)/\kappa^m$ in \meqref{eq:Phi} reduces to $r^m$ and the probe $\Phi^p_m$ is the solid harmonic $P^p_m$ itself. By \eqref{eq:staticregular} the probe field $\Nw[P^p_m]=(m+1)\nabla P^p_m$ is then a gradient. For a quantity $f$ defined on either side of a surface with unit normal $\nhat$, write
\begin{equation*}
  [f] := f^+-f^- ,
\end{equation*}
the jump across the surface, where $f^+$ is the limit of $f$ on the side into which $\nhat$ points and $f^-$ the limit on the other. On $\partial B$, with $\nhat$ the outward normal, the jump is outside less inside. Integrating by parts, with $\nabla\cdot(\sigma \vE)=0$ in the object and $[\sigma\vE\cdot\nhat]=0$, the natural condition of \meqref{eq:weak}, gives
\begin{equation*}
  \int_B\Jimp\cdot\Nw[\Phi^p_m]\dd\vxi\,\Big|_{\kappa=0}
  = (m+1)\,\sigs\Bigl(\oint_{\partial B}P^p_m\,[\vE\cdot\nhat]\dd S + \int_B P^p_m\,\nabla\cdot \vE\dd\vxi\Bigr) .
\end{equation*}
The right side carries the factor $\sigs$. Under a TE drive, whose electric field $e_{\TE}\Mw[\Phi^q_n]$ carries no factor $1/\sigs$, it therefore vanishes in a non-conducting host. The $\kappa^2$ term of \meqref{eq:integrandTM}, the current induced by the magnetised object, vanishes as well, with $\kappa^2\propto\sigs$. For $\sigs=0$ the block $\Tens_{\TM,\TE}$ is therefore zero, despite a conductivity contrast. 

The blocks $\Tens_{\TE,\TM}$ and $\Tens_{\TM,\TM}$ respond to a TM drive, whose electric field grows as $1/\sigs$, and need not vanish. They are never excited, however, because a non-conducting host supplies no TM drive. By \meqref{eq:closure} the curl of the ambient field \meqref{eq:ambient} is
\begin{equation*}
  \curl\vH_0 = \sum_{n,q}\Bigl[\kappa^2\,d^{(n)}_{\TE,q}\,\Mw[\Phi^q_n] + d^{(n)}_{\TM,q}\,\Nw[\Phi^q_n]\Bigr] = \alpha\sigs\vE_0 ,
\end{equation*}
the current density of the host, which carries the TM coefficients directly and the TE coefficients only through $\kappa^2$. For $\sigs=0$ both this curl and $\kappa^2$ vanish, and with them every $d^{(n)}_{\TM,q}$.

The block $\Tens_{\TE,\TE}$ carries no factor of $\sigs$ in either of its terms, and no block carries a factor that vanishes with the permeability of the host, so only the conductivity of the host decides which blocks exist.

\subsection{Symmetry under Exchange of Probe and Drive}
\label{sec:s-symmetry}
The probes of \meqref{eq:reaction} are the same modes as the basis drives. The total fields $(\vE_1,\vH_1)$ and $(\vE_2,\vH_2)$ of the basis drives $(c,m,p)$ and $(c',n,q)$ are each the regular mode with unit coefficient plus the perturbation it produces. Both satisfy \meqref{eq:scaledpair} with the same $\sigma$ and $\mur$, which are scalars, so
\begin{equation*}
  \nabla\cdot(\vE_1\times\vH_2-\vE_2\times\vH_1)
  = \vH_2\cdot\curl\vE_1-\vE_1\cdot\curl\vH_2-\vH_1\cdot\curl\vE_2+\vE_2\cdot\curl\vH_1 = 0
\end{equation*}
pointwise, the terms in $\ii\omega\mu_0\mur\alpha\,\vH_1\cdot\vH_2$ and in $\alpha\sigma\vE_1\cdot\vE_2$ cancelling in pairs. The tangential fields are continuous, $[\nhat\times\vE]=[\nhat\times\vH]=\bm0$ on $\partial B$, so the divergence theorem over the ball $r\le R_0$ gives
\begin{equation*}
  \oint_{r=R_0}(\vE_1\times\vH_2-\vE_2\times\vH_1)\cdot\xihat\dd S = 0 .
\end{equation*}
On the sphere each field is its regular mode plus outgoing modes. Regular parts pair to zero with each other, as do outgoing parts (Section~\ref{sec:s-extraction}), which leaves two cross terms. By \eqref{eq:pairing}, the outgoing part of field 1 paired with the regular mode of field 2 gives $(e_{c'}/\beta_n)\,\Tens^{(n,m)}_{c',c}[q,p]$. The regular mode of field 1 paired with the outgoing part of field 2 gives $-(e_c/\beta_m)\,\Tens^{(m,n)}_{c,c'}[p,q]$, the sign because the order of the fields in the pairing is reversed. The two sum to zero,
\begin{equation*}
  \frac{e_c}{\beta_m}\,\Tens^{(m,n)}_{c,c'}[p,q] = \frac{e_{c'}}{\beta_n}\,\Tens^{(n,m)}_{c',c}[q,p] ,
\end{equation*}
which is \meqref{eq:symmetry}.
\section{Propagation Through Layered Media}
\label{sec:propagation}
This section derives the statements of Section~\mref{sec:surface} that the main paper gives without proof: the field of a horizontal mode from its two vertical scalars, their separate reflection and the reflection coefficients, the vertical components of the modes and the polynomials $Q$, the layer factor, the drive by reciprocity, and the entries of the reflection matrix, its block of degree one, its symmetries and its size.

\subsection{The Two Vertical Scalars}
\label{sec:s-modes}
A layer is unchanged by horizontal translation, so a source-free field in it is a superposition of horizontal modes, each varying across the layer as $e^{\ii\bm k\cdot\vx_\rho}$. The point source of the Sommerfeld identity \meqref{eq:sommerfeld} is symmetric about the vertical through its centre, so its modes of one $k_\rho=|\bm k|$ enter averaged over the direction of $\bm k$. By Bessel's integral~\cite[eq.~(10.9.1)]{olverNISTHandbookMathematical2010} this average is $J_0(k_\rho\rho)$, and the identity takes the form
\begin{equation}
  u(\vx,\vz) = \frac{1}{2\pi}\int_{\mathbb{R}^2}\frac{1}{\gams}\,e^{\ii\bm k\cdot(\vx_\rho-\vz_\rho)}\,e^{-\gams|x_3-z_3|}\dd^2\bm k ,
  \label{eq:planewaves}
\end{equation}
with $\gams$ evaluated at $k_\rho=|\bm k|$. On a mode of horizontal wavenumber $k_\rho$, in either form, every component $F$ satisfies $\nabla_\rho^2F=-k_\rho^2F$, so for $k_\rho\ne0$ the horizontal Laplacian can be inverted. The horizontal part $\vH_\rho=(H_1,H_2,0)$ of the magnetic field then splits into a gradient and a rotated gradient,
\begin{equation*}
  \vH_\rho = \nabla_\rho\chi + \nabla_\rho\psi\times\ehat_3 ,
\end{equation*}
the first carrying the horizontal divergence of $\vH_\rho$ and the second its vertical curl. Each is fixed by one scalar. Since $\nabla\cdot\vH=0$,
\begin{equation*}
  \nabla_\rho\cdot\vH_\rho = \nabla_\rho^2\chi = -k_\rho^2\chi = -\partial_{x_3}H_3 ,
\end{equation*}
and since the vertical component of Amp\`ere's law is $(\curlx\vH)_3=\sigma E_3$,
\begin{equation*}
  \ehat_3\cdot\nabla_\rho\times\vH_\rho = -\nabla_\rho^2\psi = k_\rho^2\psi = \sigma E_3 .
\end{equation*}
Hence $\chi=k_\rho^{-2}\partial_{x_3}H_3$ and $\psi=k_\rho^{-2}\sigma E_3$, and
\begin{equation*}
  \vH = \ehat_3H_3 + \frac{1}{k_\rho^2}\Bigl[\nabla_\rho\,\partial_{x_3}H_3 + \nabla_\rho(\sigma E_3)\times\ehat_3\Bigr] ,
\end{equation*}
which is \meqref{eq:fromscalars}.

\subsection{Reflection at the Interfaces}
\label{sec:s-interfaces}
At every interface, with the jump of Section~\ref{sec:s-nonconducting} taken for the normal $\ehat_3$, so that it is above minus below,
\begin{equation}
  [\mur H_3] = 0 ,
  \qquad
  [\ehat_3\times\vH] = \bm0 ,
  \qquad
  [\ehat_3\times\vE] = \bm0 .
  \label{eq:interface}
\end{equation}
The second and third conditions each fix two horizontal components, so \eqref{eq:interface} holds five scalar conditions. Within a layer $\mur$ and $\sigma$ are constant, and $\nabla\cdot\vH=\nabla\cdot\vE=0$. The horizontal divergence of $[\ehat_3\times\vH]=\bm0$ then makes $\partial_{x_3}H_3=\partial_{x_3}(\mur H_3)/\mur$ continuous, and its horizontal curl makes $(\curlx\vH)_3=\sigma E_3$ continuous. The horizontal divergence of $[\ehat_3\times\vE]=\bm0$ makes $\partial_{x_3}E_3=\partial_{x_3}(\sigma E_3)/\sigma$ continuous, and its horizontal curl, $(\curlx\vE)_3=\ii\omega\mu_0\mur H_3$, repeats the first condition. Each of the two scalars is therefore continuous across an interface together with its vertical derivative divided by its own material constant,
\begin{equation}
  [\mur H_3] = 0 ,
  \qquad
  \Bigl[\frac{\partial_{x_3}(\mur H_3)}{\mur}\Bigr] = 0 ,
  \qquad
  [\sigma E_3] = 0 ,
  \qquad
  \Bigl[\frac{\partial_{x_3}(\sigma E_3)}{\sigma}\Bigr] = 0 ,
  \label{eq:interfacescalars}
\end{equation}
which are the four independent conditions.

In layer $j$ a scalar of horizontal wavenumber $k_\rho$ varies as $e^{\pm\gamma_jx_3}$, so its vertical derivative divided by $\mur$ or $\sigma$ is $\pm Y_j$ or $\pm Y'_j$ times the scalar, with the ratios of \meqref{eq:admittance}. The conditions on $\mur H_3$ involve only $Y$ and those on $\sigma E_3$ only $Y'$, so the two scalars are reflected independently.

At the surface the rising scalar of the host is reflected and transmitted,
\begin{equation*}
  H_3 = e^{-\gams x_3}+\Gamma_0\,e^{\gams x_3} \quad (x_3<0) ,
  \qquad
  H_3 = \tau\,e^{-k_\rho x_3} \quad (x_3>0) ,
\end{equation*}
with $\mur=1$ and $Y=k_\rho$ in the air. The first two conditions of \eqref{eq:interfacescalars} at $x_3=0$ give
\begin{equation*}
  \murs(1+\Gamma_0) = \tau ,
  \qquad
  \gams(1-\Gamma_0) = k_\rho\tau ,
\end{equation*}
whose solution, with both coefficients referenced to the surface, is
\begin{equation*}
  \Gamma_0 = \frac{\gams-\murs k_\rho}{\gams+\murs k_\rho} ,
  \qquad
  \tau = \frac{2\murs\gams}{\gams+\murs k_\rho} ,
\end{equation*}
which is \meqref{eq:RT}.

Beneath, $\Gamma_j$ is the reflection of a falling scalar in layer $j$, referenced to the bottom of that layer. The scalar transmitted into layer $j+1$ returns from the bottom of that layer with $\Gamma_{j+1}$ and the factor $e^{-2\gamma_{j+1}h_{j+1}}$ of the round trip. Summing its multiple reflections between the two boundaries of layer $j+1$, together with the direct reflection $\Gamma_{j,j+1}$ at the interface, gives the standard recursion~\cite{wardElectromagneticTheoryGeophysical1988}
\begin{equation*}
  \Gamma_j = \frac{\Gamma_{j,j+1}+\Gamma_{j+1}\,e^{-2\gamma_{j+1}h_{j+1}}}{1+\Gamma_{j,j+1}\Gamma_{j+1}\,e^{-2\gamma_{j+1}h_{j+1}}} ,
  \qquad
  \Gamma_{j,j+1} = \frac{Y_j-Y_{j+1}}{Y_j+Y_{j+1}} ,
\end{equation*}
which is \meqref{eq:recursion}. The half space $j=N$ has no lower boundary, so $\Gamma_N:=0$, and the recursion starts from $\Gamma_{N-1}=\Gamma_{N-1,N}$ at the deepest interface.

In the static limit $\gamma_j\to k_\rho$, a single half space of conductivity $\sigma_2$ beneath the host gives $\Gamma'_1\to(\sigma_2-\sigs)/(\sigma_2+\sigs)$, so a resistive layer beneath turns the vertical current back almost wholly, as the surface does.

\subsection{The Vertical Components of the Modes}
\label{sec:s-vertical}
From \meqref{eq:MN}, $\Mw[\psi]=\curl(\vxi\psi)=-\vxi\times\nabla\psi$ and $\Nw=\curl\Mw$, so
\begin{equation}
  \bigl(\Mw[\psi]\bigr)_3 = -\ehat_3\cdot(\vxi\times\nabla\psi) ,
  \qquad
  \bigl(\Nw[\psi]\bigr)_3 = -\ehat_3\cdot\curl(\vxi\times\nabla\psi) .
  \label{eq:vertical}
\end{equation}
The operator $\vxi\times\nabla$ differentiates along rotations about the centre and vanishes on any function of $r$ alone. On $\Psi^p_m$, a multiple of $P^p_m(\nabla)f(r)$, it therefore acts on the polynomial only,
\begin{equation*}
  \vxi\times\nabla\bigl[P^p_m(\nabla)f(r)\bigr] = (\vxi\times\nabla P^p_m)(\nabla)\,f(r) ,
\end{equation*}
and each component of $\vxi\times\nabla P^p_m$ is again a solid harmonic of degree $m$.

A polynomial evaluated at $\nabla$ is the operator obtained by replacing each $\xi_i$ with $\partial_i$, and a product of polynomials becomes the product of these operators, which commute. A further derivative is therefore a further factor of $\vxi$, $\partial_a\bigl[V(\nabla)f\bigr]=(\xi_aV)(\nabla)f$, and the curl of a vector $\bm{V}$ of such operators is
\begin{equation*}
  \curl\bigl[\bm{V}(\nabla)f\bigr] = (\vxi\times\bm{V})(\nabla)\,f .
\end{equation*}
With $\bm{V}=\vxi\times\nabla P^p_m$, the curl in $\Nw=\curl\Mw$ becomes a second factor $\vxi\times$.

In physical coordinates $e^{\ii\kappa r}/r=\alpha u$ and $\nabla=\alpha\nabla_x$ by \meqref{eq:scaling}, and $P^p_m(\nabla)=\alpha^mP^p_m(\nabla_x)$ since $P^p_m$ is homogeneous of degree $m$. Then \meqref{eq:Psi} gives $\Psi^p_m=c_m\,\alpha^{m+1}P^p_m(\nabla_x)\,u$ with $c_m=(-1)^m/(2m-1)!!$. The rotation $\vxi\times\nabla$ is unchanged by the scaling, and the curl is $\curl=\alpha\curlx$, so \eqref{eq:vertical} gives the vertical fields as $Q^{(m)}_{c,p}(\nabla_x)\,u$, with the polynomials
\begin{equation*}
  Q^{(m)}_{\TM,p}(\vxi) = -c_m\,\alpha^{m+1}\,\ehat_3\cdot\bigl(\vxi\times\nabla P^p_m(\vxi)\bigr) ,
  \qquad
  Q^{(m)}_{\TE,p}(\vxi) = -c_m\,\alpha^{m+2}\,\ehat_3\cdot\bigl(\vxi\times(\vxi\times\nabla P^p_m(\vxi))\bigr) ,
\end{equation*}
which is \meqref{eq:Q}.

Both are explicit. The vertical component of $\vxi\times\nabla P$ is the derivative in azimuth, which on the harmonics \eqref{eq:schmidt} exchanges the cosine and sine of each order,
\begin{equation*}
  \ehat_3\cdot(\vxi\times\nabla P) = \xi_1\partial_2P-\xi_2\partial_1P = \partial_\phi P ,
  \qquad
  \partial_\phi P^p_m = -p\,P^{-p}_m .
\end{equation*}
For the TE row, the vector triple product gives
\begin{equation*}
  \vxi\times(\vxi\times\nabla P) = \vxi\,(\vxi\cdot\nabla P) - r^2\,\nabla P ,
\end{equation*}
and by Euler's relation (Section~\ref{sec:s-vectormodes}) $\vxi\cdot\nabla P=m\,P$, so
\begin{equation*}
  \ehat_3\cdot\bigl(\vxi\times(\vxi\times\nabla P)\bigr) = m\,\xi_3P-r^2\,\partial_3P .
\end{equation*}
Hence
\begin{equation}
  Q^{(m)}_{\TM,p}(\vxi) = c_m\,\alpha^{m+1}\,p\,P^{-p}_m(\vxi) ,
  \qquad
  Q^{(m)}_{\TE,p}(\vxi) = -c_m\,\alpha^{m+2}\,\bigl(m\,\xi_3P^p_m(\vxi)-r^2\,\partial_3P^p_m(\vxi)\bigr) .
  \label{eq:Qexplicit}
\end{equation}
These act on $u$ as $Q^{(m)}_{c,p}(\nabla_x)$. By the same rule, with $\partial_{x_i}$ in place of $\partial_i$, the scalar $r^2$ becomes the Laplacian,
\begin{equation*}
  r^2(\nabla_x) = \partial_{x_1}^2+\partial_{x_2}^2+\partial_{x_3}^2 = \nabla_x^2 ,
\end{equation*}
and $r^2\partial_3P^p_m$ becomes $\nabla_x^2$ applied after $(\partial_3P^p_m)(\nabla_x)$. The polynomials act on the point source $u$, which satisfies $(\nabla_x^2+\ks^2)u=0$ away from $\vz$, so on $u$ the Laplacian is the constant $-\ks^2$ and
\begin{equation*}
  \bigl(r^2\,\partial_3P^p_m\bigr)(\nabla_x)\,u = -\ks^2\,(\partial_3P^p_m)(\nabla_x)\,u .
\end{equation*}
At degree one, with $c_1=-1$ and the orders $q=1,2,3$ of $\xi_q$,
\begin{equation*}
  Q^{(1)}_{\TM}(\nabla_x) = \alpha^2\,(-\partial_2,\ \partial_1,\ 0) ,
  \qquad
  Q^{(1)}_{\TE,q}(\nabla_x) = \alpha^3\bigl(\partial_3\partial_q+\ks^2\delta_{q3}\bigr) \quad\text{on } u ,
\end{equation*}
the vertical fields of a horizontal current dipole and of a magnetic dipole. The degrees two and three follow from Table~\ref{tab:harmonics}.

The primed polynomials multiply $\vU'$, whose vertical current is $u$, so they are read off $(\curlx\vH)_3$. By \meqref{eq:closure},
\begin{equation*}
  \TM:\quad \curlx\vH = \alpha^{-1}\Nw ,
  \qquad
  \TE:\quad \curlx\vH = \alpha^{-1}\curl\Nw = \alpha^{-1}\kappa^2\Mw = \alpha\ks^2\,\Mw ,
\end{equation*}
so the vertical current of a TM mode is $\alpha^{-1}$ times the vertical field of the TE mode on the same $\psi$, and that of a TE mode is $\alpha\ks^2$ times the vertical field of the TM mode. This gives
\begin{equation*}
  {Q'}^{(m)}_{\TM,p}(\vxi) = \alpha^{-1}\,Q^{(m)}_{\TE,p}(\vxi) ,
  \qquad
  {Q'}^{(m)}_{\TE,p}(\vxi) = \alpha\ks^2\,Q^{(m)}_{\TM,p}(\vxi) ,
\end{equation*}
which is \meqref{eq:Qprime}.

\subsection{The Layer Factor}
\label{sec:s-propagator}
Each horizontal mode of $u$ has a rising half and a falling half, which reach the surface with the amplitudes
\begin{equation*}
  \frac{k_\rho}{\gams}J_0(k_\rho\rho)\,e^{\gams z_3} ,
  \qquad
  \frac{k_\rho}{\gams}J_0(k_\rho\rho)\,\Gamma_1\,e^{-\gams(2\hs+z_3)} ,
\end{equation*}
the second after returning from the interface beneath with $\Gamma_1$. Both then reflect between the surface, with $\Gamma_0$, and the interface beneath, with $\Gamma_1$, which multiplies the amplitude at the surface by
\begin{equation*}
  \sum_{\ell\ge0}\bigl(\Gamma_0\Gamma_1e^{-2\gams\hs}\bigr)^\ell = \frac{1}{1-\Gamma_0\Gamma_1e^{-2\gams\hs}} .
\end{equation*}
The sum of the two halves at the surface, per unit $\frac{k_\rho}{\gams}J_0(k_\rho\rho)$, is therefore
\begin{equation*}
  L(k_\rho,z_3) = \frac{e^{\gams z_3}+\Gamma_1\,e^{-\gams(2\hs+z_3)}}{1-\Gamma_0\Gamma_1\,e^{-2\gams \hs}} ,
\end{equation*}
which is the layer factor \meqref{eq:layerfactor}.

\subsection{The Drive}
\label{sec:s-drive}
The transmitter is a coil at $\vy$, small against its height above the surface, with magnetic dipole moment $\vm_\mathrm{T}$. The argument uses two fields of all space, air and ground together: the field $(\vE_1,\vH_1)$ of the transmitter, and the field $(\vE_2,\vH_2)$ of the outgoing mode of type $c$, degree $n$ and order $q$ radiated from $\vz$, including its transmission into $\air$ and its reflections from the layers. Both are fields of the same layered medium: at each point the two see the same $\sigma$ and $\mur$, though these change from layer to layer, and both satisfy the conditions \eqref{eq:interface} at every interface. In the air, $\vE$ is defined only through $\curl\vE=\ii\omega\mu_0\vH$, with $\curl\vH=\bm0$, so there too
\begin{equation*}
  \nabla\cdot(\vE_1\times\vH_2-\vE_2\times\vH_1) = \ii\omega\mu_0\,(\vH_2\cdot\vH_1-\vH_1\cdot\vH_2) = 0 ,
\end{equation*}
as in the conductors. The flux of $\vE_1\times\vH_2-\vE_2\times\vH_1$ through an interface involves only the components of the fields tangential to it, which are continuous by \eqref{eq:interface}, so the divergence theorem applies in each layer and the contributions of the two sides of every interface cancel. Far from the sources the fields decay exponentially in the ground, while in the air $\vH=O(|\vx|^{-3})$, as for a dipole, and $\vE=O(|\vx|^{-2})$ through $\curl\vE=\ii\omega\mu_0\vH$, so the integral of $\vE\times\vH$ over a large sphere vanishes. With a small ball removed about $\vy$ and another about $\vz$, the latter of radius less than $\min(D,\hs-D)$ so that it lies in the host layer, both fields are source-free in what remains, and the divergence theorem there gives
\begin{equation*}
  \oint_{\vy}\bigl(\vE_1\times\vH_2-\vE_2\times\vH_1\bigr)\cdot\nhat\dd S
  + \oint_{\vz}\bigl(\vE_1\times\vH_2-\vE_2\times\vH_1\bigr)\cdot\nhat\dd S = 0 ,
\end{equation*}
where $\oint_{\vy}$ and $\oint_{\vz}$ are the integrals over the spheres about $\vy$ and $\vz$, and $\nhat$ points out of each ball.

Within the ball about $\vy$ the mode is regular and the transmitter is a point magnetic source, which in the scaled form \meqref{eq:withsources} is
\begin{equation*}
  \cK_1 = \ii\omega\mu_0\alpha\,\vm_\mathrm{T}\,\delta(\vx-\vy) ,
  \qquad
  \delta(\vx-\vy) = \alpha^{-3}\,\delta(\vxi-\vxi_y) ,
\end{equation*}
so by \meqref{eq:reciprocity}
\begin{equation*}
  \oint_{\vy} = \int\vH_2\cdot\cK_1\dd\vxi = \frac{\ii\omega\mu_0}{\alpha^2}\,\vm_\mathrm{T}\cdot\vH_2(\vy) .
\end{equation*}
By \meqref{eq:propentries} the field of the mode at the coil is column $q$ of the propagator block with the sensor placed there, $[\Prop^{(n)}_c(\vy,\vz)]_q$. On the sphere about $\vz$ the field of the transmitter is the regular expansion \meqref{eq:ambient}, and the reflections of the mode are regular as well. Two regular fields contribute nothing. By \eqref{eq:pairing} the outgoing mode pairs only with the term of \meqref{eq:ambient} of its own type, degree and order, and with the regular field standing first in the pairing,
\begin{equation*}
  \oint_{\vz} = -\frac{e_c}{\beta_n}\,d^{(n)}_{c,q} .
\end{equation*}
Since the two integrals sum to zero,
\begin{equation}
  d^{(n)}_{c,q} = \frac{\ii\omega\mu_0\,\beta_n}{\alpha^2e_c}\;\vm_\mathrm{T}\cdot\bigl[\Prop^{(n)}_c(\vy,\vz)\bigr]_q .
  \label{s-drivereciprocity}
\end{equation}
Collected over the orders,
\begin{equation*}
  \vd^{(n)}_c = \frac{\ii\omega\mu_0\,\beta_n}{\alpha^2e_c}\;\Prop^{(n)}_c(\vy,\vz)^{\top}\,\vm_\mathrm{T} ,
\end{equation*}
which is \meqref{eq:drivereciprocity}. Evaluated independently, by the Sommerfeld integral of the transmitted field and its extraction on a sphere about $\vz$, \eqref{s-drivereciprocity} holds to rounding error.

At degree one the coefficients are point values. At the centre $\Mw[\Phi^q_1]$ and every mode of higher degree vanish, and $\Nw[\Phi^q_1]=2\ehat_q$, so evaluating \meqref{eq:ambient} there, and its curl with \meqref{eq:closure}, gives
\begin{equation}
  \vd^{(1)}_{\TE} = \tfrac12\,\vH_0(\vz) ,
  \qquad
  \vd^{(1)}_{\TM} = \tfrac12\,\curl  \vH_0(\vz) = \tfrac12\,\alpha\sigs\,\vE_0(\vz) .
  \label{s-drive1}
\end{equation}

\subsection{The Entries of the Reflection Matrix}
\label{sec:s-return}
The reflected field $\vH^{\Gamma}_{c,p}$ of each mode is source-free throughout $\sea$, so within a distance $\min(D,\hs-D)$ of $\vz$ it expands in regular modes as \meqref{eq:ambient} does,
\begin{equation*}
  \vH^{\Gamma}_{c,p} = \sum_{n,q}\Bigl[\,\Ret^{(n,m)}_{\TE,c}[q,p]\,\Nw[\Phi^q_n] + \Ret^{(n,m)}_{\TM,c}[q,p]\,\Mw[\Phi^q_n]\Bigr] ,
\end{equation*}
and its coefficients are the entries of the reflection matrix. Each coefficient of a regular expansion is a derivative of the field at the centre.

At the centre a factor $\xi_i$ of a function can be moved onto an operator $A(\nabla)$ of Section~\ref{sec:s-vectormodes}, where it becomes a derivative. For one derivative this is the product rule, $\partial_i(\xi_i\psi)=\xi_i\partial_i\psi+\psi$. Applied to each derivative of $A(\nabla)$ in turn, it gives
\begin{equation*}
  A(\nabla)(\xi_i\psi) = \xi_i\,A(\nabla)\psi + (\partial_iA)(\nabla)\psi ,
\end{equation*}
whose first term vanishes at the centre. Applied twice to each term of $r^2=\xi_1^2+\xi_2^2+\xi_3^2$, it moves a factor $r^2$ in the same way, so that, with $|_{\bm0}$ denoting evaluation at the centre,
\begin{equation*}
  A(\nabla)(\xi_i\psi)\bigr|_{\bm0} = (\partial_iA)(\nabla)\psi\bigr|_{\bm0} ,
  \qquad
  A(\nabla)(r^2\psi)\bigr|_{\bm0} = (\nabla^2A)(\nabla)\psi\bigr|_{\bm0} ,
\end{equation*}
and the second vanishes for harmonic $A$. By \meqref{eq:Phi}, $(2n+1)!!\,j_n(\kappa r)/(\kappa r)^n$ is a power series in $r^2$ with constant term one~\cite[eq.~(10.53.1)]{olverNISTHandbookMathematical2010}, so $\Phi^q_n$ is $P^q_n$ plus terms $r^{2k}P^q_n$ with $k\ge1$. The operator $P^{q'}_{n'}(\nabla)$ at the centre retains only the part of degree $n'$, so with \eqref{eq:fischer}
\begin{equation*}
  P^{q'}_{n'}(\nabla)\,\Phi^q_n\bigr|_{\bm0} = (2n-1)!!\,\delta_{nn'}\delta_{qq'} .
\end{equation*}

The vector operator $(\nabla P^{q'}_{n'})(\nabla)\cdot$ reduces to this scalar one in two steps. First, a curl moves from the field to the operator. For a vector polynomial $\bm a$, the scalar triple product $\bm a\cdot(\nabla\times\bm V)=(\bm a\times\nabla)\cdot\bm V$ gives
\begin{equation*}
  \bm a(\nabla)\cdot\curl\bm V = (\bm a\times\vxi)(\nabla)\cdot\bm V .
\end{equation*}
Second, a factor $\vxi$ moves to the operator by the first rule, applied to each component: $\bm a(\nabla)\cdot(\vxi\psi)|_{\bm0}=(\nabla\cdot\bm a)(\nabla)\psi|_{\bm0}$. By \meqref{eq:MN}, $\Mw[\psi]=\curl(\vxi\psi)$ and $\Nw[\psi]=\curl\curl(\vxi\psi)$. Moving one curl for $\Mw$ and two for $\Nw$, and then the factor $\vxi$, gives
\begin{align*}
  (\nabla P^{q'}_{n'})(\nabla)\cdot\Mw[\psi]\bigr|_{\bm0} &= (\nabla\cdot\bm b_1)(\nabla)\psi\bigr|_{\bm0} , &
  \bm b_1 &= \nabla P^{q'}_{n'}\times\vxi ,
  \\
  (\nabla P^{q'}_{n'})(\nabla)\cdot\Nw[\psi]\bigr|_{\bm0} &= (\nabla\cdot\bm b_2)(\nabla)\psi\bigr|_{\bm0} , &
  \bm b_2 &= \bm b_1\times\vxi = n'P^{q'}_{n'}\,\vxi-r^2\,\nabla P^{q'}_{n'} .
\end{align*}
The divergence of $\bm b_1$ is zero, since $\nabla P^{q'}_{n'}$ and $\vxi$ are both curl-free. The divergence of $\bm b_2$ is $n'(n'+1)P^{q'}_{n'}$, by Euler's relation $\vxi\cdot\nabla P^{q'}_{n'}=n'P^{q'}_{n'}$ and $\nabla^2P^{q'}_{n'}=0$. The extraction from $\Nw$ is therefore $n'(n'+1)$ times the scalar one, and with $\curl\Mw=\Nw$ and \meqref{eq:closure} for the curls,
\begin{align*}
  (\nabla P^{q'}_{n'})(\nabla)\cdot\Nw[\Phi^q_n]\bigr|_{\bm0} &= n(n+1)(2n-1)!!\,\delta_{nn'}\delta_{qq'} ,
  &
  (\nabla P^{q'}_{n'})(\nabla)\cdot\Mw[\Phi^q_n]\bigr|_{\bm0} &= 0 ,
  \\
  (\nabla P^{q'}_{n'})(\nabla)\cdot\curl\Mw[\Phi^q_n]\bigr|_{\bm0} &= n(n+1)(2n-1)!!\,\delta_{nn'}\delta_{qq'} ,
  &
  (\nabla P^{q'}_{n'})(\nabla)\cdot\curl\Nw[\Phi^q_n]\bigr|_{\bm0} &= 0 .
\end{align*}
The operator $(\nabla P^q_n)(\nabla)$ applied at the centre therefore extracts the TE coefficient of degree $n$ and order $q$ from a regular field and the TM coefficient from its curl. The expansion of $\vH^{\Gamma}_{c,p}$ converges with all its derivatives within the distance $\min(D,\hs-D)$, so it may be differentiated term by term. In physical coordinates, with $\nabla=\alpha\nabla_x$, the entries are
\begin{align}
  \Ret^{(n,m)}_{\TE,c}[q,p] &= w_n\,\alpha^{n-1}\bigl[(\nabla P^q_n)(\nabla_x)\cdot\vH^{\Gamma}_{c,p}\bigr](\vz) ,
  \nonumber\\
  \Ret^{(n,m)}_{\TM,c}[q,p] &= w_n\,\alpha^{n-1}\bigl[(\nabla P^q_n)(\nabla_x)\cdot\curl\vH^{\Gamma}_{c,p}\bigr](\vz) ,
  \label{eq:retgrad}
\end{align}
with $w_n:=1/[n(n+1)(2n-1)!!]$, which is \meqref{eq:retentries}. Here $(\nabla P^q_n)(\nabla_x)$ is the gradient of the solid harmonic with each $\xi_i$ replaced by $\partial_{x_i}$, as in \meqref{eq:Psi}.

\subsection{Reduction to the Reflected Point Sources}
\label{sec:s-retintegral}
By \meqref{eq:reflectedfield} the reflected field is built from $\vU_\Gamma$ and $\vU'_\Gamma$, and their contractions with $(\nabla P^q_n)(\nabla_x)$ are the polynomials \meqref{eq:Q} of degree $n$ and order $q$. Inside the integral over $k_\rho$ the operators act on horizontal modes, on which $\nabla_\rho^2=-k_\rho^2$ and, by the Helmholtz equation, $\nabla_x^2=-\ks^2$. The contractions therefore follow from identities between polynomials, with $\vxi$ in place of $\nabla_x$, $|\vxi_\rho|^2=-k_\rho^2$ and $|\vxi|^2=-\ks^2$. Since $\vxi_\rho\cdot\nabla_\rho P^q_n=nP^q_n-\xi_3\partial_3P^q_n$ and $1=-|\vxi_\rho|^2/k_\rho^2$, \meqref{eq:Ufree} gives
\begin{align*}
  \nabla P^q_n\cdot\Bigl(\ehat_3+\frac{\xi_3\vxi_\rho}{k_\rho^2}\Bigr)
  &= \frac{n\,\xi_3P^q_n-|\vxi|^2\,\partial_3P^q_n}{k_\rho^2}
  = \frac{\ehat_3\cdot\bigl(\vxi\times(\vxi\times\nabla P^q_n)\bigr)}{k_\rho^2} ,
  &
  \nabla P^q_n\cdot\frac{\vxi_\rho\times\ehat_3}{k_\rho^2}
  &= -\frac{\ehat_3\cdot(\vxi\times\nabla P^q_n)}{k_\rho^2} .
\end{align*}
The curls exchange the two forms,
\begin{equation*}
  \vxi\times\Bigl(\ehat_3+\frac{\xi_3\vxi_\rho}{k_\rho^2}\Bigr)=\ks^2\,\frac{\vxi_\rho\times\ehat_3}{k_\rho^2} ,
  \qquad
  \vxi\times\frac{\vxi_\rho\times\ehat_3}{k_\rho^2}=\ehat_3+\frac{\xi_3\vxi_\rho}{k_\rho^2} ,
\end{equation*}
the first since $\xi_3^2=k_\rho^2-\ks^2$. With \meqref{eq:Q},
\begin{align*}
  (\nabla P^q_n)(\nabla_x)\cdot\vU_\Gamma &= -\frac{Q^{(n)}_{\TE,q}(\nabla_x)}{c_n\alpha^{n+2}k_\rho^2}\,u_\Gamma ,
  &
  (\nabla P^q_n)(\nabla_x)\cdot\vU'_\Gamma &= \frac{Q^{(n)}_{\TM,q}(\nabla_x)}{c_n\alpha^{n+1}k_\rho^2}\,u'_\Gamma ,
  \\
  (\nabla P^q_n)(\nabla_x)\cdot\curlx\vU_\Gamma &= \frac{\ks^2\,Q^{(n)}_{\TM,q}(\nabla_x)}{c_n\alpha^{n+1}k_\rho^2}\,u_\Gamma ,
  &
  (\nabla P^q_n)(\nabla_x)\cdot\curlx\vU'_\Gamma &= -\frac{Q^{(n)}_{\TE,q}(\nabla_x)}{c_n\alpha^{n+2}k_\rho^2}\,u'_\Gamma .
\end{align*}
The derivatives $Q^{(m)}_{c,p}(-\nabla_z)$ and ${Q'}^{(m)}_{c,p}(-\nabla_z)$ commute with these. Substituting in \eqref{eq:retgrad}, with $w_n=(-1)^nc_n/n(n+1)$, $\curl=\alpha\curlx$ and the relations \meqref{eq:Qprime}, the constant $c_n$ and the power $\alpha^n$ cancel and both types take one form,
\begin{equation}
  \Ret^{(n,m)}_{c',c}[q,p] = \frac{(-1)^{n+1}\eta_{c'}}{n(n+1)\alpha^3}\Bigl[Q^{(n)}_{c',q}(\nabla_x)\,Q^{(m)}_{c,p}(-\nabla_z)\,\frac{u_\Gamma}{k_\rho^2}
  -\frac{1}{\ks^2}\,{Q'}^{(n)}_{c',q}(\nabla_x)\,{Q'}^{(m)}_{c,p}(-\nabla_z)\,\frac{u'_\Gamma}{k_\rho^2}\Bigr]_{\vx=\vz} ,
  \label{eq:retQ}
\end{equation}
with $\eta_{\TE}=1$ and $\eta_{\TM}=-\kappa^2$. The derivatives act inside the integral over $k_\rho$ and are evaluated at $\rho=0$, so each entry is one integral over $k_\rho$.

At degree one $P^q_1=\xi_q$, and the polynomials \meqref{eq:Q}, with $c_1=-1$, are the components of the vector polynomials
\begin{equation*}
  \bm A(\vxi) := \vxi\times(\vxi\times\ehat_3) ,
  \qquad
  \bm B(\vxi) := \ehat_3\times\vxi ,
\end{equation*}
with $Q^{(1)}_{\TE,q}=\alpha^3A_q$ and $Q^{(1)}_{\TM,q}=\alpha^2B_q$. The entries then form the blocks
\begin{equation}
  \begin{aligned}
  \Ret^{(1,1)}_{\TE,\TE} &= \frac{\alpha^3}{2}\Bigl[\bm A\bm A^{\top}\frac{u_\Gamma}{k_\rho^2}-\ks^2\,\bm B\bm B^{\top}\frac{u'_\Gamma}{k_\rho^2}\Bigr]_{\vx=\vz} , &
  \Ret^{(1,1)}_{\TM,\TM} &= \frac{\alpha^3}{2}\Bigl[\bm A\bm A^{\top}\frac{u'_\Gamma}{k_\rho^2}-\ks^2\,\bm B\bm B^{\top}\frac{u_\Gamma}{k_\rho^2}\Bigr]_{\vx=\vz} ,
  \\
  \Ret^{(1,1)}_{\TE,\TM} &= \frac{\alpha^2}{2}\Bigl[\bm A\bm B^{\top}\frac{u_\Gamma}{k_\rho^2}-\bm B\bm A^{\top}\frac{u'_\Gamma}{k_\rho^2}\Bigr]_{\vx=\vz} , &
  \Ret^{(1,1)}_{\TM,\TE} &= -\frac{\alpha^4\ks^2}{2}\Bigl[\bm B\bm A^{\top}\frac{u_\Gamma}{k_\rho^2}-\bm A\bm B^{\top}\frac{u'_\Gamma}{k_\rho^2}\Bigr]_{\vx=\vz} ,
  \end{aligned}
  \label{eq:ret11}
\end{equation}
with the first factor of each product at $\nabla_x$ and the second at $-\nabla_z$. The blocks of one type differ only by the exchange of $u_\Gamma$ and $u'_\Gamma$.

\subsection{Symmetries of the Reflection Matrix}
\label{sec:s-retsymmetry}
The reflection matrix has the counterpart of the symmetry \meqref{eq:symmetry} of the T-matrix, by the same pairing. The pairing is taken between the fields $(\vE_1,\vH_1)$ and $(\vE_2,\vH_2)$ of the outgoing modes $(c,m,p)$ and $(c',n,q)$ of unit coefficient radiated from $\vz$, each with its transmission into the air and its reflections, in the layered ground without the object. Both are source-free everywhere except at $\vz$. A small ball about $\vz$ is removed from space, and as in Section~\ref{sec:s-drive} the divergence theorem over what remains, whose boundary is the sphere bounding the ball and the sphere at infinity, gives
\begin{equation*}
  \oint_{\vz}\bigl(\vE_1\times\vH_2-\vE_2\times\vH_1\bigr)\cdot\xihat\dd S = 0 ,
\end{equation*}
with $\xihat$ pointing out of the ball. On that sphere each field is its outgoing mode plus the regular expansion of its reflection, whose coefficients are a column of $\Ret$. The two outgoing modes pair to zero, as do the two reflections. By \eqref{eq:pairing}, the outgoing mode of field 1 with the reflection of field 2 gives $(e_c/\beta_m)\,\Ret^{(m,n)}_{c,c'}[p,q]$, and the reflection of field 1 with the outgoing mode of field 2 gives $-(e_{c'}/\beta_n)\,\Ret^{(n,m)}_{c',c}[q,p]$. Their sum vanishes, so
\begin{equation}
  \frac{e_c}{\beta_m}\,\Ret^{(m,n)}_{c,c'}[p,q] = \frac{e_{c'}}{\beta_n}\,\Ret^{(n,m)}_{c',c}[q,p] ,
  \label{eq:retsymmetry}
\end{equation}
as \meqref{eq:symmetry} for the T-matrix. A block of one type and one degree is therefore symmetric, and at one degree the two cross blocks are transposes of each other up to the factor $e_{\TE}/e_{\TM}$.

The layers are symmetric about the vertical through $\vz$, so the reflection matrix couples only orders of the same $|p|$. They are also unchanged by reflection in a vertical plane through $\vz$, which maps a harmonic varying as $\cos p\phi$ to itself and one varying as $\sin|p|\phi$ to its negative, and which reverses a TE mode relative to the TM mode on the same harmonic. The blocks of one type therefore couple each order with itself, and the cross blocks couple $p$ with $-p$ and vanish at $p=0$.

\subsection{The Size of the Reflection Matrix}
\label{sec:s-reflectionsize}
Near the object the four paths of \meqref{eq:gGamma} have lengths $2D$, $2(\hs-D)$, and $2\hs$ for the two that meet both boundaries, so each term of $u_\Gamma$ is a point source at an image of $\vz$: in the surface, in the interface beneath, or in both. Since $\operatorname{Re}\gams\ge k_\rho$, the host attenuates each image below its static value $e^{-k_\rho\lambda}$, with $\lambda$ the length of the path. A reflected mode of degree $m$ is therefore at most a multipole of degree $m$ at an image, and its re-expansion in regular modes of degree $n$ about $\vz$ is $O\bigl((\alpha/2D)^{m+n+1}\bigr)$ for the image in the surface and $O\bigl((\alpha/2(\hs-D))^{m+n+1}\bigr)$ for the image beneath, times the reflection coefficients. The images in both boundaries are farther still. The re-expansion converges over the object only if each image lies farther from $\vz$ than twice the circumscribed radius, which is the condition of Section~\mref{sec:sea}.

\section{The Assembled Model}
\label{sec:s-assembled}
This section shows that the assembled model is reciprocal, as the physical problem is, and that its truncation by order keeps every block up to a given power of the size of the object, with reciprocity preserved.

\subsection{Reciprocity}
\label{sec:s-reciprocity}
Exchanging the transmitter and a receiving coil, positions and moments, leaves the signal unchanged in the physical problem, and the model shares this property. For a receiving coil of moment $\vm_\mathrm{R}$ at $\vx$, \meqref{eq:shape} and \meqref{eq:drivereciprocity} give
\begin{equation*}
  \vm_\mathrm{R}\cdot\Delta \vH = \frac{\ii\omega\mu_0}{\alpha^2}\,\vm_\mathrm{R}^{\top}\Prop(\vx,\vz)\,\bm{S}\,\Prop(\vy,\vz)^{\top}\vm_\mathrm{T} ,
\end{equation*}
with $\bm{S}:=\Tens(\Id-\Ret\Tens)^{-1}\bLambda^{-1}$ and $\bLambda$ the diagonal matrix of the factors $e_c/\beta_m$ of \meqref{eq:symmetry}. The coils enter only through the propagators on either side of $\bm{S}$, and exchanging them transposes the expression, so the signal is unchanged if $\bm{S}$ is symmetric. Without the reflection $\bm{S}=\Tens\bLambda^{-1}$, which is symmetric by \meqref{eq:symmetry}. With the reflection,
\begin{equation*}
  \bm{S} = \sum_{\ell\ge0}\Tens(\Ret\Tens)^\ell\bLambda^{-1} ,
\end{equation*}
which is symmetric if $\bLambda\Ret$ is symmetric as $\bLambda\Tens$ is. By \eqref{eq:retsymmetry} and \meqref{eq:symmetry},
\begin{equation*}
  \Ret^{\top} = \bLambda\Ret\bLambda^{-1} ,
  \qquad
  \Tens^{\top} = \bLambda\Tens\bLambda^{-1} .
\end{equation*}
Transposing a term of $\bm{S}$ reverses its factors and replaces each by this form, and the inner $\bLambda$ and $\bLambda^{-1}$ cancel in pairs, which returns the term unchanged.

\subsection{Truncation by Order}
\label{sec:s-truncation}
The truncation $\Delta\vH^{(K)}$ applies one threshold, $\alpha^{K+1}$, to the blocks of both matrices. A block of the T-matrix is kept when its contribution in Section~\ref{sec:s-order} is at most $\alpha^{K+1}$, that is when its order is at most $K$. A block of the reflection matrix enters only along the paths $\Prop\,\Tens\,\Ret\,\Tens\cdots\vd$ of \meqref{eq:return}, where it multiplies the path by a power of $\alpha$. That power is simplest to find after a rescaling of the TM coefficients.

Written as $\alpha^{-1}\va_{\TM}$ and $\alpha^{-1}\vd_{\TM}$, the TM coefficients carry the same powers as the TE ones. The propagator and the drive then carry $\alpha^{m+2}$ and $\alpha^{n-1}$ for both types, the TE row of the table of Section~\ref{sec:s-order}. A diagonal block of the T-matrix carries no power of $\alpha$, and a cross block carries one. In \eqref{eq:retQ} the polynomials $Q$ of degree $n$ carry $\alpha^{n+1}$ or $\alpha^{n+2}$, so every block $\Ret^{(n,m)}$ carries $\alpha^{n+m+1}$, whatever its types.

A path through the blocks $\Tens^{(m_0,n_0)},\dots,\Tens^{(m_k,n_k)}$ of the T-matrix, joined by the $k$ reflections $\Ret^{(n_i,m_{i+1})}$, multiplies these powers,
\begin{equation*}
  O\Bigl(\alpha^{m_0+n_k+1}\,\alpha^{\chi}\prod_{i=0}^{k-1}\alpha^{n_i+m_{i+1}+1}\Bigr) ,
\end{equation*}
where $\chi$ counts the blocks $\Tens^{(m_i,n_i)}$ on the path that change the type, each of which carries one power of $\alpha$. The first factor is that of the propagator and the drive at the ends of the path, and each reflection multiplies the path by its own $\alpha^{n+m+1}$, at least $\alpha^3$. The truncation therefore keeps the blocks $\Ret^{(n,m)}$ with $n+m\le K$, whose factor is at most $\alpha^{K+1}$, the threshold of the T-matrix.

The order of a block depends on its degrees and types only through $m+n$ and whether the types differ, so the set of blocks kept is unchanged under the exchange of $(c,m)$ with $(c',n)$. The truncated $\bLambda\Tens$ and $\bLambda\Ret$ therefore remain symmetric, and the argument of Section~\ref{sec:s-reciprocity} applies to them unchanged: every truncated model is reciprocal.

For an object with a centre of symmetry the diagonal blocks vanish for odd $m+n$ and the cross blocks for even $m+n$ (Section~\mref{sec:cross}), so every block that remains has even order. The terms of odd order, among them the cross blocks of degree one, come from the asymmetry of the object alone, and for such an object $\Delta\vH^{(3)}=\Delta\vH^{(2)}$.

The count treats the ratios of the size of the object to the ranges of the sensor and the coil and to the distances of its images alike, and the ratio \meqref{eq:ratio} of the galvanic to the inductive response as $O(1)$.

\section{The Limits of the T-Matrix}
\label{sec:s-limits}
This section reduces the transmission problem in its two limits to the eddy-current and conduction problems that define the MPT and the electric polarizability tensor, and relates the blocks of higher degree to the generalised tensors. Both limits use the regular modes of degree one at $\kappa=0$, the solid harmonics $\Phi^q_1=\xi_q$, for which by \eqref{eq:staticregular}
\begin{equation}
  \Mw[\xi_q] = \ehat_q\times\vxi ,
  \qquad
  \Nw[\xi_q] = \curl(\ehat_q\times\vxi) = 2\ehat_q ,
  \label{eq:staticmodes}
\end{equation}
and the constant $\beta_1=1/8\pi$.

\subsection{The Inductive Limit}
\label{sec:s-mpt}
As $\eps\to0$ at fixed $\nua$, $\kappa^2=\ii\eps$ vanishes in $\sea$ while the object keeps $g=\ii\nua/\murb$. Each step below then holds to an error of $O(\eps)$, although $\kappa\propto\sqrt{\eps}$. The probes and drives contain $\kappa$ only through $j_n(\kappa r)/\kappa^n$, a series in $\kappa^2$, and the exterior only through $g=\ii\eps/\murs$. The outgoing condition brings odd powers of $\kappa$ through
\begin{equation*}
  \frac{e^{\ii\kappa R}}{R} = \frac{1}{R} + \ii\kappa - \frac{\kappa^2R}{2} - \frac{\ii\kappa^3R^2}{6} + \dots ,
\end{equation*}
but the fields depend on it only through its derivatives and through $\kappa^2$ times it, which carry no odd power below $\kappa^3$. The first correction is therefore of $O(\kappa^2)=O(\eps)$. The host is taken non-magnetic, $\murs=1$, so that its exterior is the free space in which the MPT is defined.

\emph{The drive.} By \eqref{s-drive1} the TE coefficients of degree one are $\vd^{(1)}_{\TE}=\tfrac12\vH_0(\vz)$, and by \meqref{eq:TE} and \eqref{eq:staticmodes} the field they carry is
\begin{equation*}
  \vH = \sum_q d^{(1)}_{\TE,q}\,\Nw[\xi_q] = \vH_0(\vz) ,
  \qquad
  \vE = e_{\TE}\sum_q d^{(1)}_{\TE,q}\,\Mw[\xi_q] = \tfrac12e_{\TE}\,\vH_0(\vz)\times\vxi .
\end{equation*}
The factor $2$ of $\Nw[\xi_q]$ cancels the $\tfrac12$ of the coefficients, so the magnetic field is the uniform $\vH_0(\vz)$; $\Mw[\xi_q]$ carries no such factor, and the electric field is the one whose curl Faraday's law requires. The TM coefficients $\tfrac12\alpha\sigs\vE_0$ vanish with $\sigs$ (Section~\ref{sec:s-nonconducting}).

\emph{The problem.} Outside the object $g=\ii\eps$ vanishes, so \meqref{eq:curlcurl} there reads $\curl\curl\vE=0$, and by \meqref{eq:scaledpair} $\curl\vH=0$: the exterior carries no current. Inside, \meqref{eq:curlcurl} is unchanged. The transmission problem is therefore the eddy-current problem for the object, of conductivity $\sigma_*$ and relative permeability $\murb$, in free space, driven by the uniform field $\vH_0(\vz)$. This is the problem that defines the MPT~\cite{ammariTargetDetectionCharacterization2014,ledgerCharacterizingShapeMaterial2015}.

\emph{The tensor.} The MPT is defined by $\vm=\mathcal{M}\vH_0(\vz)$, so it is found from the moment of the object, which the reaction formula gives as an integral. With the probe fields \eqref{eq:staticmodes}, $\Jimp\cdot(\ehat_p\times\vxi)=\ehat_p\cdot(\vxi\times\Jimp)$, so \meqref{eq:integrandTE} gives
\begin{equation*}
  a^{(1)}_{\TE,p} = \frac{1}{8\pi}\Bigl[\alpha\int_B(\vxi\times\Jimp)_p\dd\vxi + 2\int_B(\Mmat)_p\dd\vxi\Bigr] ,
\end{equation*}
and by \meqref{eq:response1}, with $\vx-\vz=\alpha\vxi$ and $\dd\vx=\alpha^3\dd\vxi$,
\begin{equation}
  \vm = \frac12\int_B(\vx-\vz)\times\Jimp\dd\vx + \int_B\Mmat\dd\vx ,
  \label{eq:mmoment}
\end{equation}
the magnetic moment of the eddy currents and of the magnetisation. The field in the object is linear in $\vH_0(\vz)$. In the notation of~\cite{ledgerCharacterizingShapeMaterial2015} it is
\begin{equation*}
  \vE = \tfrac12e_{\TE}\sum_iH_{0,i}\,\bigl(\bm\theta_i+\ehat_i\times\vxi\bigr) ,
\end{equation*}
where $\bm\theta_i$ solves
\begin{align*}
  \curl\bigl(\mur^{-1}\curl\bm\theta_i\bigr) - g\,\bm\theta_i &= g\,\ehat_i\times\vxi
  && \text{in $B$ and outside it} ,\\
  [\bm\theta_i\times\nhat] = \bm0 ,\quad
  \bigl[\mur^{-1}\curl\bm\theta_i\times\nhat\bigr] &= -2\bigl[\mur^{-1}\bigr]\,\ehat_i\times\nhat
  && \text{on $\partial B$} ,\\
  \bm\theta_i &= O(|\vxi|^{-1})
  && \text{as $|\vxi|\to\infty$} ,
\end{align*}
with $\mur$ equal to $\murb$ in $B$ and $1$ outside, $g=0$ and $\nabla\cdot\bm\theta_i=0$ outside $B$, and $[\,\cdot\,]$ the jump of Section~\ref{sec:s-nonconducting}, outside less inside. This is the problem above for the uniform field $\ehat_i$, written for the reaction $\bm\theta_i$ to the incident field $\ehat_i\times\vxi$, whose curl $2\ehat_i$ gives the jump term, and multiplied through by $\mu_0$ it is their problem for $B$ itself. In the object $\Jimp=\sigma_*\vE$, and by Faraday's law
\begin{equation*}
  \vH = \frac{1}{\murb}\sum_iH_{0,i}\,\bigl(\ehat_i+\tfrac12\curl\bm\theta_i\bigr) ,
\end{equation*}
so \eqref{eq:mmoment} gives
\begin{equation}
  \mathcal{M}_{mi} = \frac{\ii\omega\mu_0\alpha^5}{4}\,\ehat_m\cdot\int_B\sigma_*\,\vxi\times(\bm\theta_i+\ehat_i\times\vxi)\dd\vxi
  + \alpha^3\int_B\Bigl(1-\frac{1}{\murb}\Bigr)\Bigl(\delta_{mi}+\tfrac12\ehat_m\cdot\curl\bm\theta_i\Bigr)\dd\vxi ,
  \label{eq:mptintegral}
\end{equation}
the first term from the eddy currents and the second from the magnetisation. For a homogeneous object these are $-\check{\mathcal{C}}_{mi}$ and $\mathcal{N}_{mi}$ of~\cite{ledgerCharacterizingShapeMaterial2015}, whose sum is the MPT there. Only $\Tens_{\TE,\TE}$ is driven, so
\begin{equation*}
  \vm = 4\pi\alpha^3\,\Tens^{(1,1)}_{\TE,\TE}\,\tfrac12\vH_0(\vz) ,
  \qquad
  \mathcal{M} = 2\pi\alpha^3\,\Tens^{(1,1)}_{\TE,\TE}\big|_{\eps=0} ,
\end{equation*}
which is \meqref{eq:mpt}.

\subsection{The Static Limit}
\label{sec:s-pst}
As $\omega\to0$ at a fixed contrast $\zs$, for a homogeneous object of conductivity $\sigma_*$, $\kappa^2$ vanishes in the object and in $\sea$ together.

\emph{The drive.} By \eqref{s-drive1} the TM coefficients of degree one are $\vd^{(1)}_{\TM}=\tfrac12\alpha\sigs\vE_0(\vz)$, and by \meqref{eq:TM} and \eqref{eq:staticmodes} the field they carry is
\begin{equation*}
  \vE = e_{\TM}\sum_q d^{(1)}_{\TM,q}\,\Nw[\xi_q] = \vE_0(\vz) ,
  \qquad
  \vH = \sum_q d^{(1)}_{\TM,q}\,\Mw[\xi_q] = \tfrac12\alpha\sigs\,\vE_0(\vz)\times\vxi ,
\end{equation*}
since $e_{\TM}=1/\alpha\sigs$ and the factor $2$ of $\Nw[\xi_q]$ cancel the coefficients' $\tfrac12\alpha\sigs$. The electric field is uniform, and the magnetic field is the one whose curl Amp\`ere's law requires. The TE drive $\tfrac12\vH_0$ reaches the TM response only through $\Tens^{(1,1)}_{\TM,\TE}=\kappa^2\bigl(\Tens^{(1,1)}_{\TE,\TM}\bigr)^{\!\top}$, by \meqref{eq:symmetry}, which vanishes.

\emph{The problem.} Both terms of \meqref{eq:weak} that carry $g$ vanish as $\omega\to0$, but they cannot simply be dropped, because they carry the conservation of current. A gradient test field $\vw=\nabla\phi$, with $\phi$ vanishing on $\partial\Omega$, removes the curl terms, and dividing what remains by $\ii\omega\mu_0\alpha^2$ leaves
\begin{equation*}
  \int_\Omega\sigma\vE\cdot\nabla\phi\dd\vxi = 0 ,
\end{equation*}
since the host current $\sigs\vE_0$ is conserved on its own. This is the weak form of $\nabla\cdot(\sigma\vE)=0$ with $[\sigma\vE\cdot\nhat]=0$, and it holds at every frequency. Faraday's law in \meqref{eq:scaledpair}, $\curl\vE=\ii\omega\mu_0\mur\alpha\vH$, then makes the total field a gradient in the limit,
\begin{equation*}
  \vE = \nabla u ,
\end{equation*}
with $[u]=0$. Since $\vE=\vE_0+\Delta\vE$, with $\vE_0$ the uniform drive above and $\Delta\vE$ the perturbation, which decays away from the object, $u-\vE_0(\vz)\cdot\vxi$ decays as well. Together these are the conduction problem $\nabla\cdot(\sigma\nabla u)=0$ that defines the polarization tensor~\cite{polyaIsoperimetricInequalitiesMathematical1951,ammariPolarizationMomentTensors2007}.

\emph{The tensor.} The electric polarizability tensor is defined by $\vp=\mathcal{P}\sigs\vE_0(\vz)$, so it is found from the current moment of the object, which the reaction formula gives as an integral. At $\kappa=0$ only the first term of \meqref{eq:integrandTM} remains, and with \eqref{eq:staticmodes}
\begin{equation*}
  a^{(1)}_{\TM,p} = \frac{\alpha}{4\pi}\int_B(\Jimp)_p\dd\vxi ,
\end{equation*}
so by \meqref{eq:response1}
\begin{equation}
  \vp = \int_B\Jimp\dd\vx = (\sigma_*-\sigs)\int_B\nabla u\dd\vx ,
  \label{eq:pmoment}
\end{equation}
the excess current of the object. The potential is linear in $\vE_0(\vz)$. In the notation of~\cite{ammariPolarizationMomentTensors2007} it is, up to a constant,
\begin{equation*}
  u = \sum_jE_{0,j}(\vz)\,v_j ,
\end{equation*}
where $v_j$ solves
\begin{align*}
  \nabla\cdot\bigl((\sigma/\sigs)\nabla v_j\bigr) &= 0
  && \text{in $B$ and outside it} ,\\
  [v_j] = 0 ,\quad
  \bigl[(\sigma/\sigs)\,\partial v_j/\partial n\bigr] &= 0
  && \text{on $\partial B$} ,\\
  v_j-\xi_j &= O(|\vxi|^{-2})
  && \text{as $|\vxi|\to\infty$} ,
\end{align*}
with $\sigma/\sigs$ equal to $\sigma_*/\sigs$ in $B$ and $1$ outside, and $[\,\cdot\,]$ the jump as above. This is the problem above for the uniform field $\ehat_j$, written for the total potential, and divided through by $\sigs$ it is their problem for $B$ itself. Since $\dd\vx=\alpha^3\dd\vxi$, \eqref{eq:pmoment} gives
\begin{equation}
  \mathcal{P}_{ij} = \alpha^3(\zs-1)\int_B\partial_iv_j\dd\vxi .
  \label{eq:pstintegral}
\end{equation}
The integral is the polarization tensor of $B$ at the contrast $\zs$, written $M_{ij}(k,B)$ with $k=\zs$ in~\cite{ammariPolarizationMomentTensors2007}, so $\mathcal{P}=\alpha^3M$: the tensor scales with the volume of the object. Only $\Tens_{\TM,\TM}$ is driven, so
\begin{equation*}
  \vp = 4\pi\alpha^2\,\Tens^{(1,1)}_{\TM,\TM}\,\tfrac12\alpha\sigs\vE_0(\vz) ,
  \qquad
  \mathcal{P} = 2\pi\alpha^3\,\Tens^{(1,1)}_{\TM,\TM}\big|_{\omega=0} ,
\end{equation*}
which is \meqref{eq:pst}.

\subsection{Higher Degrees}
\label{sec:s-higher}
In both limits $\kappa$ vanishes in the host, so the basis drives and probes of higher degree are, by \eqref{eq:staticregular}, gradients of solid harmonics. In the inductive limit the object keeps its eddy currents, as in~\cite{ledgerGeneralisedMagneticPolarizability2018}; only the fields outside it are static. The blocks of higher degree are the components in the harmonic basis of the generalised tensors. Below, $D^k$ denotes the tensor of $k$-th partial derivatives, as in~\cite{ledgerGeneralisedMagneticPolarizability2018}; for a polynomial $P$ of degree $m$, $D^mP$ is constant, the symmetric tensor of its coefficients.

\emph{The inductive blocks.} In the host at $\eps=0$ the TE basis drive $(n,q)$ is the field $\vH_0=(n+1)\nabla P^q_n$. In physical coordinates it and its derivatives at the centre are
\begin{equation*}
  \vH_0(\vx) = (n+1)\,\alpha^{1-n}\,(\nabla P^q_n)(\vx-\vz) ,
  \qquad
  D^{n-1}\vH_0(\vz) = (n+1)\,\alpha^{1-n}\,D^nP^q_n .
\end{equation*}
Outside the object, by \eqref{eq:staticregular}, the response \meqref{eq:response} is $\Delta\vH=-\sum_{m,p}m\,a^{(m)}_{\TE,p}\nabla\Psi^p_m$, and with $1/r=4\pi\alpha G$ and $G(\vx,\vz)=1/4\pi|\vx-\vz|$,
\begin{equation*}
  \Delta\vH = \sum_{m,p}\frac{(-1)^{m+1}4\pi m\,\alpha^{m+2}}{(2m-1)!!}\,a^{(m)}_{\TE,p}\,\nabla_xP^p_m(\nabla_x)\,G .
\end{equation*}
Ledger and Lionheart~\cite{ledgerGeneralisedMagneticPolarizability2018} write the same field, summed over the degrees and the multi-indices $K(m)$ and $J(n)$, as
\begin{equation*}
  (\Delta\vH)_i = \sum\bigl(D_x^{m+1}G\bigr)_{i,K(m)}\,\mathfrak{M}_{K(m)J(n)}\,\bigl(D^{n-1}\vH_0(\vz)\bigr)_{J(n)} .
\end{equation*}
This is the gradient of $\mathcal{U}(\nabla_x)G$ for the polynomial
\begin{equation*}
  \mathcal{U}(\vxi) = \sum_K\mathcal{U}_K\,\vxi_K ,
  \qquad
  \mathcal{U}_K = \sum_J\mathfrak{M}_{KJ}\,\bigl(D^{n-1}\vH_0\bigr)_J .
\end{equation*}
Only the harmonic part of $\mathcal{U}$ survives on $G$, and by \eqref{eq:fischer} its component along $P^p_m$ is
\begin{equation*}
  \frac{P^p_m(\nabla)\,\mathcal{U}}{(2m-1)!!} = \frac{1}{(2m-1)!!}\sum_K\bigl(D^mP^p_m\bigr)_K\,\mathcal{U}_K .
\end{equation*}
Equating the components,
\begin{equation}
  \Tens^{(m,n)}_{\TE,\TE}[p,q]\big|_{\eps=0} = \frac{(-1)^{m+1}(n+1)}{4\pi m\,\alpha^{m+n+1}}\sum_{K,J}\bigl(D^mP^p_m\bigr)_K\,\mathfrak{M}_{K(m)J(n)}\,\bigl(D^nP^q_n\bigr)_J ,
  \label{eq:gmpt}
\end{equation}
in which the factor $\alpha^{m+n+1}$ of the definition of $\mathfrak{M}$ cancels. At degree one $\mathfrak{M}$ is the MPT, and
\begin{equation*}
  \Tens^{(1,1)}_{\TE,\TE}\big|_{\eps=0} = \frac{\mathcal{M}}{2\pi\alpha^3} ,
\end{equation*}
which is \meqref{eq:mpt}.

\emph{The galvanic blocks.} The TM basis drive $(n,q)$ carries $\vE=e_{\TM}(n+1)\nabla P^q_n$. As in Section~\ref{sec:s-pst} the field in the object is then $e_{\TM}(n+1)\nabla v^q_n$, with $v^q_n$ the solution of the conduction problem there with $P^q_n$ in place of $\xi_j$, and since $e_{\TM}\alpha\sigs=1$,
\begin{equation*}
  \alpha\Jimp = \Bigl(\frac{\sigma_*}{\sigs}-1\Bigr)(n+1)\,\nabla v^q_n .
\end{equation*}
With the probe $\Nw[P^p_m]=(m+1)\nabla P^p_m$, \meqref{eq:integrandTM} at $\kappa=0$ gives
\begin{equation*}
  \Tens^{(m,n)}_{\TM,\TM}[p,q]\big|_{\omega=0} = (m+1)(n+1)\,\beta_m\Bigl(\frac{\sigma_*}{\sigs}-1\Bigr)\int_B\nabla P^p_m\cdot\nabla v^q_n\dd\vxi .
\end{equation*}
In~\cite{ammariPolarizationMomentTensors2007} the problem is posed for each monomial $\vxi^{\bm a}=\xi_1^{a_1}\xi_2^{a_2}\xi_3^{a_3}$, of degree $|\bm a|=a_1+a_2+a_3$, with solution $u_{\bm a}$. With $\varphi_{\bm a}=[\partial u_{\bm a}/\partial n]$ the jump of its normal derivative, the transmission condition gives the generalised polarization tensor as
\begin{equation*}
  M_{\bm a\bm b} = \int_{\partial B}\vxi^{\bm b}\varphi_{\bm a}\dd S = \Bigl(\frac{\sigma_*}{\sigs}-1\Bigr)\int_B\nabla\vxi^{\bm b}\cdot\nabla u_{\bm a}\dd\vxi .
\end{equation*}
Writing the harmonics in monomials,
\begin{equation*}
  P^q_n = \sum_{|\bm a|=n}A^q_{\bm a}\,\vxi^{\bm a} ,
  \qquad
  P^p_m = \sum_{|\bm b|=m}B^p_{\bm b}\,\vxi^{\bm b} ,
\end{equation*}
the solution for $P^q_n$ is $v^q_n=\sum_{\bm a}A^q_{\bm a}u_{\bm a}$, so
\begin{equation}
  \Tens^{(m,n)}_{\TM,\TM}[p,q]\big|_{\omega=0} = (m+1)(n+1)\,\beta_m\sum_{\bm a,\bm b}A^q_{\bm a}B^p_{\bm b}\,M_{\bm a\bm b} .
  \label{eq:gpt}
\end{equation}
At degree one $(m+1)(n+1)\beta_1=1/2\pi$ and the sum is the polarization tensor $M$ of $B$, so with \eqref{eq:pstintegral}
\begin{equation*}
  \Tens^{(1,1)}_{\TM,\TM}\big|_{\omega=0} = \frac{M}{2\pi} = \frac{\mathcal{P}}{2\pi\alpha^3} ,
\end{equation*}
which is \meqref{eq:pst}.

\section{The Range Laws and the Buried MPT}
\label{sec:s-rangelaws}
This section derives the range laws \meqref{eq:rangelaw} of the propagator and the ratio \meqref{eq:ratio} of the galvanic to the inductive response, and compares the transmitted point source with the kernel of the buried MPT \meqref{eq:buried}.

Within a skin depth of the object the integrand of $u_\tau$ is dominated by $k_\rho\sim1/R\gg1/\dels$. There, by \meqref{eq:admittance} with $\dels=\sqrt{2/\omega\mu_0\murs\sigs}$,
\begin{equation*}
  \gams = \sqrt{k_\rho^2-\ii\omega\mu_0\murs\sigs}
  = \sqrt{k_\rho^2-\frac{2\ii}{\dels^2}}
  \approx k_\rho .
\end{equation*}
For non-magnetic layers whose skin depths also exceed $R$, every $\gamma_j$ and every admittance $Y_j$ of \meqref{eq:admittance} is then close to $k_\rho$, and the layers are invisible at these wavenumbers. By \meqref{eq:RT}, $\Gamma_0\to0$ and $\tau\to1$, and every interface coefficient of \meqref{eq:recursion} vanishes, so that $\Gamma_1\to0$. The layer factor \meqref{eq:layerfactor} reduces to $L\to e^{k_\rho z_3}$, and with $k_\rho/\gams\to1$ the transmitted point source \meqref{eq:transmittedsource} becomes the Sommerfeld integral of the static point source,
\begin{equation*}
  u_\tau \to \int_0^\infty J_0(k_\rho\rho)\,e^{-k_\rho(x_3-z_3)}\dd k_\rho = \frac{1}{R} ,
\end{equation*}
since $x_3>z_3$. This is homogeneous of degree $-1$, so each derivative costs one power of $R$. A layer beneath with a smaller skin depth leaves $\Gamma_1$ finite, and its term in $L$ is the image of the object in that interface, which is negligible where $R\ll2(\hs-D)$. By \meqref{eq:propentries} only $Q$ enters the propagator. By \eqref{eq:Qexplicit}, $Q^{(m)}_{\TM}$ has degree $m$ and the factor $\alpha^{m+1}$, and $Q^{(m)}_{\TE}$ has degree $m+1$ and the factor $\alpha^{m+2}$. Acting on $u_\tau\to1/R$, these give
\begin{equation*}
  \bigl|\Prop^{(m)}_{\TM}\bigr| \sim \Bigl(\frac{\alpha}{R}\Bigr)^{m+1} ,
  \qquad
  \bigl|\Prop^{(m)}_{\TE}\bigr| \sim \Bigl(\frac{\alpha}{R}\Bigr)^{m+2} ,
\end{equation*}
which is \meqref{eq:rangelaw}.

By \meqref{eq:drivereciprocity} the drive is the propagator at the coil divided by $e_c$, so with \meqref{eq:rangelaw} at the range $R_\mathrm{T}$,
\begin{equation*}
  \frac{|\vd^{(1)}_{\TM}|}{|\vd^{(1)}_{\TE}|}
  = \Bigl|\frac{e_{\TE}}{e_{\TM}}\Bigr|\,\frac{|\Prop^{(1)}_{\TM}(\vy,\vz)|}{|\Prop^{(1)}_{\TE}(\vy,\vz)|}
  \sim |\kappa^2|\,\frac{R_\mathrm{T}}{\alpha} = \frac{\eps R_\mathrm{T}}{\alpha} .
\end{equation*}
With the diagonal blocks of degree one replaced by their limits, $\Tens^{(1,1)}_{\TE,\TE}=\mathcal{M}/2\pi\alpha^3$ and $\Tens^{(1,1)}_{\TM,\TM}=\mathcal{P}/2\pi\alpha^3$, the galvanic term stands to the inductive one as the product of the ratios of the propagators, the blocks and the drives,
\begin{equation*}
  \frac{|\Delta\vH_{\TM}|}{|\Delta\vH_{\TE}|}
  \sim \frac{R}{\alpha}\cdot\frac{|\mathcal{P}|}{|\mathcal{M}|}\cdot\frac{\eps R_\mathrm{T}}{\alpha}
  = \frac{|\mathcal{P}|}{|\mathcal{M}|}\,\frac{2R_\mathrm{T}R}{\dels^2} ,
\end{equation*}
with $\eps=2\alpha^2/\dels^2$, which is \meqref{eq:ratio}. Beyond a skin depth the contributing wavenumbers satisfy $k_\rho\dels\ll1$, where $\gams^2=k_\rho^2-2\ii/\dels^2$ gives $\gams\approx(1-\ii)/\dels$.

The buried MPT \meqref{eq:buried} of~\cite{ledgerCharacterisingBuriedObjects2025} is derived for an object in a conducting half space, which is the layered ground with nothing beneath the host, $\Gamma_1=0$. Its kernel is then the transmitted point source of the model. With $\Gamma_1=0$ the layer factor \meqref{eq:layerfactor} is $e^{\gams z_3}$, and with $\tau$ of \meqref{eq:RT}
\begin{equation*}
  \frac{u_\tau}{4\pi} = \frac{\murs}{2\pi}\int_0^\infty\frac{k_\rho\,J_0(k_\rho\rho)}{\gams+\murs k_\rho}\,e^{\gams z_3-k_\rho x_3}\dd k_\rho .
\end{equation*}
The kernel $G_\mathrm{s}$ of~\cite{ledgerCharacterisingBuriedObjects2025}, their (41) with the exponent of their (42b), is a two-dimensional Fourier integral over the horizontal wave vector $\bm k=(u,v)$, which for an integrand depending on $k_\rho=|\bm k|$ alone is a Hankel transform,
\begin{equation*}
  G_\mathrm{s} = \frac{1}{(2\pi)^2}\iint\frac{e^{\gams z_3-k_\rho x_3}}{\gams+\murs k_\rho}\,e^{\ii\bm k\cdot(\vx-\vz)_\rho}\dd u\dd v
  = \frac{1}{2\pi}\int_0^\infty\frac{k_\rho\,J_0(k_\rho\rho)}{\gams+\murs k_\rho}\,e^{\gams z_3-k_\rho x_3}\dd k_\rho .
\end{equation*}
Hence $u_\tau/4\pi=\murs G_\mathrm{s}$. The factor $\murs$ is that between the excess magnetisation \meqref{eq:densities} and their $(\mu_{\mathrm{r},*}-\murs)\vH$, so the two kernels give the same field. Their result \meqref{eq:buried} drops the factor, taking $|\murs-1|=O(\alpha/D)$, and for the non-magnetic hosts of the main paper $u_\tau/4\pi=G_\mathrm{s}$ exactly.

The model and the buried MPT \meqref{eq:buried} differ in where the vertical derivative of the dipole is taken. For the half space, $\partial_{z_3}$ in the propagator acts on $e^{\gams z_3}$ and inserts $\gams$, while $\partial_{x_3}$ at the sensor acts on $e^{-k_\rho x_3}$ and inserts $-k_\rho$; the horizontal derivatives agree. For a horizontal magnetic moment, $Q^{(1)}_{\TE,q}(\nabla_x)=\alpha^3\partial_3\partial_q$ carries one vertical derivative, and the two forms differ by the factor
\begin{equation*}
  \frac{\gams}{k_\rho} = 1-\frac{\ii}{(k_\rho\dels)^2}+O\bigl((k_\rho\dels)^{-4}\bigr) ,
\end{equation*}
a relative change of $O\bigl((R/\dels)^2\bigr)$ for $k_\rho\sim1/R$. For the vertical moment the source side carries $(\partial_{z_3}^2+\ks^2)\to\gams^2+\ks^2=k_\rho^2$ and the sensor side $\partial_{x_3}^2\to k_\rho^2$, the same, so only the horizontal moments change.

\section{The Reference Solution and the Drive}
\label{sec:reference}
This section describes the finite element reference against which the model is tested: how it solves for the perturbation, how the drive of the transmitting coil is computed, and how far the reference and the T-matrix have converged. Every factor of the model \meqref{eq:shape} is evaluated from the formulas of the main text: the T-matrix by \meqref{eq:reaction} from its own transmission problem \meqref{eq:weak}, the propagator by \meqref{eq:propentries}, the drive by \meqref{eq:drivereciprocity} and the reflection matrix by \meqref{eq:retentries}. None is taken from the reference solution, so the comparison tests all of them together.

\subsection{Solving for the Perturbation}
The reference introduces a magnetic vector potential $\vA$ with $\vH=(\mu_0\mur)^{-1}\curl\vA$, so that the eddy-current system of the main text becomes
\begin{equation}
  \curl\bigl(\mur^{-1}\curl\vA\bigr)-\ii\omega\mu_0\sigma\vA=\mu_0\vJ^\mathrm{s} ,
  \label{eq:vectorpotential}
\end{equation}
It is solved in a box $\Omega$ of side $\ell_0=28D$, with $\nhat\times\vA=\bm{0}$ on its boundary in place of the decay condition. Testing with $\vw$ in $\Hcurl$, with $\nhat\times\vw=\bm0$ on the boundary, and integrating the curl term by parts gives the weak form
\begin{equation}
  \bigl(\nu\curl\vA,\curl\vw\bigr)-\ii\,(s\vA,\vw) = \mu_0\,(\vJ^\mathrm{s},\vw) ,
  \qquad
  s = \begin{cases}
    \omega\mu_0\sigma , & \sigma>0 ,\\
    \varepsilon_{\mathrm{reg}}/\ell_0^2 , & \sigma=0 ,
  \end{cases}
  \label{eq:Aweak}
\end{equation}
with $\nu=\mur^{-1}$ and $(\bm a,\bm b)=\int_\Omega\bm a\cdot\bm b\dd\vx$. Where the medium conducts, $s$ gives the conduction term of \eqref{eq:vectorpotential}. Where it does not, that term is absent and the gradients lie in the kernel of the curl term, so that $\vA$ would be fixed there only up to a gradient and the discrete system would be singular. Rather than impose a Coulomb gauge there, $s$ takes the small value $\varepsilon_{\mathrm{reg}}/\ell_0^2$, an artificial conductivity with the phase of the conduction term. It is written as a relative perturbation: measured against the curl term for a field varying on the scale $\ell_0$ of the box, its size is exactly $\varepsilon_{\mathrm{reg}}$. The high-order gradient basis functions are dropped where the medium is non-conducting, and the box is discretised by second-order edge elements. One mesh serves every material configuration; only the coefficients change.

The model is compared with the perturbation of $\vA$ by the object. At metal-detector scale this is of order $10^{-4}$ of the total field and may be taken as the difference of two solutions, with and without the object. For an object at $50$\,m it is of order $10^{-7}$, and such a difference is set by the conditioning of the regularised system, scaling with $\varepsilon_{\mathrm{reg}}$.

The two weak forms are therefore subtracted before discretising rather than after. Writing $\vA=\vA_0+\Delta\vA$ for the background solution and the perturbation, the difference of the two forms \eqref{eq:Aweak} is
\begin{equation}
  \bigl(\nu\curl\Delta\vA,\curl\vw\bigr)-\ii\bigl(s\,\Delta\vA,\vw\bigr)
  =-\bigl((\nu-\nu_0)\curl\vA_0,\curl\vw\bigr)
   +\ii\bigl((s-s_0)\vA_0,\vw\bigr) ,
  \label{eq:perturbationweak}
\end{equation}
where $\nu$ and $s$ are the material maps with the object and $\nu_0$ and $s_0$ those without it. The regularisation is part of $s$, so its difference enters the right-hand side with that of the conductivity, and \eqref{eq:perturbationweak} is the exact difference of the two discrete systems. The right-hand side is supported within the object, where the two maps differ, so $\Delta\vA$ is computed without cancellation. It costs one solve beyond the background solution on the same mesh, and the field compared with the model is $\Delta\vA$ itself.

\subsection{The Drive of the Coil}
The drive is the transposed propagator \meqref{eq:drivereciprocity}. The coil, of radius $0.5$\,m at an altitude of $50$\,m, is a vertical magnetic dipole at its centre with the moment of its winding; its finite size changes the drive by less than $10^{-4}$. To degree $N$ the drive is the propagator of this dipole differentiated up to $N-1$ times in the position of the source, with the derivatives taken under the Hankel integral.

The drive agrees with the background solution $\vA_0$ of the reference. Both are sampled at $32$ points on the ellipsoid of the semi-axes of the object and fitted there by a linear function of the offset, and the two fits agree to within $0.2\%$ in $\vH_0$ and $\vE_0$ and to within $0.3\%$ in their gradients in every case.

\subsection{Convergence of the Reference and the T-Matrix}
\label{sec:s-budget}
The references are computed on the high-performance computing facility of the Queensland University of Technology, whose largest nodes have $1.5$\,TB of memory, and the direct factorisation of each sets its memory. The references of the main text have $0.7$ to $3.1$ million unknowns and need about $70$ to $500$\,GB, and $150$\,GB at the base case.

The reference meshes the object at four second-order elements per skin depth of the object, with prismatic layers at its surface, and the top layer near the object at three elements per skin depth of that layer, in a box of side $28D$; the interfaces below lie within the refined part of the mesh. The transmission problem \meqref{eq:weak} of the T-matrix is solved in the scaled coordinates on a ball of radius $9$ with $\nhat\times\vE=\bm0$ on its boundary, with second-order elements of size $0.14$ in the object, growing to $2.7$ at the boundary. With these settings the error of $\Delta\vH^{(4)}$ at the base case in the sea is $0.76\%$ for the permeable box and $0.75\%$ for the non-magnetic box.

To find how much of this is the error of the reference, the base case is refined towards the limit of a node. In the reference the box is enlarged to $56D$, the elements in the object are reduced by a factor of $1.5$, and those of the top layer and the air near the object by a factor of $2$, which gives $3.7$ million unknowns and needs about $0.6$\,TB; the same settings at $\nua=300$ would need about $1.9$\,TB, beyond the largest node. In the transmission problem the ball is enlarged to radius $18$, the elements in the object are reduced to $0.10$ and the order is raised to three. With both refined, the error of $\Delta\vH^{(4)}$ falls from $0.76\%$ to $0.20\%$ for the permeable box and from $0.75\%$ to $0.12\%$ for the non-magnetic one.

The errors of the main text are therefore set by the reference rather than by the model. Each order beyond $\Delta\vH^{(4)}$ changes the field by less than $0.1\%$, so the higher orders may be dropped.

The model is also compared, for the non-magnetic box, along a survey line close to the ground, with sensors at an altitude of $0.1D$ from above the object to an offset of $4D$. The mesh of the campaign is refined about the object and the ring, not along the line, where the field varies on the scale of the sensor height. For the line the reference is therefore refined in three steps: the refined part of the mesh is widened to cover the whole line, and a slab along the line is then meshed at $0.16D$ and at $0.08D$, with each sensor read as the mean over a small sphere. The last step changes the reference on the line by at most $1.0\%$, directly above the object.

\begin{figure}[!t]
\centering
\includegraphics{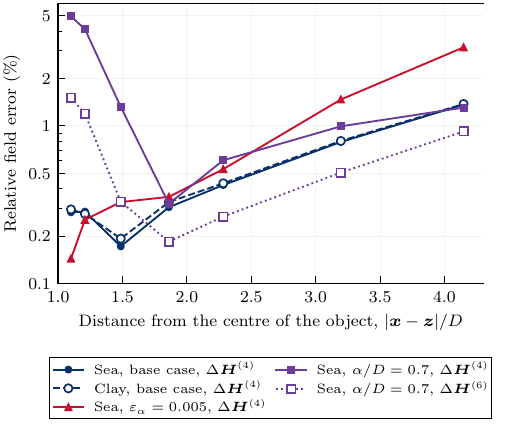}
\caption{Relative field error along the survey line for the non-magnetic box against the reference refined along it, at each sensor's distance from the centre of the object. The sensors are at an altitude of $0.1D$, from directly above the object to an offset of $4D$. The T-matrix is refined as in Section~\ref{sec:s-budget}, except at $\eps=0.005$.}
\label{fig:line}
\end{figure}

Fig.~\ref{fig:line} shows the error of the model along the line against the refined reference. At the base case, in both grounds, it is below $0.5\%$ within a distance $2D$ of the centre of the object, and rises to $1.4\%$ at the far end of the line, where the perturbation is weakest. Against the reference of the campaign the same model has errors of $2\%$ to $7\%$ with no trend along the line, the noise of a point reading in the coarser mesh.

Two cases depart from the base case. At $\eps=0.005$ the error at the far end is $3.2\%$, and it does not change with the order of the model. For the largest object the sensor directly above it is only $1.3$ circumscribed radii from its centre, where the expansion in degree converges slowly. The error there is $5.0\%$ at $\Delta\vH^{(4)}$ and $1.5\%$ at $\Delta\vH^{(6)}$.

\section{The Ladder from the Baseline to the Model}
\label{sec:s-ladder}
Table~\ref{tab:ladder} adds the ingredients of the model one at a time for both boxes in the sea, at the base case and for the largest object; on land the errors follow those in the sea (Section~\mref{sec:numerics}). From the buried MPT, the galvanic response is added first through its static limit $\mathcal{P}$, then through the $\eps$-retained diagonal blocks $\Tens_{\TE,\TE}$ and $\Tens_{\TM,\TM}$. The reflection then gives $\Delta\vH^{(2)}$, and the last two rows are the orders three and four.

Adding the galvanic response, even through $\mathcal{P}$, takes the non-magnetic box at the base case from $46\%$ to $3.4\%$: at this depth the galvanic response carries most of its perturbation. For the permeable box, whose inductive response carries most of the perturbation, the same step goes from $6.4\%$ to $4.0\%$. For the largest object the static pair reaches only $15\%$ and $17\%$. Retaining $\eps$ in both diagonal blocks lowers these to $11\%$ and $8.1\%$: at $\eps=0.27$ the induction of the host across the object is no longer small, so $\mathcal{P}$ misjudges the host current the object diverts. It also lowers the permeable box at the base case to $1.7\%$, since its galvanic block departs there from $\mathcal{P}$ by $22\%$.

Over the near interface the reflection lowers the dipole model of the non-magnetic box from $2.9\%$ to $2.4\%$, and its order four from $0.9\%$ to $0.8\%$. For the permeable box it raises the dipole model from $1.7\%$ to $2.1\%$, the error of the omitted reflection having partly cancelled that of truncation, but lowers the order four from $1.2\%$ to $0.8\%$. Almost all of the gain comes from the block $\Ret^{(1,1)}$.

The order three lowers each column by at most $1.3$ percentage points; its terms arise from the notch alone. The largest object stays near $10\%$ until the order four, which adds the cross blocks of total degree three, carrying $13\%$ of its perturbation for the non-magnetic box and $4\%$ for the permeable one, and the diagonal blocks of total degree four. $\Delta\vH^{(4)}$ brings every column to $1.4\%$ or below. The further orders change the field by less than the error of the reference (Section~\ref{sec:s-budget}).

\begin{table}[!htb]
\caption{Relative field error over the ring, in percent, as the ingredients of the model are added, for the non-magnetic and the permeable box in the sea at $\nua=30$: at the base case, and for the largest object, $\alpha/D=0.7$ and $\eps=0.27$.}
\label{tab:ladder}
\centering
\setlength{\tabcolsep}{4pt}
\begin{tabular}{@{}lrrrr@{}}
\toprule
 & \multicolumn{2}{c}{Non-magnetic} & \multicolumn{2}{c}{Permeable}\\
\cmidrule(lr){2-3}\cmidrule(l){4-5}
Model & base & $\alpha/D=0.7$ & base & $\alpha/D=0.7$\\
\midrule
Buried MPT~\cite{ledgerCharacterisingBuriedObjects2025} & 46.2 & 40.9 & 6.4 & 15.9\\
Free-space MPT and $\mathcal{P}$ & 3.4 & 15.0 & 4.0 & 16.5\\
$\eps$-retained $\Tens_{\TE,\TE}$ and $\Tens_{\TM,\TM}$ & 2.9 & 11.4 & 1.7 & 8.1\\
With reflection, $\Delta\vH^{(2)}$ & 2.4 & 10.6 & 2.1 & 10.1\\
Order three, $\Delta\vH^{(3)}$ & 1.9 & 9.6 & 1.6 & 8.9\\
Order four, $\Delta\vH^{(4)}$ & 0.8 & 1.2 & 0.8 & 1.4\\
\bottomrule
\end{tabular}
\end{table}

\printbibliography